%% file: tb_main.tex
\documentclass[fleqn,usenatbib]{mnras}

\usepackage{mathptmx}

\usepackage[T1]{fontenc}

\DeclareRobustCommand{\VAN}[3]{#2}
\let\VANthebibliography\thebibliography
\def\thebibliography{\DeclareRobustCommand{\VAN}[3]{##3}\VANthebibliography}

\usepackage{graphicx}	
\usepackage{amsmath}	
\usepackage{amssymb}	
\usepackage{verbatim}
\usepackage[normalem]{ulem}

\usepackage{color} 
\usepackage{xcolor}

\title[Brightness temperature gradients along AGN jets]{MOJAVE -- XXII. Brightness temperature distributions and geometric profiles along parsec-scale AGN jets}

\author[E. V. Kravchenko et al.]{
E. V. Kravchenko,$^{1,2}$\thanks{E-mail: evgenia.v.kravchenko@gmail.com}
I. N. Pashchenko,$^{1}$ 
D. C. Homan,$^{3}$ 
Y. Y. Kovalev,$^{4}$ 
M. L. Lister,$^{5}$ 
\newauthor
A. B. Pushkarev,$^{6,1}$ 
E. Ros$^{4}$ 
and T. Savolainen$^{7,8,4}$ 
\\
$^{1}$Astro Space Center, Lebedev Physical Institute, Profsouznaya 84/32, Moscow 117997, Russia\\
$^{2}$Moscow Institute of Physics and Technology, Institutsky per. 9, Moscow region, Dolgoprudny, 141700, Russia\\
$^{3}$Department of Physics, Denison University, Granville, OH 43023, USA\\ 
$^{4}$Max-Planck-Institut f{\"u}r Radioastronomie, Auf dem H{\"u}gel 69, D-53121 Bonn, Germany\\ 
$^{5}$Department of Physics and Astronomy, Purdue University, 525 Northwestern Avenue, West Lafayette, IN 47907, USA\\ 
$^{6}$Crimean Astrophysical Observatory, 298409 Nauchny, Crimea\\ 
$^{7}$Aalto University Mets{\"a}ovi Radio Observatory, Mets{\"a}hovintie 114, FI-02540 Kylm{\"a}l{\"a}, Finland \\
$^{8}$Aalto University Department of Electronics and Nanoengineering, PL 15500, FI-00076 Aalto, Finland
}

\date{Accepted 2025 February 20. Received 2024 February 20; in original form 2023 May 03}
\pubyear{2025}

\begin{document}
\label{firstpage}
\pagerange{\pageref{firstpage}--\pageref{lastpage}}
\maketitle

\begin{abstract}
Radial gradients of the brightness temperatures along the parsec-scale jets of Active Galactic Nuclei (AGN) can be used to infer the energy balance and to estimate the parameter range of physical conditions in these regions. In this paper, we present a detailed study of the brightness temperature gradients and geometry profiles of relativistic jets of 447 AGN based on 15\,GHz Very Long Baseline Array observations performed between 1994 and 2019.
We used models of the jet structure using two-dimensional Gaussian components and analysed variations in their brightness temperatures and sizes along the jets.
The size of the jet components, $R$, increases with projected distance from the jet base, $r$, as $R\propto r^{1.02\pm0.03}$, i.e., typically following a conically expanding streamline and therefore indicating that the size of jet components is a good tracer of jet geometry.
The brightness temperature gradients along the jets typically follow a power-law $T_\mathrm{b} \propto r^{-2.82\pm0.07}$. 
Half of the sample sources show non-monotonic $R(r)$ or $T_\mathrm{b}(r)$ profiles and their distributions were characterised by a double power-law model.
We found at least six scenarios to explain the enhancement of the brightness temperature by a presence of inhomogeneities (shocks, jet recollimation) or curvature effects (helical structures, helical magnetic field, non-radial motion, bent jets).
Our results are consistent with the scenario that the jet features can be simplified as optically thin moving blobs.
In the sources demonstrating transition from a conical to parabolic jet shape, the gradient of the $T_{\rm b}(R)$ changes at the position of the break consistent with the model of magneto-hydrodynamic acceleration.
\end{abstract}


\input{Sect_Intro}


\section{Results}
\label{sec:results}

It is noteworthy that for many sources, the $R(r)$ and $T_{\rm b}(r)$ gradients are not monotonic and exhibit multiple extrema, breaks or repeating patterns which resemble zigzag or lightning bolt profiles.
The visual inspection shows that there are mainly three types of the brightness temperature and the component size profiles: simple power-law dependence (e.g. 1741$+$196, Fig.~\ref{fig:tbr_1741}), single break (e.g. M87, Fig.~\ref{fig:tbgrad}) and zigzag pattern (e.g. 1716$+$686 and BL~Lac). 
We discuss the most complex cases in Section~\ref{sec:complex} and the most interesting or peculiar cases in detail in Appendix~\ref{sec:individual} and show that a number of phenomena can explain the observed zoo of distributions.
We note that due to the simplified assumption we used to fit profiles (i.e., single and double power laws, equation~\ref{eq:tbbreak}), the automated fit does not always adequately describe the observational data. 
All fitted parameters are summarized in Table~\ref{tab:results}.

\begin{table*}
  \centering
  \caption{Collimation and the brightness temperature evolution parameters. For the cases which strongly favour the double power-laws model with the break, we provide corresponding parameters (see equations~\ref{eq:Rr}, \ref{eq:tbbreak} and \ref{eq:tbrfit}). Otherwise, the results by a single power-law model are given. The columns are as follows: (1) B1950 name; (2) power-law index either of a single power-law fit or of a first slope in a double power-law fit of $R(r)$; (3) power-law index of a second slope to $R(r)$; (4) Position of the break; (5)  power-law index either of a single power-law fit or of a first slope in a double power-law fit of $T_{\rm b}(r)$; (6) power-law index of a second slope to $T_{\rm b}(r)$; (7) position of the break; (8)  power-law index either of a single power-law fit or of a first slope in a double power-law fit of $T_{\rm b}(R)$; (9) power-law index of a second slope to $T_{\rm b}(R)$; (10) Separation of the brake from the jet apex. This table is available in its entirety in machine-readable forms from CDS. A portion containing five random sources is shown here for guidance regarding its form and content.}
  \label{tab:results}
  $\begin{array}{llllllllll}
  \hline
   \text{Source} & d_1 & d_2 & r_{\rm b, R} & s_1 & s_2 & r_{\rm b,T} &  \hat{s}_1 & \hat{s}_2 & r_{\rm b, TR} \\
    & & & \text{(mas)} & & & \text{(mas)} & & & \text{(mas)}\\
   (1) & (2) & (3) & (4) & (5) & (6) & (7) & (8) & (9) & (10)\\
  \hline
0003-066 & 1.03\pm0.22&0.38\pm0.03  &0.82\pm0.03  &-2.52\pm0.59 & -0.68\pm0.04 & 0.87\pm0.02&-3.60\pm0.12 & -3.11\pm0.18 & 0.69\pm0.03\\
0003+380 & 1.15\pm0.10& \ldots    & \ldots    &-4.30+0.13& \ldots    & \ldots    &-3.07\pm0.14& \ldots    & \ldots    \\
1611+343^{\ast} & 0.54\pm0.05 & 0.17\pm0.03 & 3.05\pm0.02 & -3.60\pm0.18 &-5.33\pm0.27 & 2.71\pm0.03 &-3.60\pm0.18 &-2.73\pm0.08 & 0.50\pm0.02\\
1623+569 & 1.23\pm0.06 & \ldots    & \ldots    & -5.22\pm0.51 &-3.46\pm0.28 & 1.90\pm0.30 &-3.12\pm0.19 &-2.94\pm0.22 & 0.55\pm0.03\\
1633+382^{\ast} & 1.05\pm0.06 &-0.01\pm0.07 & 1.39\pm0.09 & -3.45\pm0.13 &-5.23\pm0.28 & 2.88\pm0.03 &-3.57\pm0.09 &-3.19\pm0.17 & 0.83\pm0.02\\
  \hline
  \end{array}$
  \flushleft $^{\ast}$ Distributions have complex behaviour.
\end{table*}

\begin{table*}
    \centering
    \caption{Median values of the fitted power-law indices for different optical classes and a different range of redshifts. The columns are as follows: (1) Optical class; (2) Number of corrseponding sources; (3) Number of sources at redshift $z<0.5$; (4) Number of sources at redshift $z<0.1$; (5) Median value of the power-law index $d$ in $R\propto r^{d}$; (6) Median value of the power-law index $d$ for sources at $z<0.5$; (7) Median value of the power-law index $d$ for sources at $z<0.1$; (8) Median value of the power-law index $s$ in $T_\mathrm{b} \propto r^{s}$; (9) Median value of the power-law index $s$ for sources at $z<0.5$; (10) Median value of the power-law index $s$ for sources at $z<0.1$; (11) Median value of the power-law index $\hat{s}$ in $T_\mathrm{b} \propto R^{\hat{s}}$; (12) Median value of the power-law index $\hat{s}$ for sources at $z<0.5$; (13) Median value of the power-law index $\hat{s}$ for sources at $z<0.1$.}
    \label{tab:fit_results}
    \begin{tabular}{lrrrccccccccc}
        \hline
        Opt. class & $N$& $N_\mathrm{z<0.5}$ & $N_\mathrm{z<0.1}$& $d$ & $d_\mathrm{z<0.5}$ & $d_\mathrm{z<0.1}$ & $s$ & $s_\mathrm{z<0.5}$ & $s_\mathrm{z<0.1}$ & $\hat{s}$ & $\hat{s}_\mathrm{z<0.5}$ & $\hat{s}_\mathrm{z<0.1}$\\
        (1) & (2) & (3) & (4) & (5) & (6) & (7) & (8) & (9) & (10) & (11) & (12) & (13)\\
        \hline
        Quasars        &271&32&\ldots& 0.95 & 0.92 &\ldots& $-2.83$ & $-3.11$ &\ldots & $-2.99$ & $-3.13$ &\ldots\\
        BL~Lac objects &135&53&17    & 1.12 & 1.08 & 0.97 & $-2.90$ & $-2.66$ &$-2.57$& $-2.59$ & $-2.56$ & $-2.45$\\
        Radio galaxies &25 &23&18    & 0.77 & 0.84 & 0.81 & $-2.59$ & $-2.77$ &$-2.68$& $-2.83$ & $-2.90$ & $-3.03$\\
        NLSY1          & 5 & 2&\ldots& 1.16 & 0.87 &\ldots&$-3.04$ & $-2.84$ &\ldots & $-2.78$ & $-2.97$ &\ldots\\
        Unidentified   &11 &\ldots&\ldots& 0.91 &\ldots&\ldots& $-2.69$ & \ldots  &\ldots & $-2.68$ & \ldots  &\ldots\\
        \hline
        All            &447&\ldots&\ldots& 1.02 &\ldots&\ldots& $-2.82$ &\ldots&\ldots& $-2.87$ &\ldots&\ldots\\
        \hline
    \end{tabular}
\end{table*}

\subsection{Jet geometry}
\subsubsection{Single power-law fits}

In Fig.~\ref{fig:s_grad_simple}, we show the dependence of the jet component size versus the radial distance of the 15\,GHz core for the sources which are better fit by a single power-law dependence $R\propto r^{d}$.
The resultant distribution of the $d$-indices of all sources is shown in Fig.~\ref{fig:fit_R_r}.
The median $d$-values for quasars, BL~Lacs and NLSY1 are $\approx1$ (Table~\ref{tab:fit_results}), indicating that their jet shapes are close to conical, i.e. expanding freely.
Radio galaxies are characterised by lower values of $d$, suggesting that their jets are more collimated and follow a quasi-parabolic streamline. 
The Anderson-Darling (AD) test rejects the null hypothesis that the quasar and BL~Lac distributions of $d$ are drawn from the same distribution (with p-value $<0.001$).

\begin{table*}
	\centering
	\caption{Known sources with a geometry transition at 15\,GHz-VLBA scales and their collimation parameters. The columns are as follows: (1) B1950 name; (2) other name; (3) optical class; (4) redshift; (5) and (6) positions of the break defined from $R(r)$ and $T_{\rm b}(r)$, respectively; (7) location of the break defined in other studies, given in (8). }
	\label{tab:transjets}
	\begin{tabular}{llllllll} 
	\hline
	 Source & Alias & Opt & Redshift & $r_{\rm b, R}$ & $r_{\rm b, T}$ & $r_{\rm b, ref.}$ & Reference\\
	  &  & class & & (mas) & (mas) & (mas) & \\
	 (1) & (2) & (3) & (4) & (5) & (6) & (7) & (8)\\
	\hline
 0111$+$021 &  UGC 00773 & B &0.047 &2.0$\pm$1.4$^{\dag}$  & 1.79$\pm$0.07 & 2.5$\pm$0.3 &\citet{2020MNRAS.495.3576K}\\
 0238$-$084 &  NGC 1052  & G &0.005 &3.1$^{\dag}$& 2.9$^{\dag}$ & 2.9$\pm$0.6&\citet{2004AA...426..481K, 2020MNRAS.495.3576K}\\
 0321$+$340 &  1H 0323+342& N &0.061 & 1.46$\pm$0.16 & 4.66$\pm$1.1 & $\thicksim7$ &\citet{2018ApJ...860..141H}\\
 0415$+$379 &  3C 111     & G&0.0491 &7.81$\pm$0.03& 4.18$\pm$0.02 & 7.2$\pm$0.2&\citet{2020MNRAS.495.3576K}\\
 0430$+$052 &  3C 120     &G &0.033 & 12.47$\pm$0.10 & 5.0$\pm$0.4 & 2.7$\pm$0.4&\citet{2020MNRAS.495.3576K}\\
 0815$-$094 & TXS 0815-094&B &\ldots   &1.20$\pm$0.16$^{\dag}$& 1.25$\pm$0.15 & 1.4$\pm$0.3&\citet{2020MNRAS.495.3576K}\\
 1133$+$704 &  Mrk 180 & B&0.045278 &1.0$\pm$0.3$^{\dag}$& 1.17$\pm$0.03 & 1.39$\pm$0.09&\citet{2020MNRAS.495.3576K}\\
 1142$+$198 & 3C 264   & G & 0.022 & 1.67$\pm$0.14& 3.6$\pm$0.3 & $\thicksim4$ & \citet{2019AA...627A..89B}\\
 1226$+$023 & 3C 273  & Q    &0.1576 & 4.7$\pm$0.2 &5.22$\pm$0.03& $\thicksim13$&\citet{2022ApJ...940...65O}\\
 1514$+$004 & PKS 1514+00 & G &0.052 & 4.8$\pm$0.4$^{\dag}$ & 1.42$\pm$0.09 & 3.1$\pm$0.2&\citet{2020MNRAS.495.3576K}\\
 1637$+$826 & NGC 6251  & G    &0.024 & 3.4$\pm$0.2 & 3.67$\pm$0.05 & 1.9$\pm$0.3 & \citet{2020MNRAS.495.3576K}\\
 1807$+$698 & 3C~371  & B    &0.051 & 1.03$\pm0.08^{\dag}$ & 0.80$\pm$0.01 & 1.5$\pm$0.3 & \citet{2020MNRAS.495.3576K}\\
 1928$+$738 & 4C$+$73.18 & Q & 0.302 & 3.24$\pm$0.11 & 5.17$\pm$0.05 & 4.72$\pm$0.72 & \citet{2024AA...688A..94Y}\\
 1957$+$405 & Cygnus~A  & G  &0.0561 &1.68$\pm$0.09 & 2.72$\pm$0.03$^{\dag}$ & $\thicksim2$; $\thicksim5$ & \citet{2016AA...585A..33B}, \citet{1998AA...329..873K}\\
 2013$+$370 & TXS 2013+370& Q &0.859 & 0.5$\pm0.2^{\dag}$  & 0.91$\pm$0.14 & $\thicksim0.5$&\citet{2020AA...634A.112T}\\
 2200$+$420 & BL Lac  & B    &0.0686 & 2.43$\pm$0.02 & 1.33$\pm$0.03 &2.45$\pm$0.10&\citet{2020MNRAS.495.3576K}\\ 
 \hline
 \end{tabular}
 \flushleft
 $^{\dag}$Single power-law model fits the data better; fit with the break is not significant. \\
 $^{\ddag}$Position of the break defined in other works is not seen in the MOJAVE data.  
\end{table*}

The typical range of distances where the jet maintains a parabolic streamline is found to be within $\thicksim10^6~r_\mathrm{g}$ of the core \citep[c.f.][]{2020MNRAS.495.3576K}.
The MOJAVE sources with different optical classes have different redshift distributions (median redshifts are 1.05 for quasars, 0.27 for BL Lac objects and 0.06 for radio galaxies; see Fig.~\ref{fig:z_distr}), thus linear resolution may have an impact on the derived parameters.
We investigated if the jets in the radio galaxies appear to be more collimated because of probed shorter projected linear distances and, therefore, have resolved jets compared to those in more distant BL~Lacs and quasars.
We considered sources in a ranges of redshift (i) $z<0.5$, chosen so that the number of sources in sub-classes is approximately the same, and (ii) $z<0.1$, for which linear resolution of 15\,GHz VLBA observations is better than 1~pc enabling detecting the transition zone \citep{2020MNRAS.495.3576K}. The resultant median values of the power-law indices are summarised in Table~\ref{tab:fit_results}.
For (i) case, the AD-test does not reject the null hypothesis that quasars and BL~Lac objects $d$ are drawn from the same distribution; the corresponding p-value is 0.057.
For (ii) case, the AD-test shows that BL~Lac objects and radio galaxies are drawn from the same distribution.

The distribution of the fitted parameter $r_0$ in equation~\ref{eq:Rr}, which is assumed to be the separation of the 15\,GHz VLBI core from the true jet base, has a median of 109\,$\mu$as.
This distance is connected to the core shift $\Delta r$ between 15 and 8~GHz as $r_0 \approx 1.1 \Delta r$ assuming a conical jet and, therefore, $k_\mathrm{r}=1$ in the core position frequency dependence $r \propto \nu^{-1/k_{\rm r}}$.
Thus, the derived $r_0$ value roughly agrees with the core shift estimates measured between 8 and 15\,GHz \citep[median=128~$\mu$as, 160 sources,][]{2012A&A...545A.113P}.
However, for the collimated jets, $k_{\rm r}$ could be less than unity \citep{Porth_etal11, 2024MNRAS.528.2523N}. Moreover, \cite{2022MNRAS.509.1899N} showed that for the accelerating jet with a toroidal magnetic field in the jet region with $\Gamma < 1/\theta$ the exponent $k_r = 4d/3$.  
We estimated $k_{\rm r}$ comparing the median separation of the 15\,GHz VLBI core obtained from our $R(r)$ fits with the value expected from the median core shift measured between 8 and 15\,GHz by \citet{2012A&A...545A.113P}.
This yields $k_{\rm r} = 0.83\pm0.03$, which favours the scenario of an accelerating jet in the core region and corresponds to $d = 0.62\pm0.02$.
It is noteworthy that the bias of the core shift measurements found in \citet{2020MNRAS.499.4515P} does not change this result significantly, because both the measured core shifts and the positions of the VLBI cores are biased upward. This result is consistent with \citet{2017ApJ...834...65A}, who found an indication of a quasi-parabolic geometry in the core regions of 56 radio-loud AGN using multi-frequency core size data. The same conclusion was made by \citet{2018A&A...614A..74A} from the analysis of the core shapes at 15 and 43\,GHz in the complete S5 polar cap sample. 
\citet{2022MNRAS.509.1899N} obtained the same result using the universal MHD acceleration profile and employing the speeds measured in the VLBI core during the radio flares in 11 radio-loud AGN by \citet{2019MNRAS.486..430K}.

\begin{figure*}
\centering
    \includegraphics[width=0.5\columnwidth]{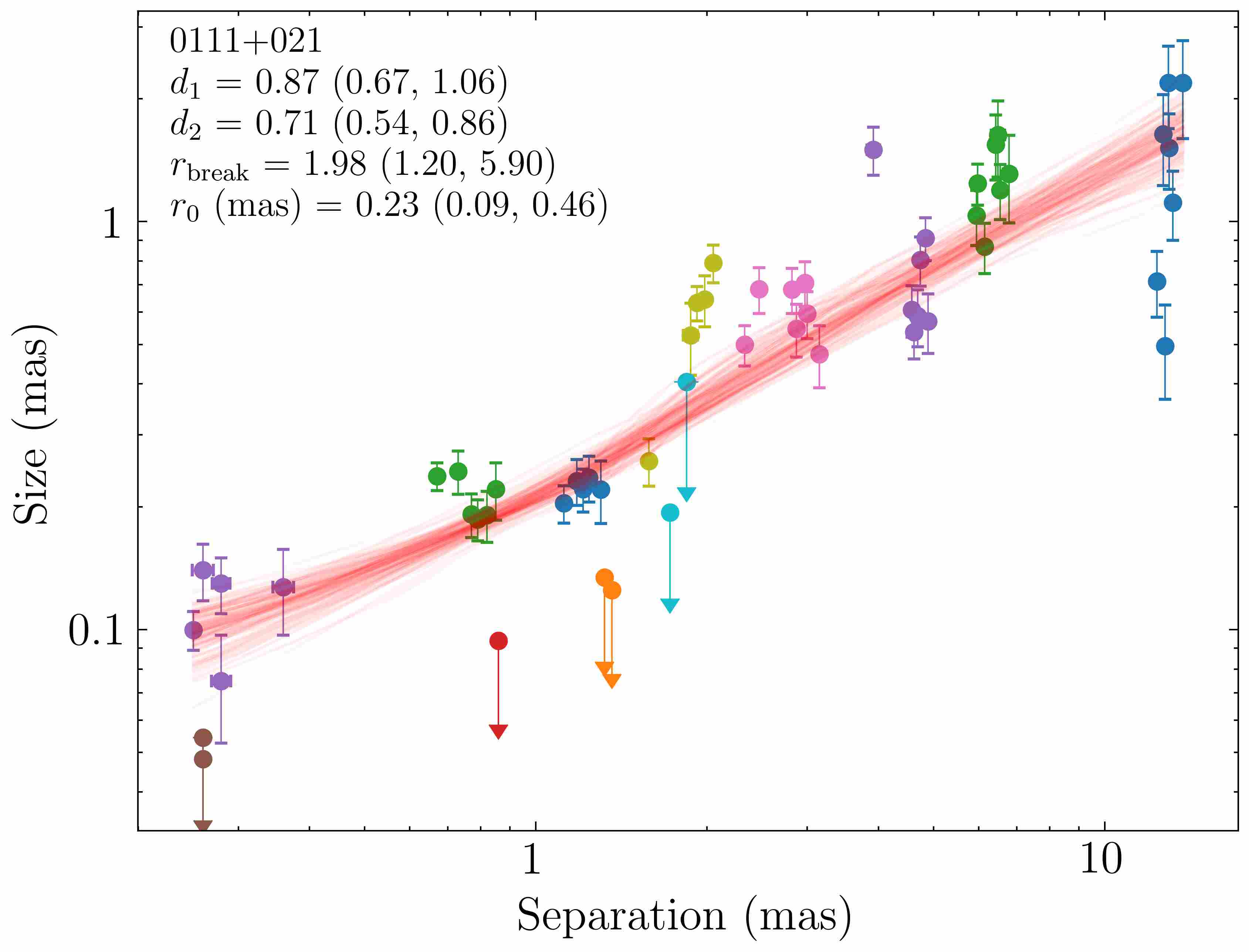}
    \includegraphics[width=0.5\columnwidth]{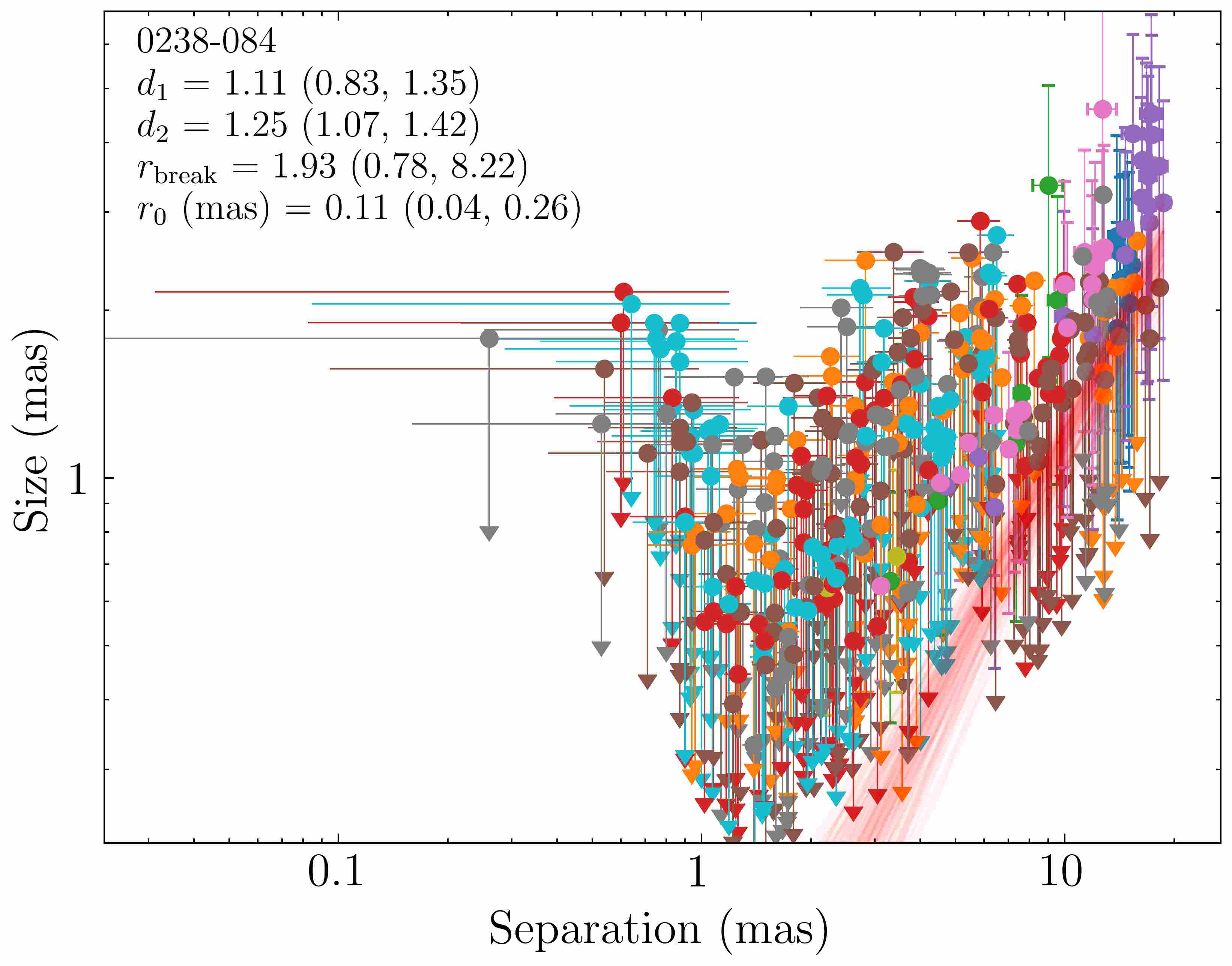}
    \includegraphics[width=0.5\columnwidth]{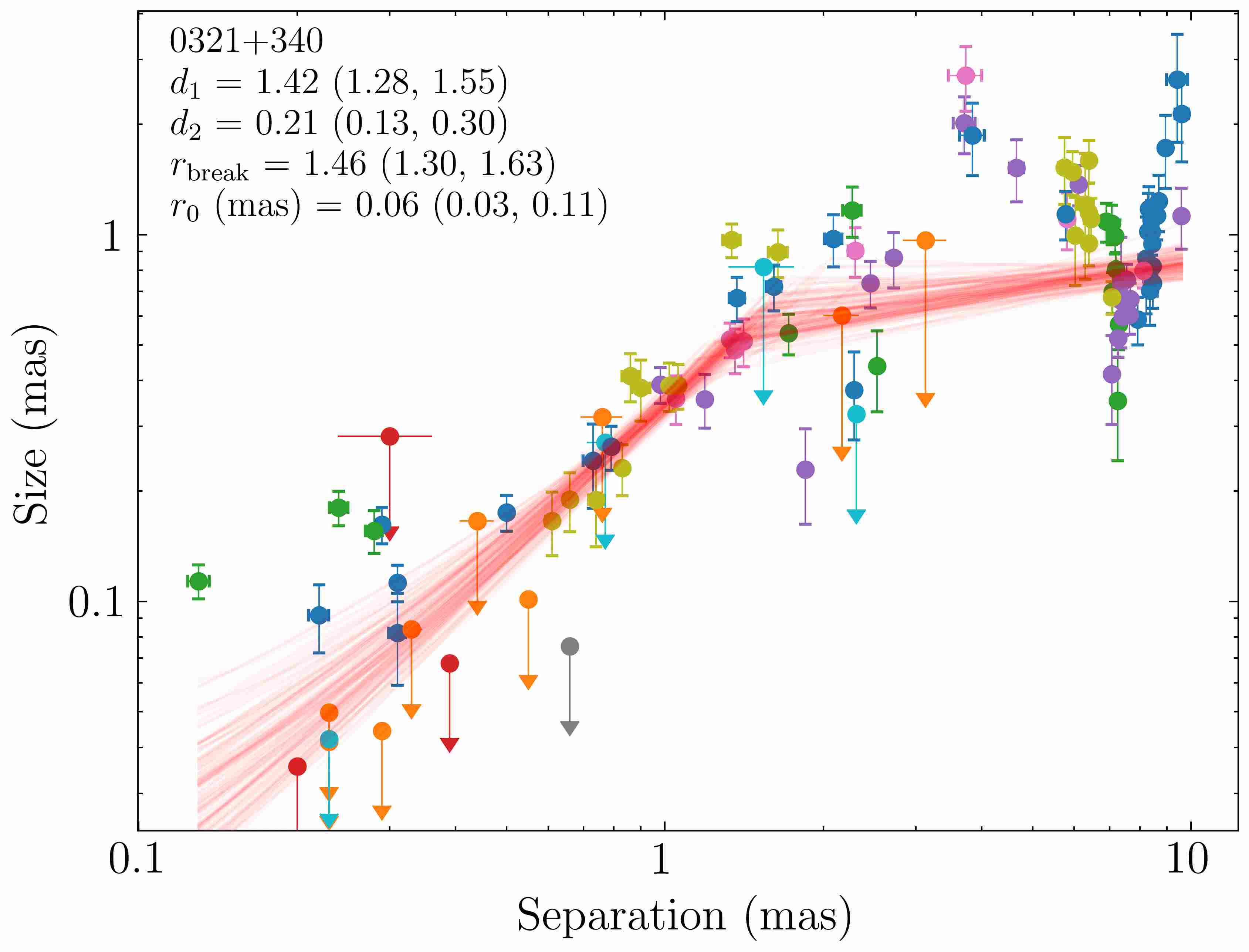}
    \includegraphics[width=0.5\columnwidth]{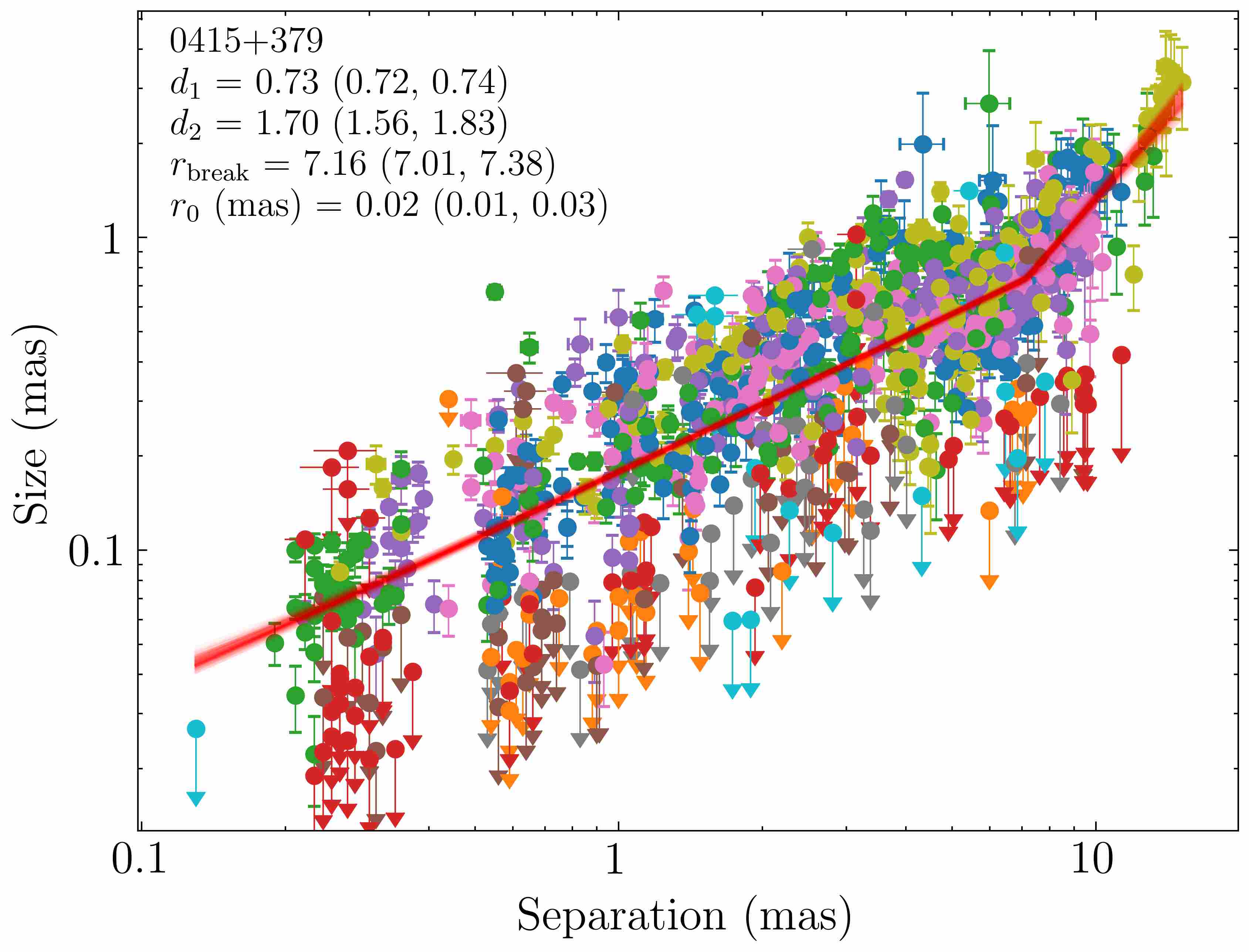}\\
    \includegraphics[width=0.5\columnwidth]{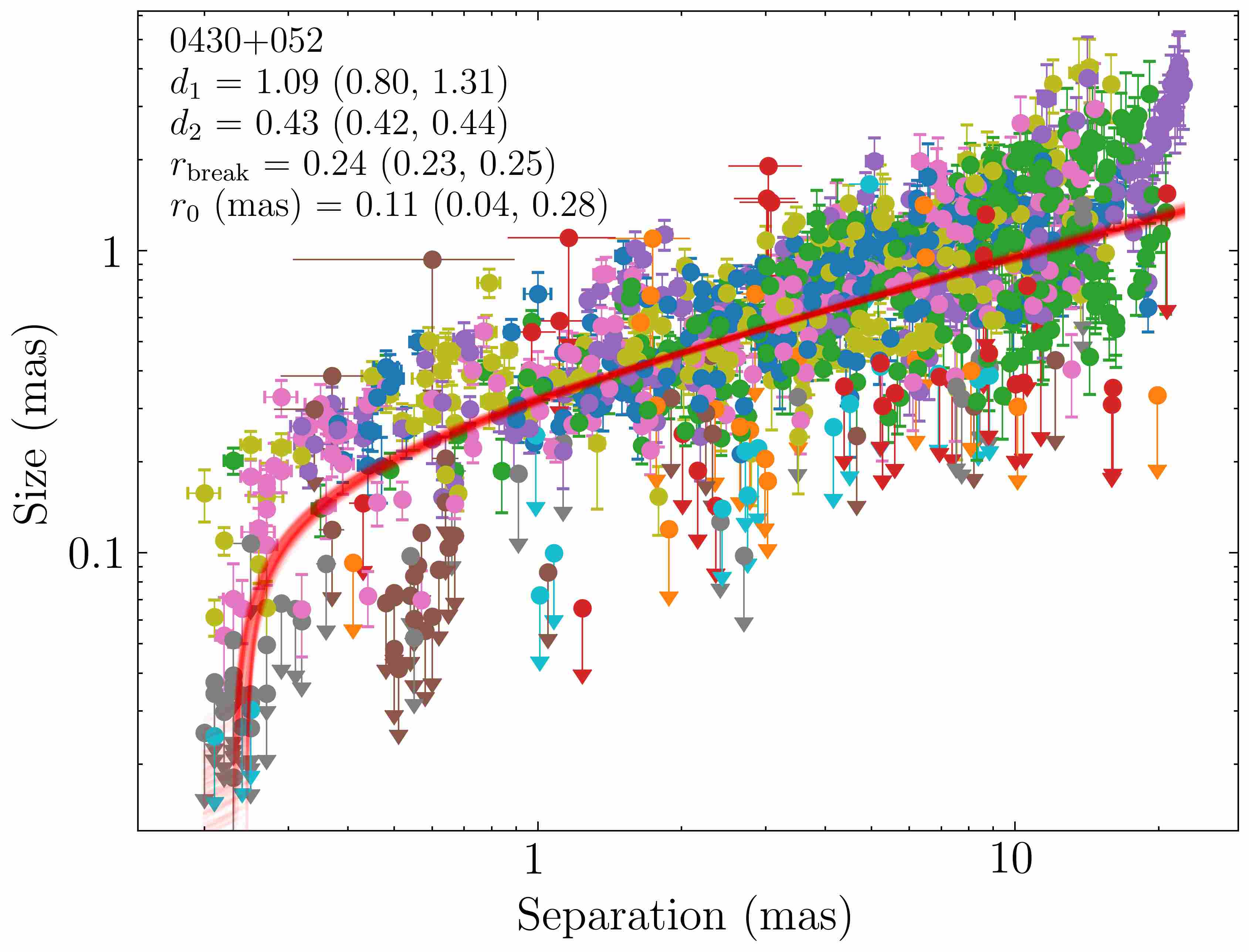}
    \includegraphics[width=0.5\columnwidth]{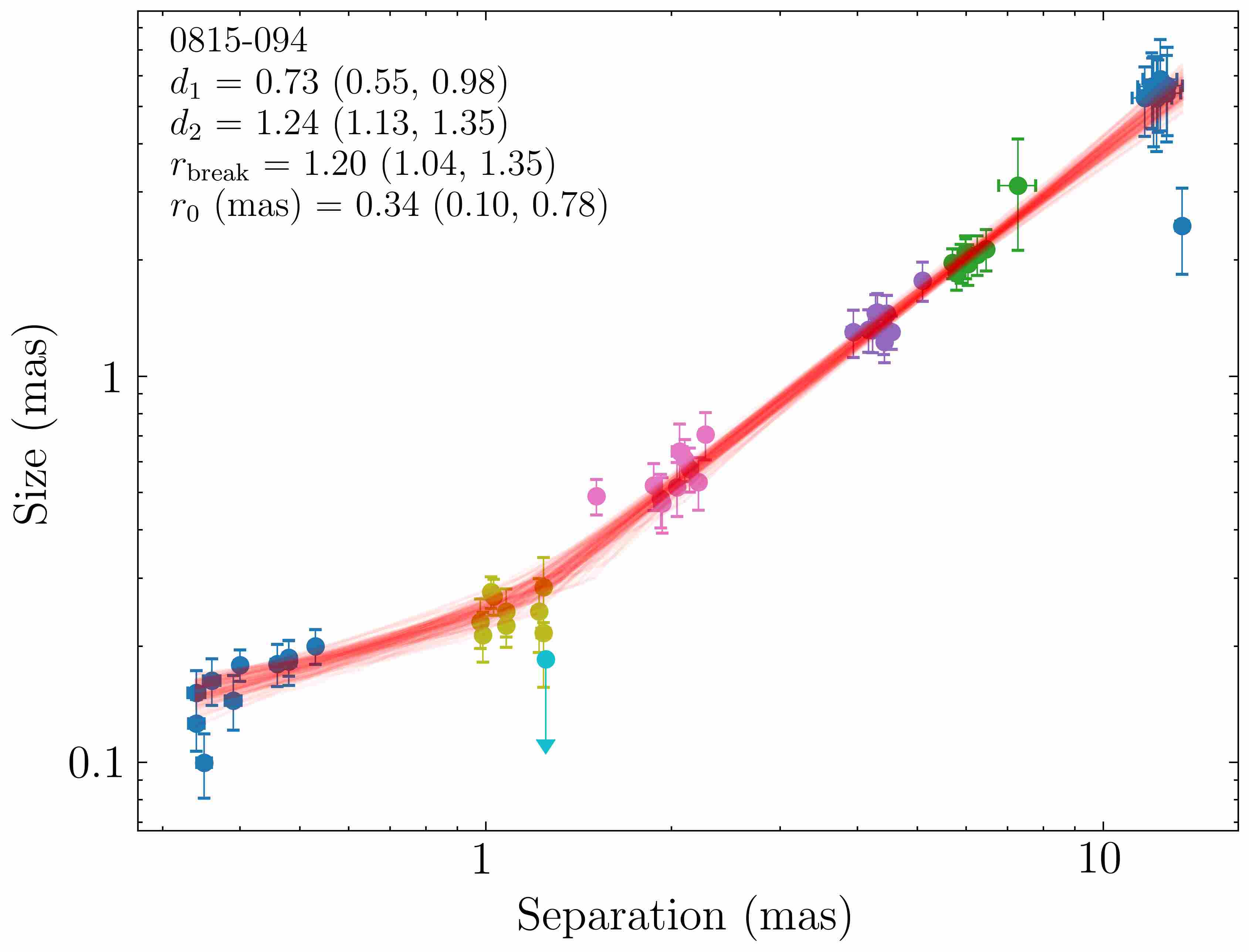}
    \includegraphics[width=0.5\columnwidth]{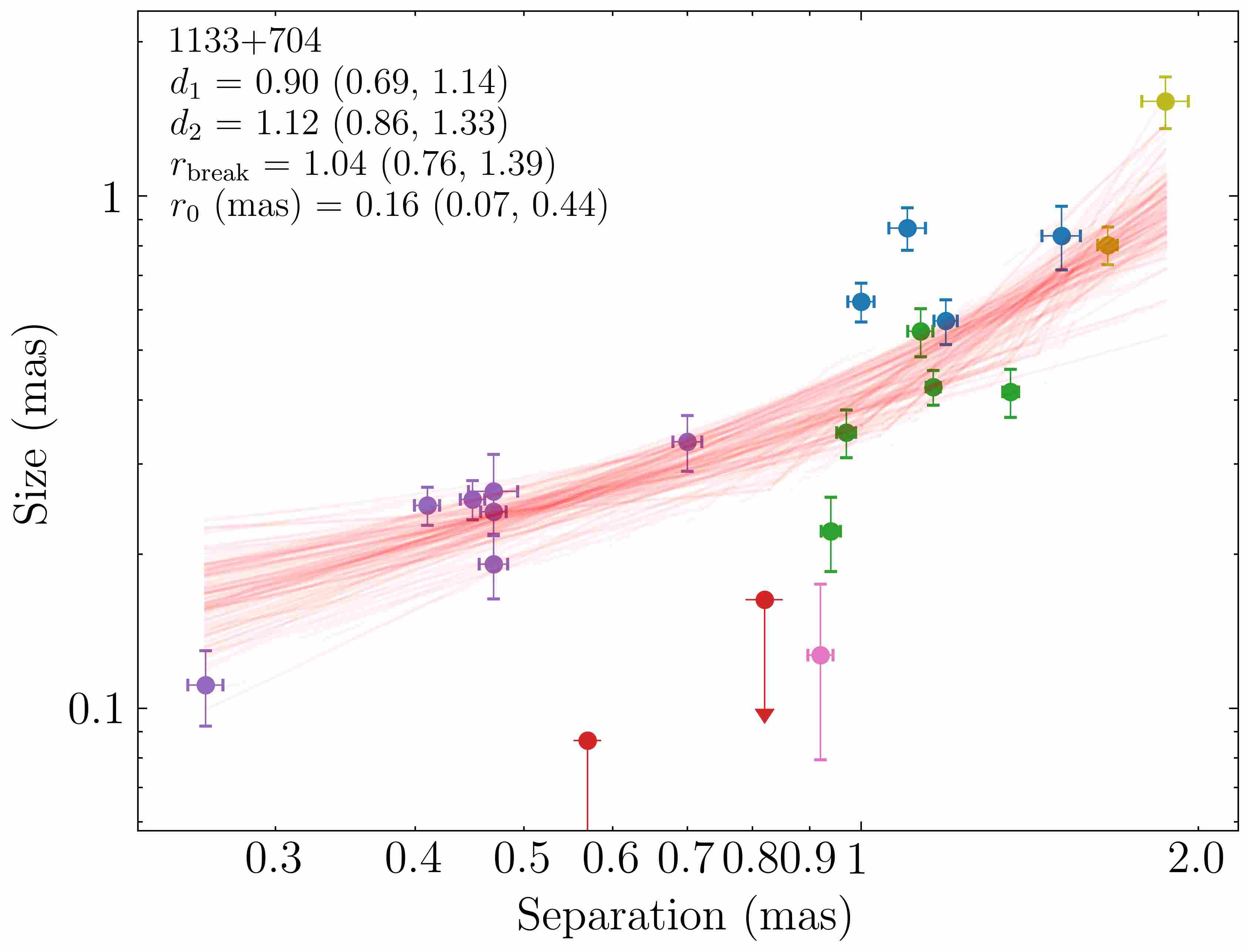}
    \includegraphics[width=0.5\columnwidth]{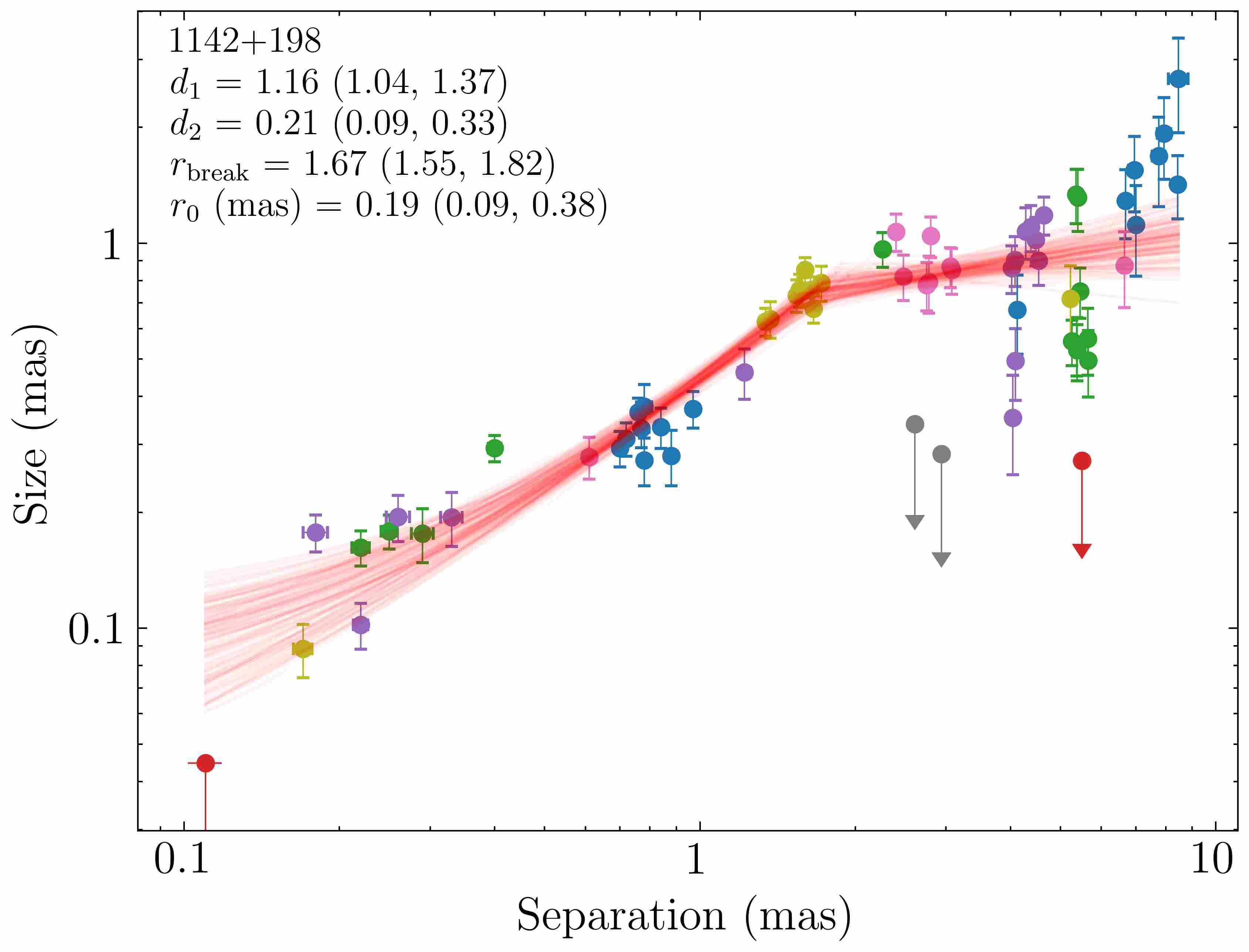}\\
    \includegraphics[width=0.5\columnwidth]{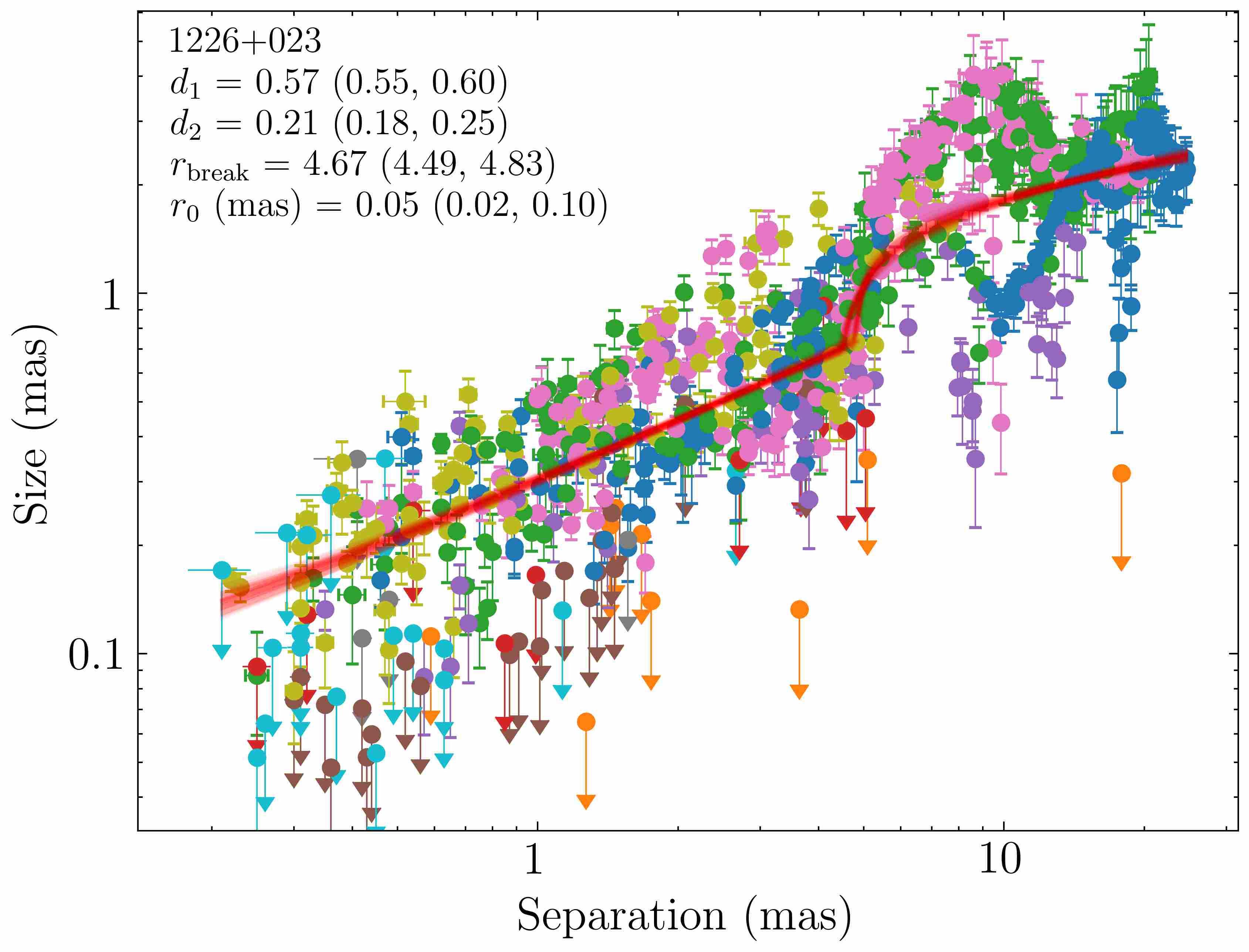}
    \includegraphics[width=0.5\columnwidth]{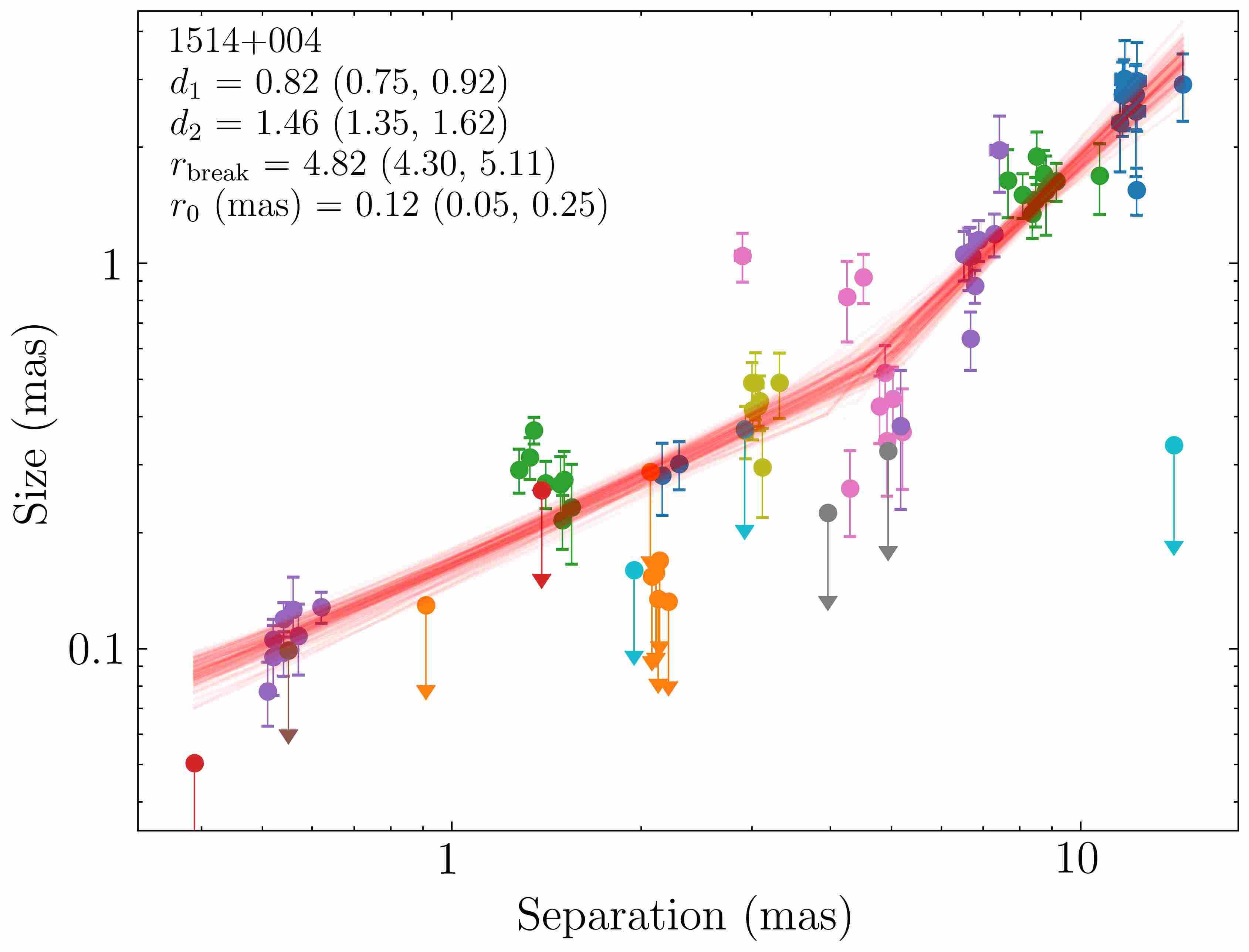}
    \includegraphics[width=0.5\columnwidth]{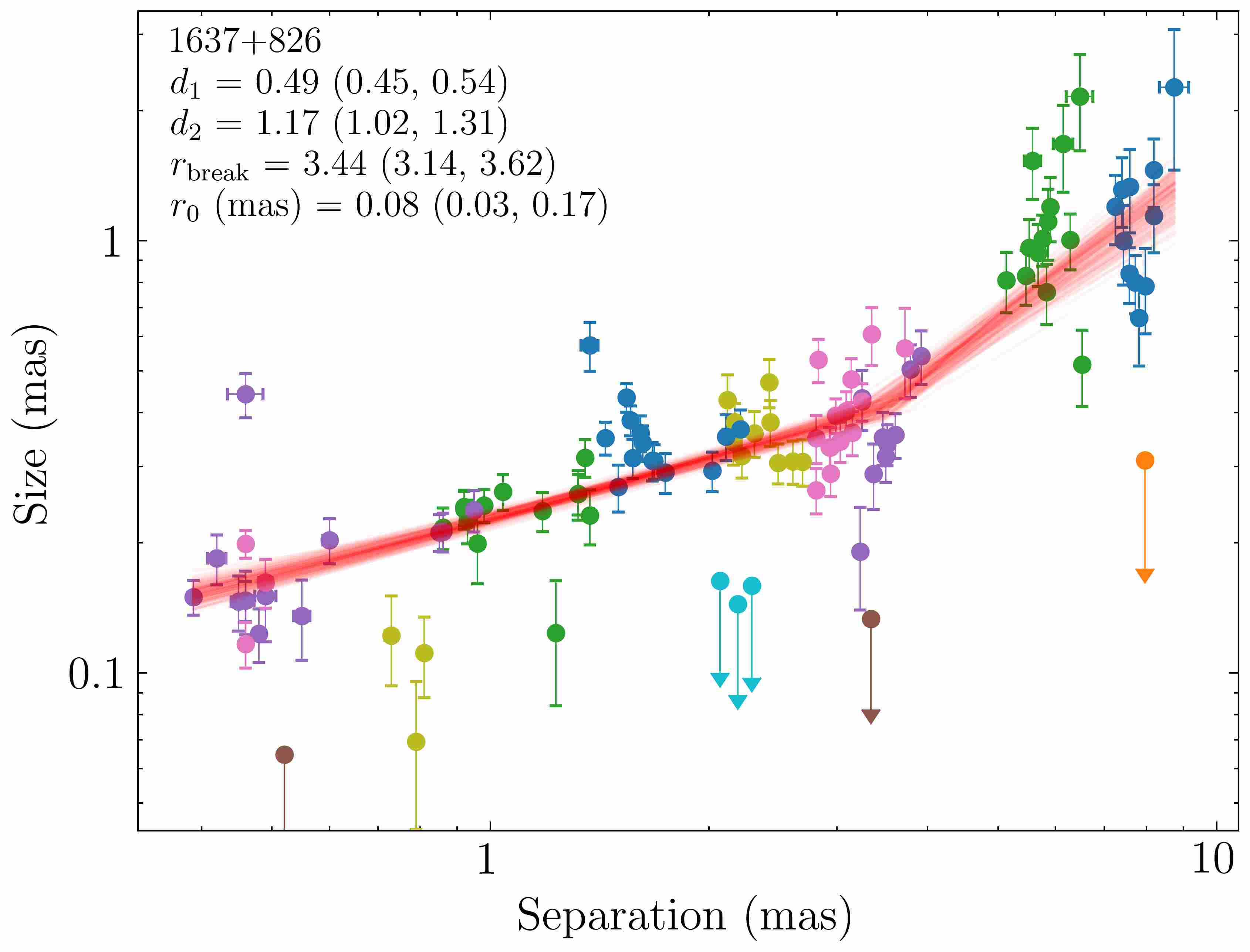}    \includegraphics[width=0.5\columnwidth]{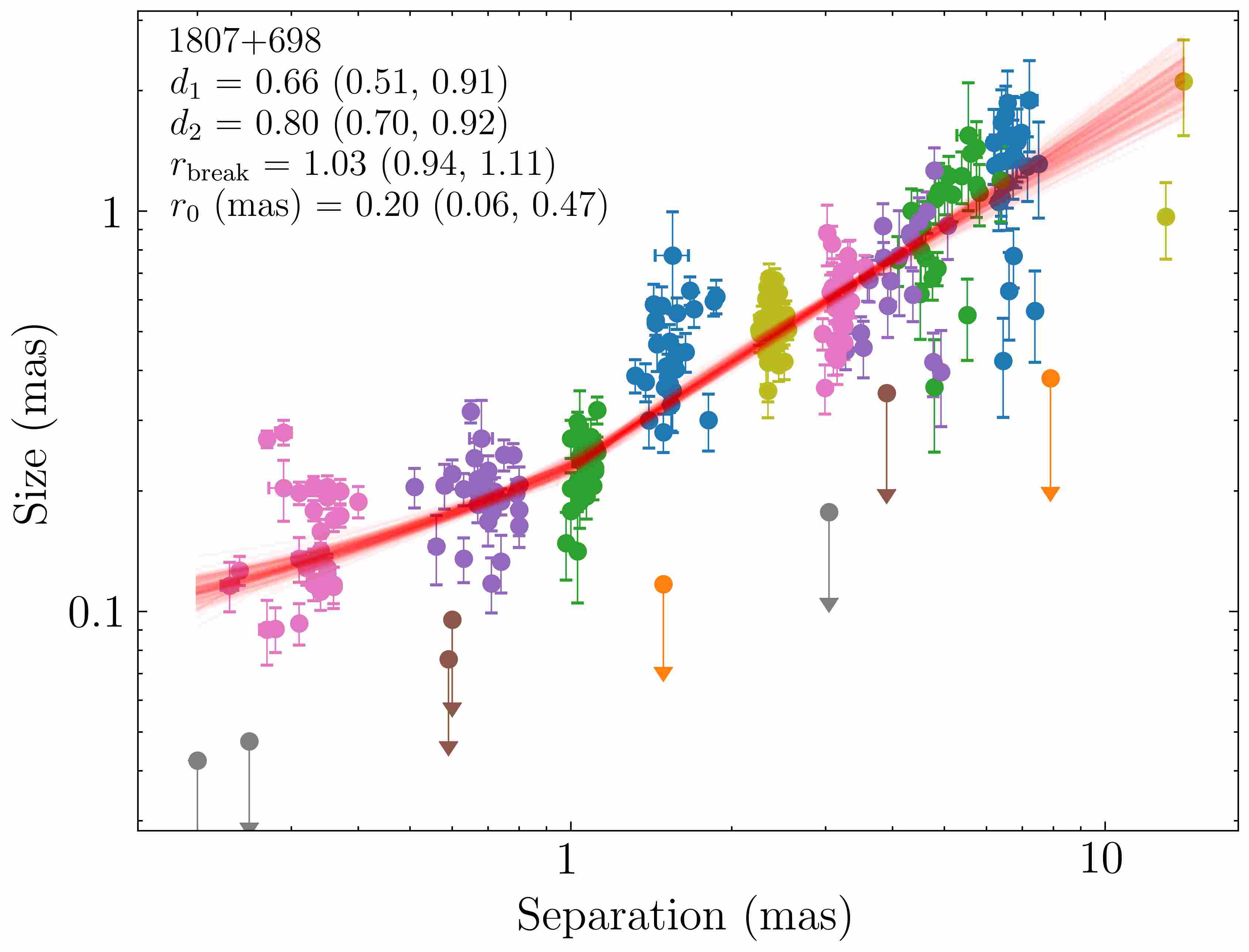}\\
    \includegraphics[width=0.5\columnwidth]{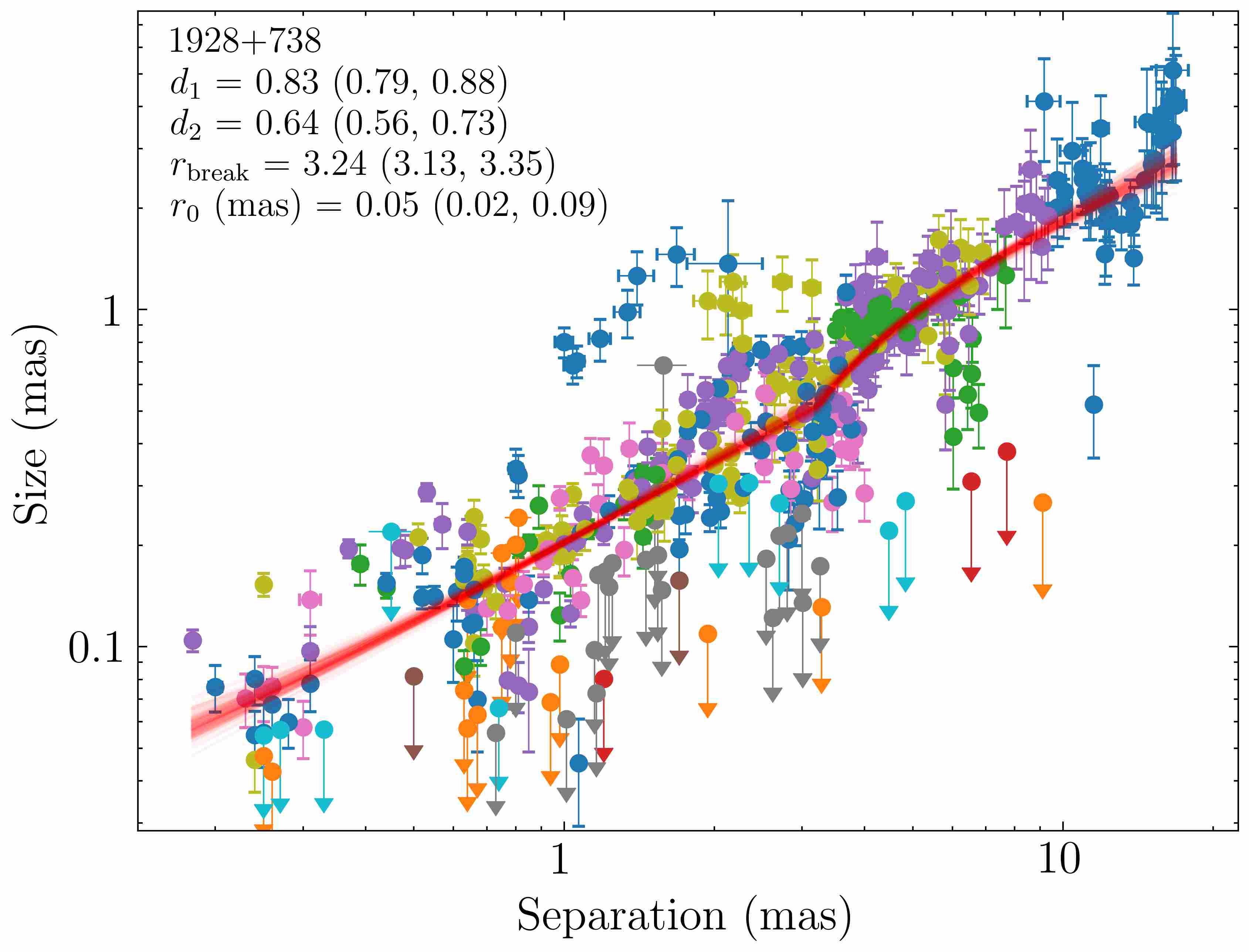}
    \includegraphics[width=0.5\columnwidth]{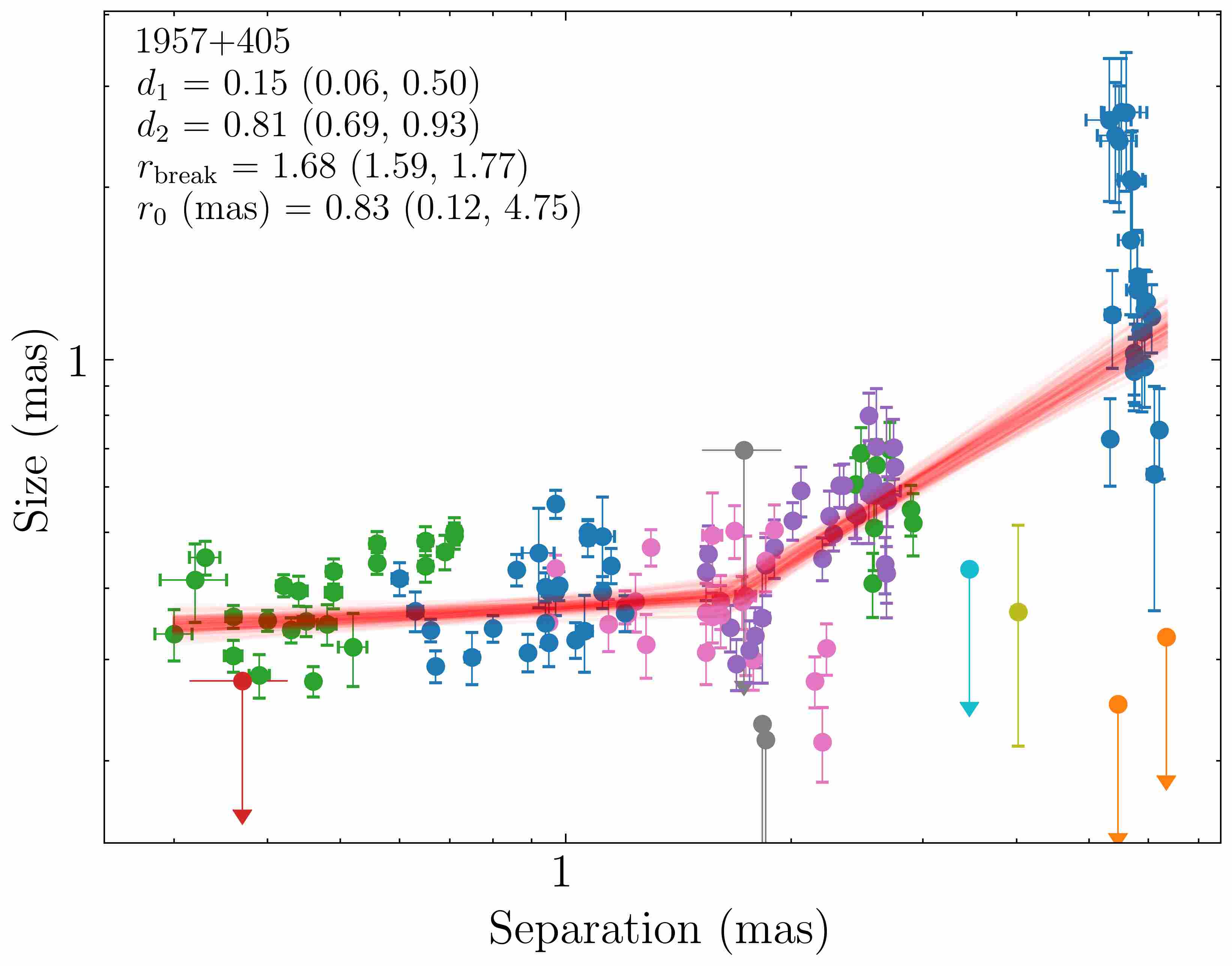}
    \includegraphics[width=0.5\columnwidth]{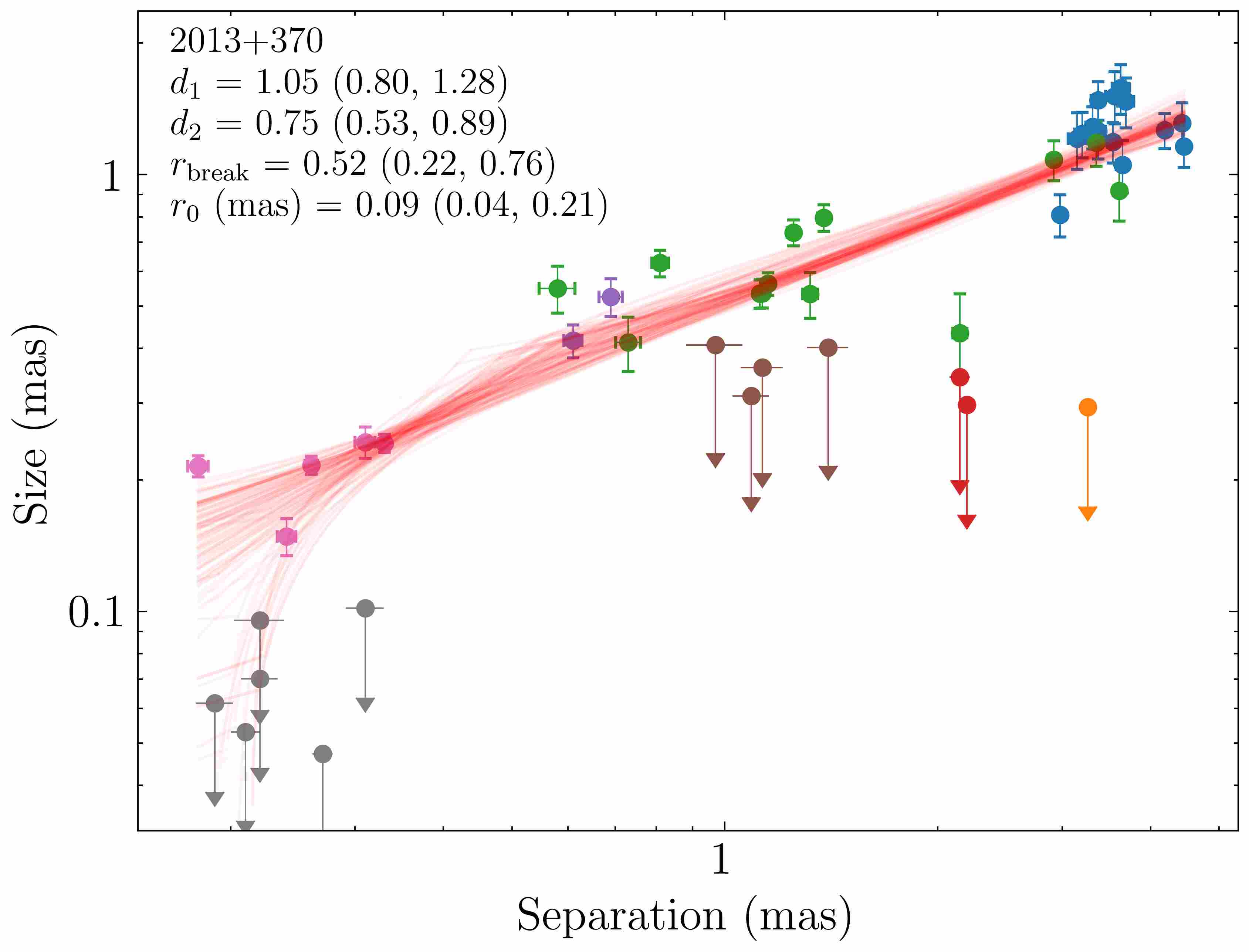}
    \includegraphics[width=0.5\columnwidth]{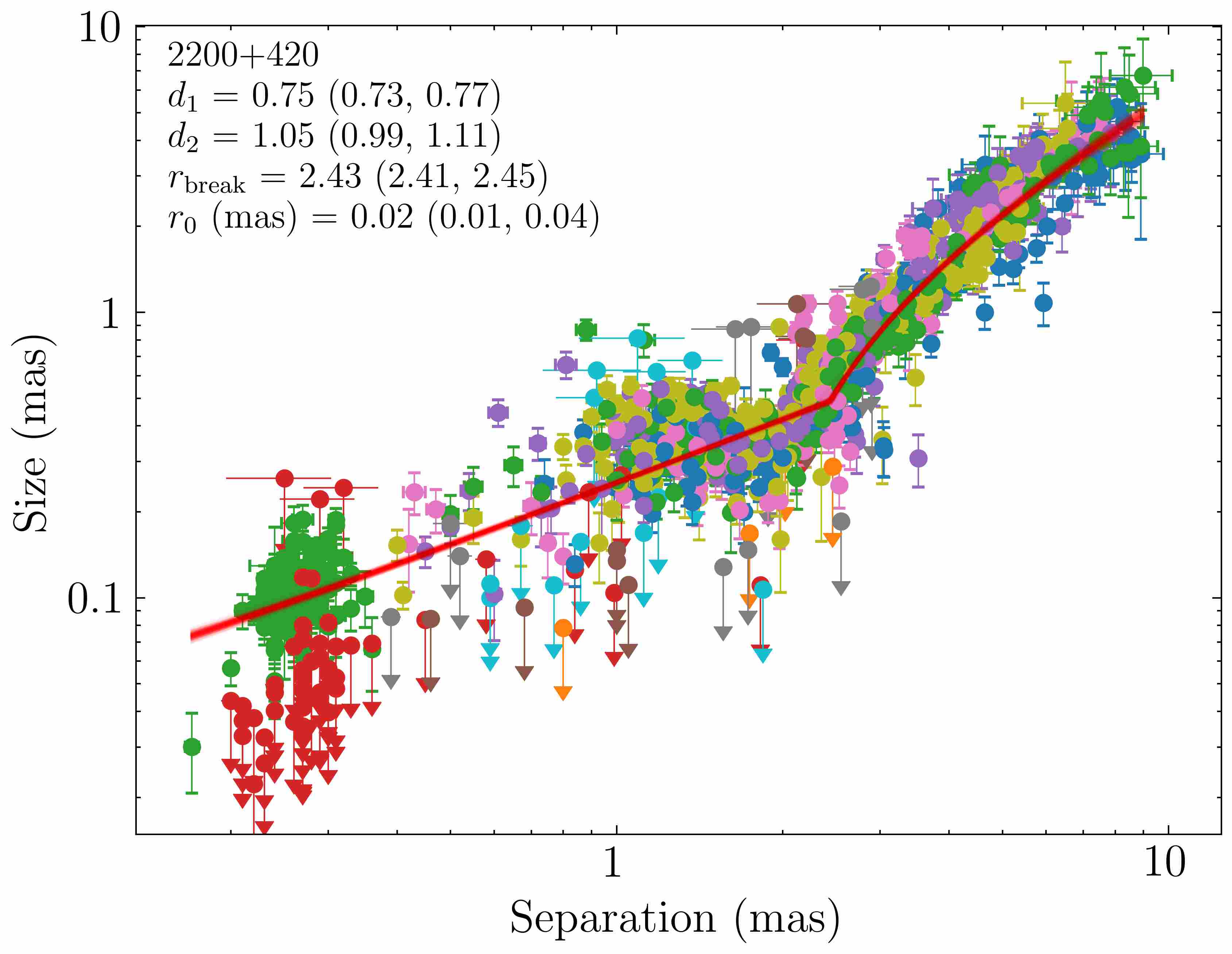}\\
    \caption{FWHM of the jet components vs the radial distance from the 15\,GHz core for the detected jet shape transition in previous studies at VLBA scales (see Table~\ref{tab:transjets}). Different colours denote individual jet components. The arrows indicate the upper limits, equation~\ref{eq: lobanov_limit}. The solid red lines indicate the result of a fit with a double power-law model.}
    \label{fig:jet_breaks}
\end{figure*}

\subsubsection{Geometry with a break}
\label{sec:rd_break}

We find 117 sources whose dependence of the jet width on the projected distance from the core is non-linear and is better fit by a model with a break ($\Delta (\log{~Z})>5$, Sec.~\ref{sec:parametrization}). The $R(r)$ profiles for these sources are shown in Fig.~\ref{fig:srgrad}, and the fitted parameters are given in Table~\ref{tab:results}.
The corresponding distribution of the power-law indices $d_1$ and $d_2$ is given in Appendix~\ref{sec:apndx_distr_break}.
In many cases with the break, the behaviour of the jet component size around $r_{\rm break}$ is not monotonic and is accompanied by a sharp compression and then rapid expansion. 
The same behaviour was observed before at the position of the jet shape transition accompanied by the formation of a shock, for example, in 1H~0323$+$342  \citep{2018ApJ...857L...6D} and in Cygnus~A \citep{1998AA...329..873K, 2016AA...585A..33B}.
The interaction models of a relativistic jet with an external medium indeed predict very strong reduction in the transverse size of the shocked jet region accompanied by subsequent rapid re-expansion sideways \citep{2008IJMPD..17.1603L, 2009ApJ...699.1274B, 2018A&A...609A.122B}.
Therefore, such a $R(r)$ profile may point to the shocked region. We discuss the locations in the break positions in the jet further in Section~\ref{sec:drbreak}.

Searches for a change in the jet geometry using the evolving size of the jet features with their distance from the core was also conducted by \citet{2017A&A...606A.103H} for 161 AGNs (part of our sample). \citet{2017A&A...606A.103H} found a change in the jet geometry in 36 sources.
In 19 out of these cases, we confirm the broken power-law profiles, yielding a median difference of 0.34\,mas between their and our estimates of $r_{\rm break}$.
The discrepancy between the results obtained by \citet{2017A&A...606A.103H} and in this paper, is that only epochs prior to 2016 were employed in their analysis. Also, they used a model $R(r)\propto a(r)^{d}$, assuming the core has a zero size (i.e. skipping the $r_0$ in Equation~\ref{eq:Rr}). That could bias the estimates \citep{2020MNRAS.495.3576K}.

Among the 16 sources where the geometry transition was previously detected on the VLBA scales (listed in Table~\ref{tab:transjets} and shown in Fig.~\ref{fig:jet_breaks}), we confirmed 12 cases which show the position of the break to be consistent with those determined in the above mentioned papers.
We attribute a small discrepancy between the estimates of the positions of the break in our study and other studies to a complex profile $R(r)$ at the location of $r_{\rm break}$.
In the case of 1H~0323$+$342 and 3C~273, this discrepancy is significantly larger due to considerably complex profiles.
Two cases show the preferred models with a single slope, however the broken power-law fit yields a break position consistent with those published.
In 3C~120, the position of the break differs from the one defined in \citet{2020MNRAS.495.3576K}. We suggest that this is due to a large spread in the position angle of different components located at the same jet position we see in multi-epoch observations. 
In section~\ref{sec:complex}, we show that there is a number of such sources where significant position angle (PA) variations of individual jet components are observed \citep{2013AJ....146..120L}. When the individual components in $R(r)$ and the $T_{\rm b}(r)$ distributions are highlighted, the break is clearly visible (see Fig.~\ref{fig:pa}).

Below, we show that the non-monotonic distributions can be produced by a number of phenomena other than jet recollimation, which yields a zoo of different profiles.

\begin{figure*}
\centering
    \includegraphics[width=0.67\columnwidth]{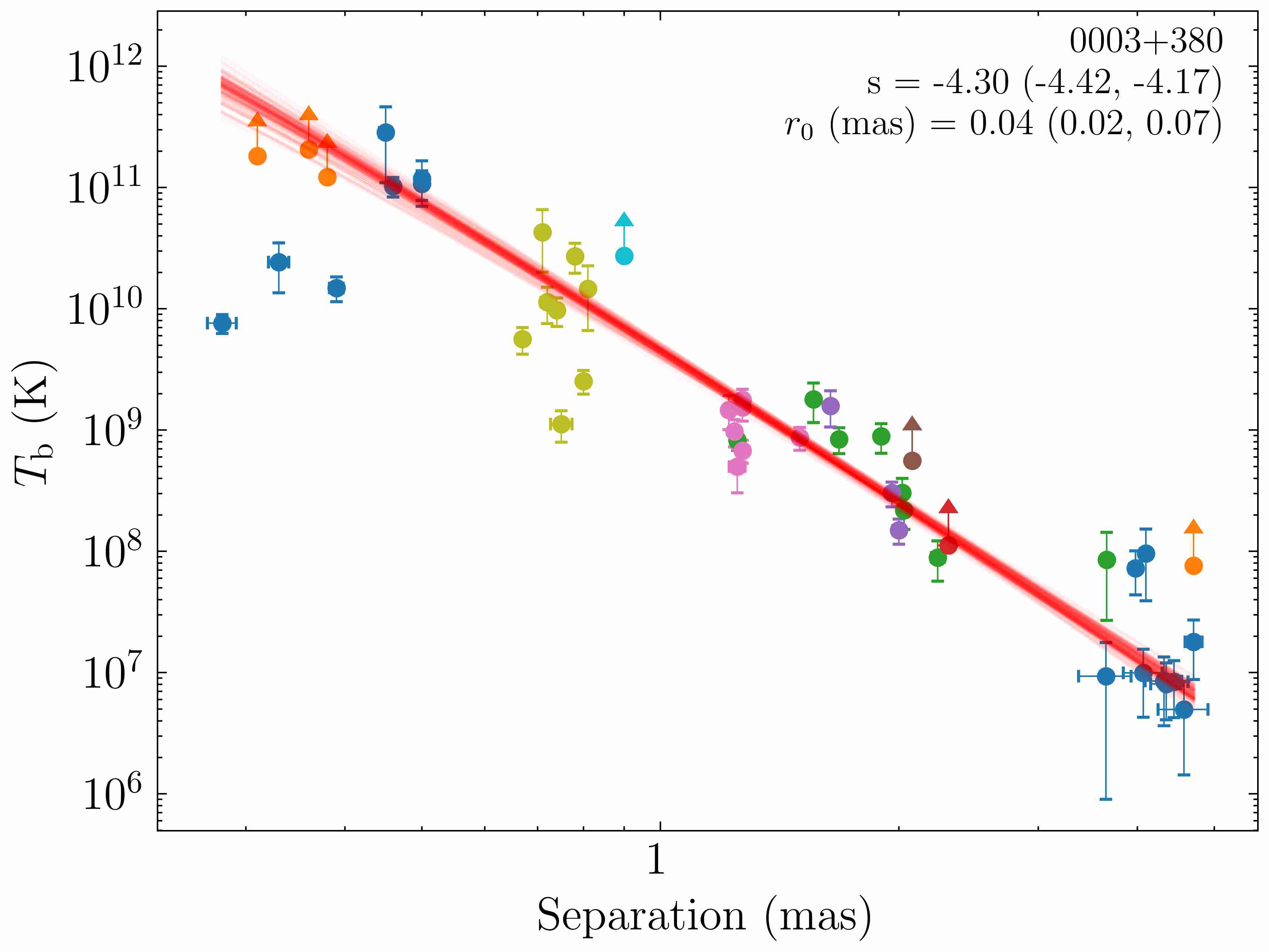}
    \includegraphics[width=0.67\columnwidth]{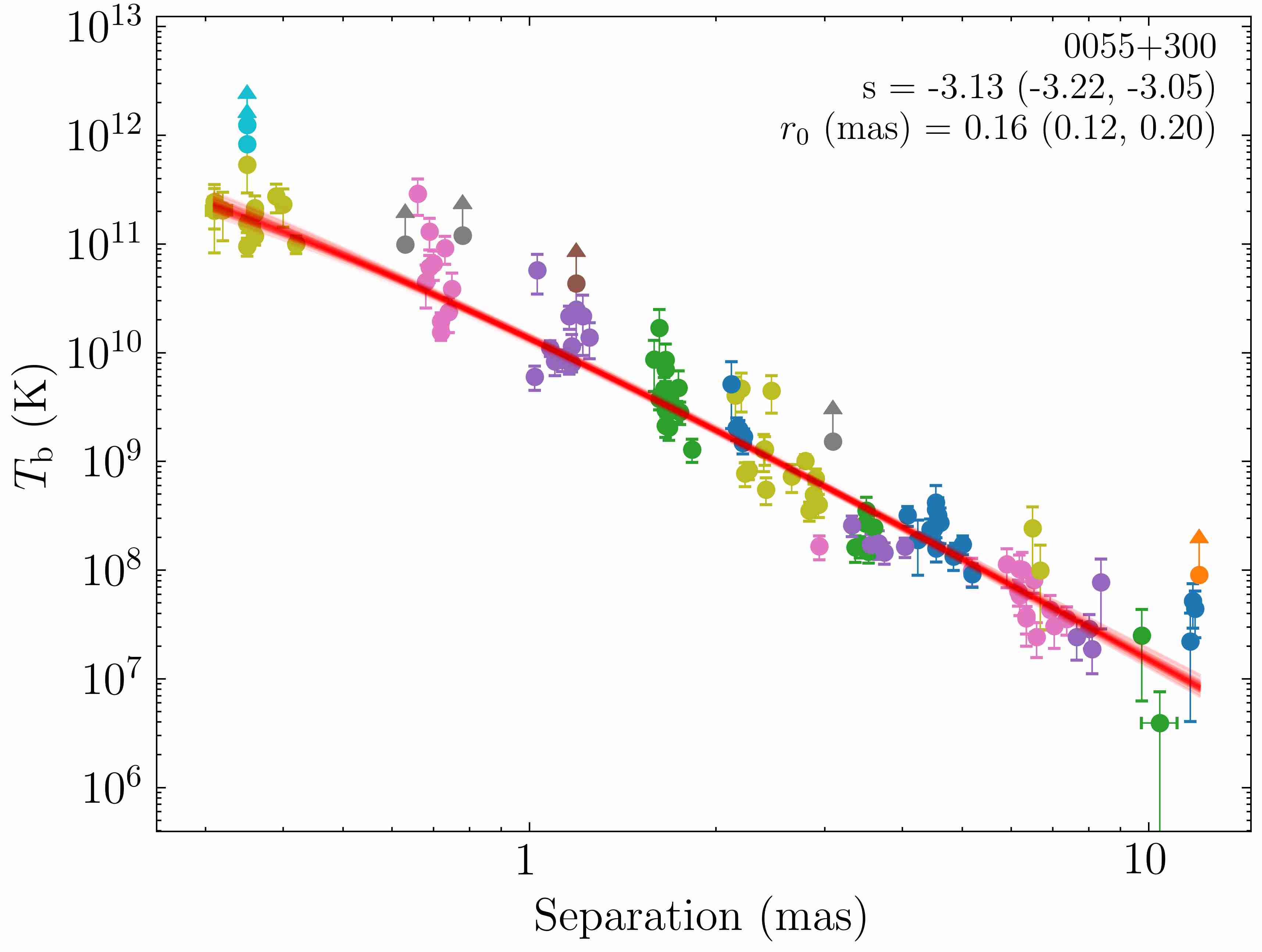}
    \includegraphics[width=0.67\columnwidth]{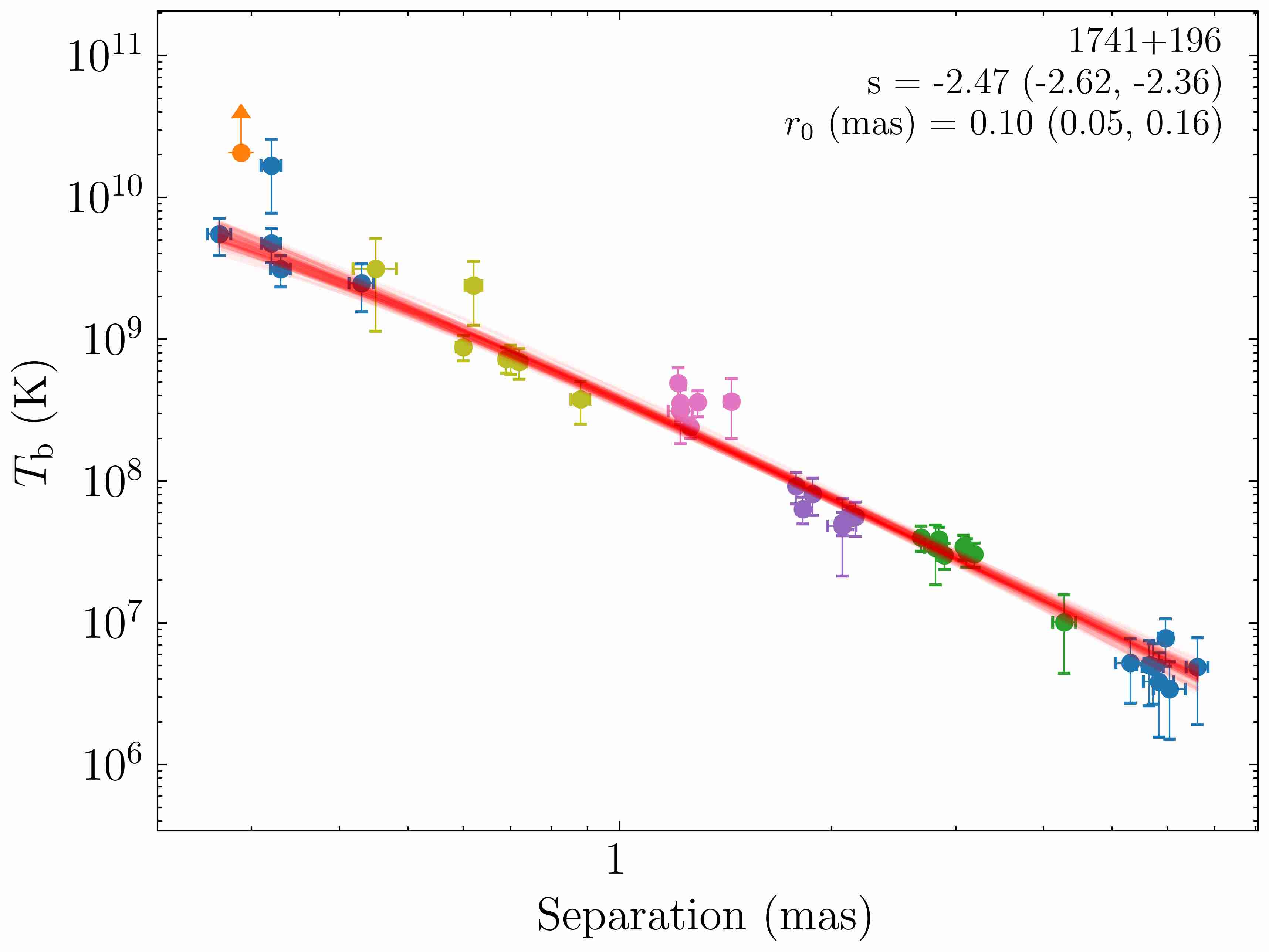}
    \caption{Brightness temperature vs the radial distance from from the 15\,GHz core for 0003$+$380, 0055$+$300 and 1741$+$196. Different colours denote individual jet components. The arrows indicate the lower limits on $T_{\rm b}$, obtained from the upper limits on $R$ with equation~\ref{eq: lobanov_limit}. The solid red lines indicate the results of the automated fit assuming a single power-law slope $T_{\rm b}(r) = a_0 (r +r_0)^s$. The plots for all of the sources are available on-line from the journal as supplementary material. The dependencies for the sources showing a break in their distributions are given in Fig.~\ref{fig:tbgrad}.}
    \label{fig:tbr_1741}
\end{figure*}

\subsection{Brightness temperature gradient along the jet}

\begin{figure}
    \centering
	\includegraphics[width=0.9\columnwidth]{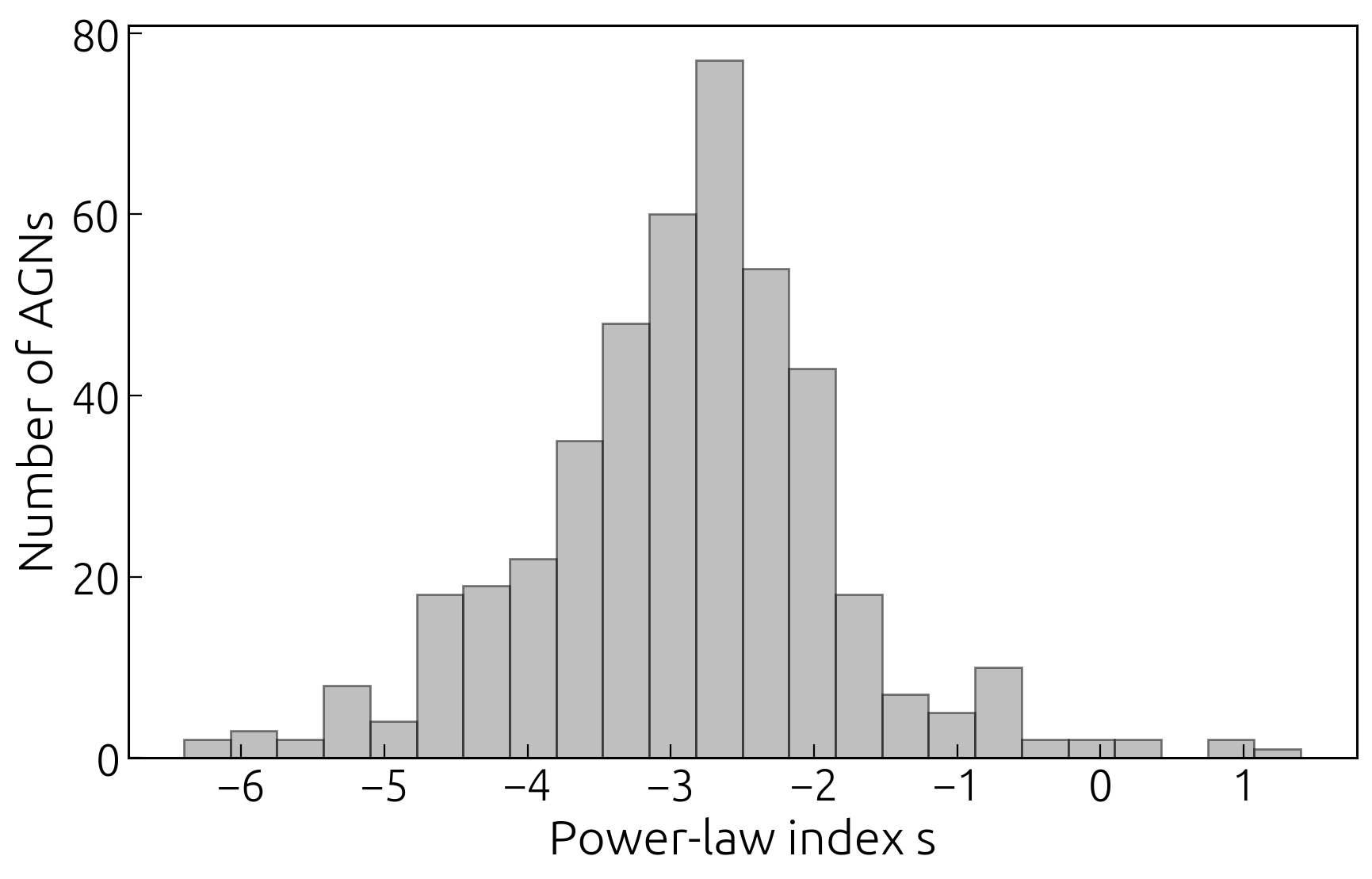}
    \caption{Histogram of the power-law indices $s$ in $T_{\rm b}\propto r^s$ for all sources. The quasar 0923$+$392 with $s=2.63$ and the radio galaxy 1509$+$054 with $s=-8.57$ are not shown.}
    \label{fig:tbindx}
\end{figure}

Figure~\ref{fig:tbr_1741} shows the dependence $T_{\rm b}(r)$ along the jets which are better fit by a single power-law model than a double one (see section~\ref{sec:bestfit}).
Figure~\ref{fig:tbindx} and Table~\ref{tab:fit_results} summarise the distribution of the power-law indices $s$, which has a median of $-2.82$.
The redshift distributions of QSOs, BL~Lacs and RGs are different, Fig.~\ref{fig:z_distr}. Comparing these classes could cause problems because more distant sources from the flux limited sample are more luminous due to the Malmquist bias, and the linear resolution degrades with the redshift up to $z\approx 1$, most rapidly at small $z$.
To test if the different linear resolution may have an impact on the derived parameters due to a significant difference in redshifts of different source optical classes, we calculate medians for the distribution of sources at $z<0.1$ and $z<0.5$, see Table~\ref{tab:fit_results}. 
The AD-tests cannot reject the null hypothesis that quasars and BL~Lacs, and RGs and BL~Lacs are drawn from the same distribution.

\begin{figure*}
\includegraphics[width=0.67\columnwidth]{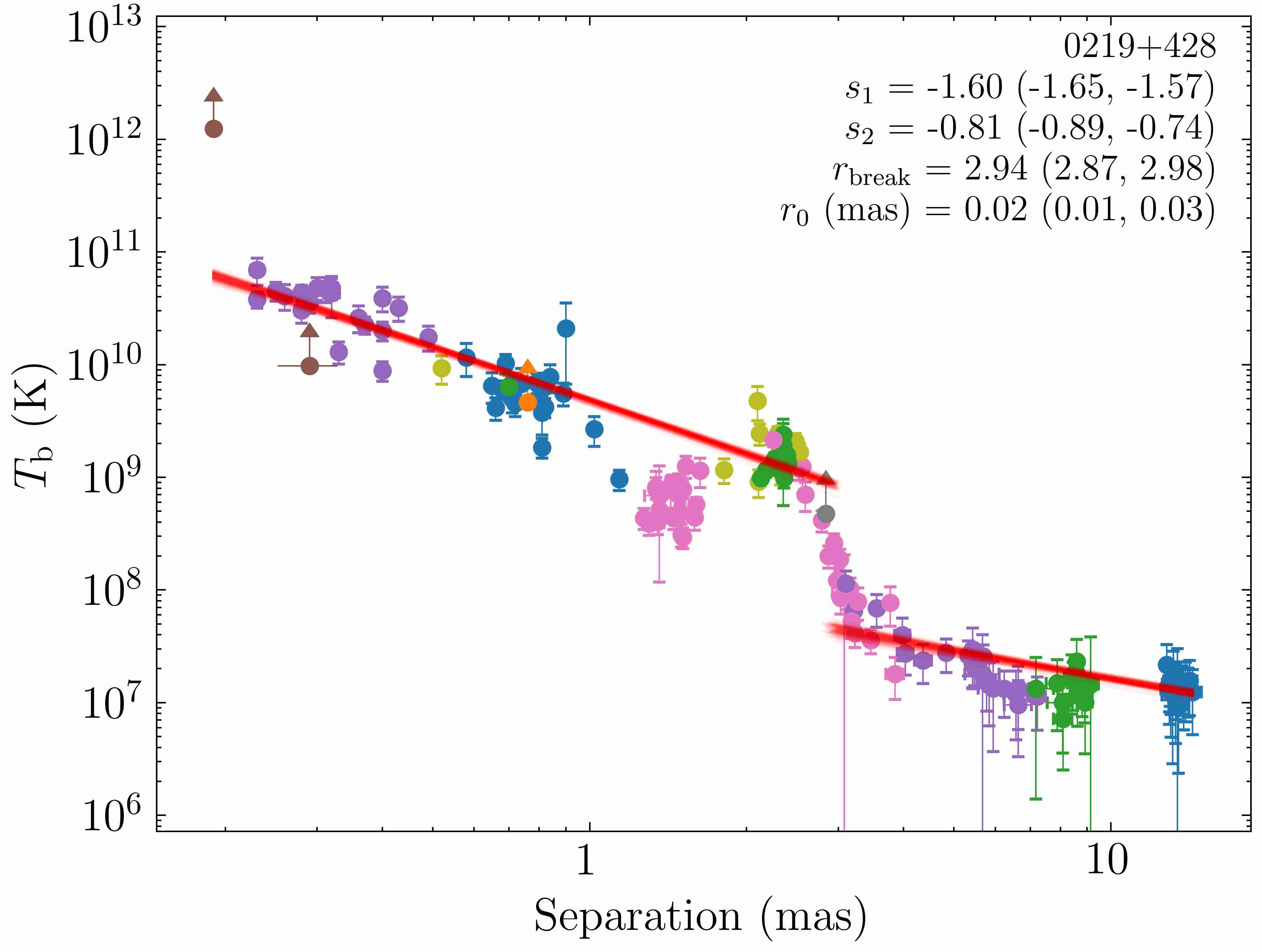}
\includegraphics[width=0.67\columnwidth]{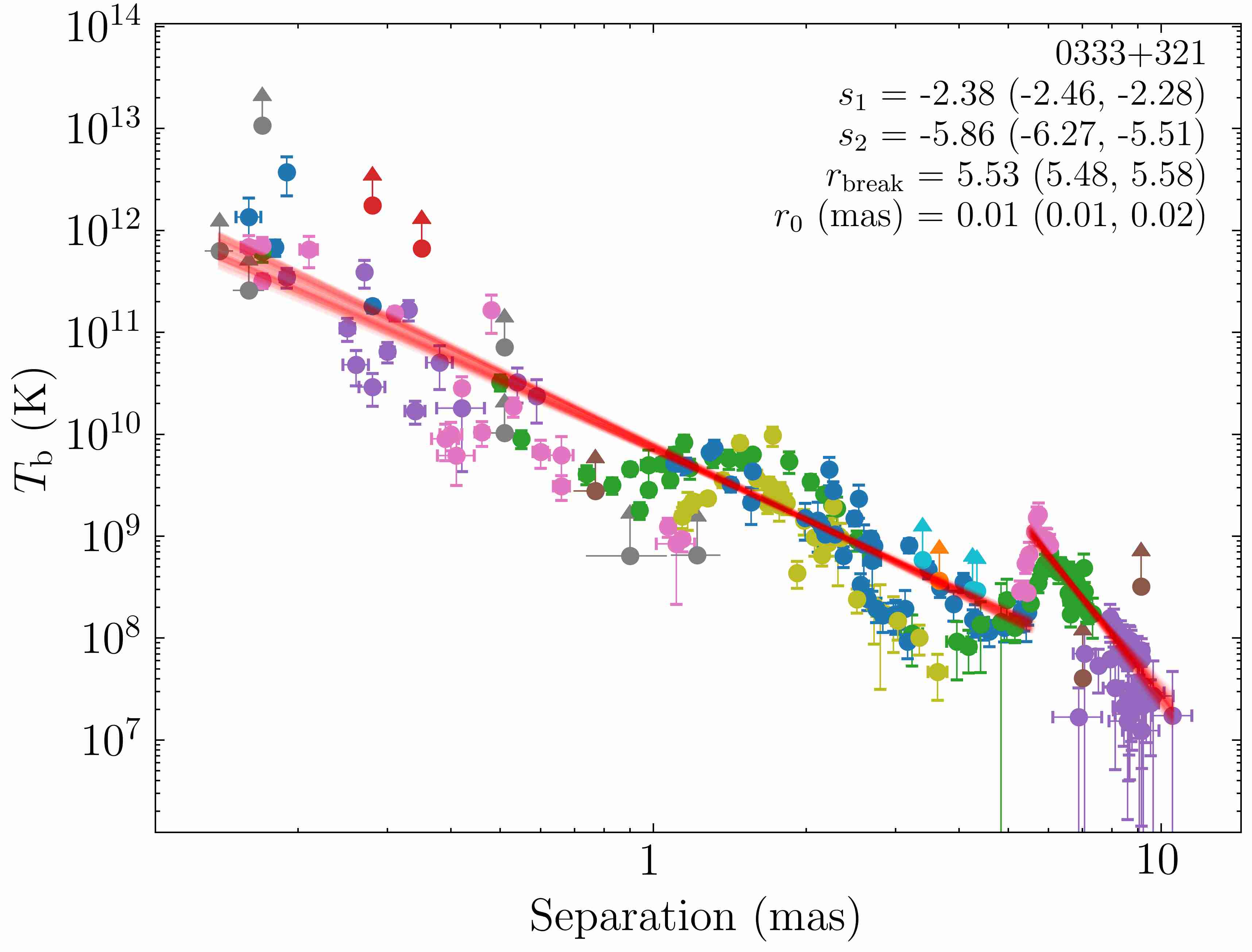}
\includegraphics[width=0.67\columnwidth]{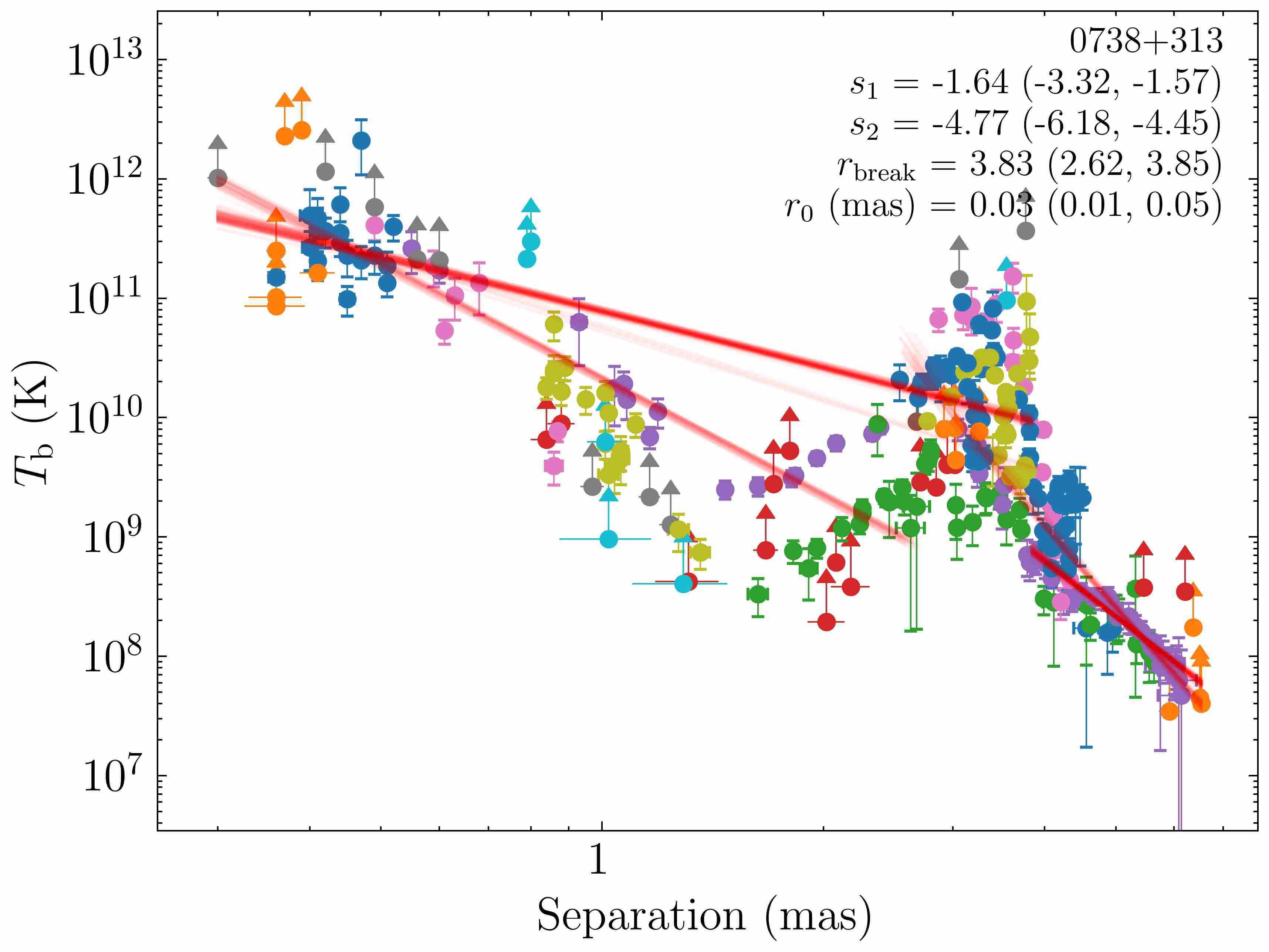}\\
\includegraphics[width=0.67\columnwidth]{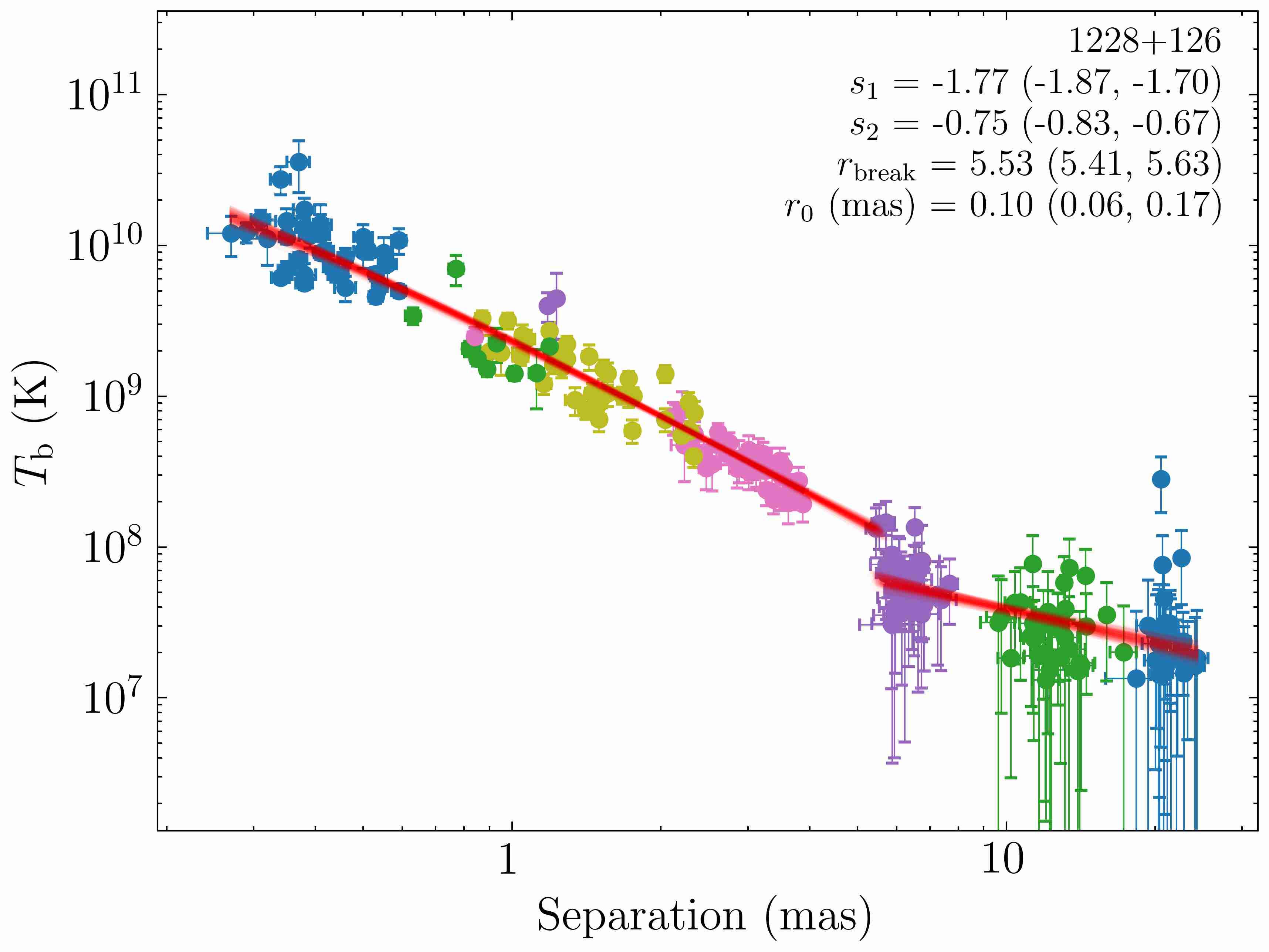}
\includegraphics[width=0.67\columnwidth]{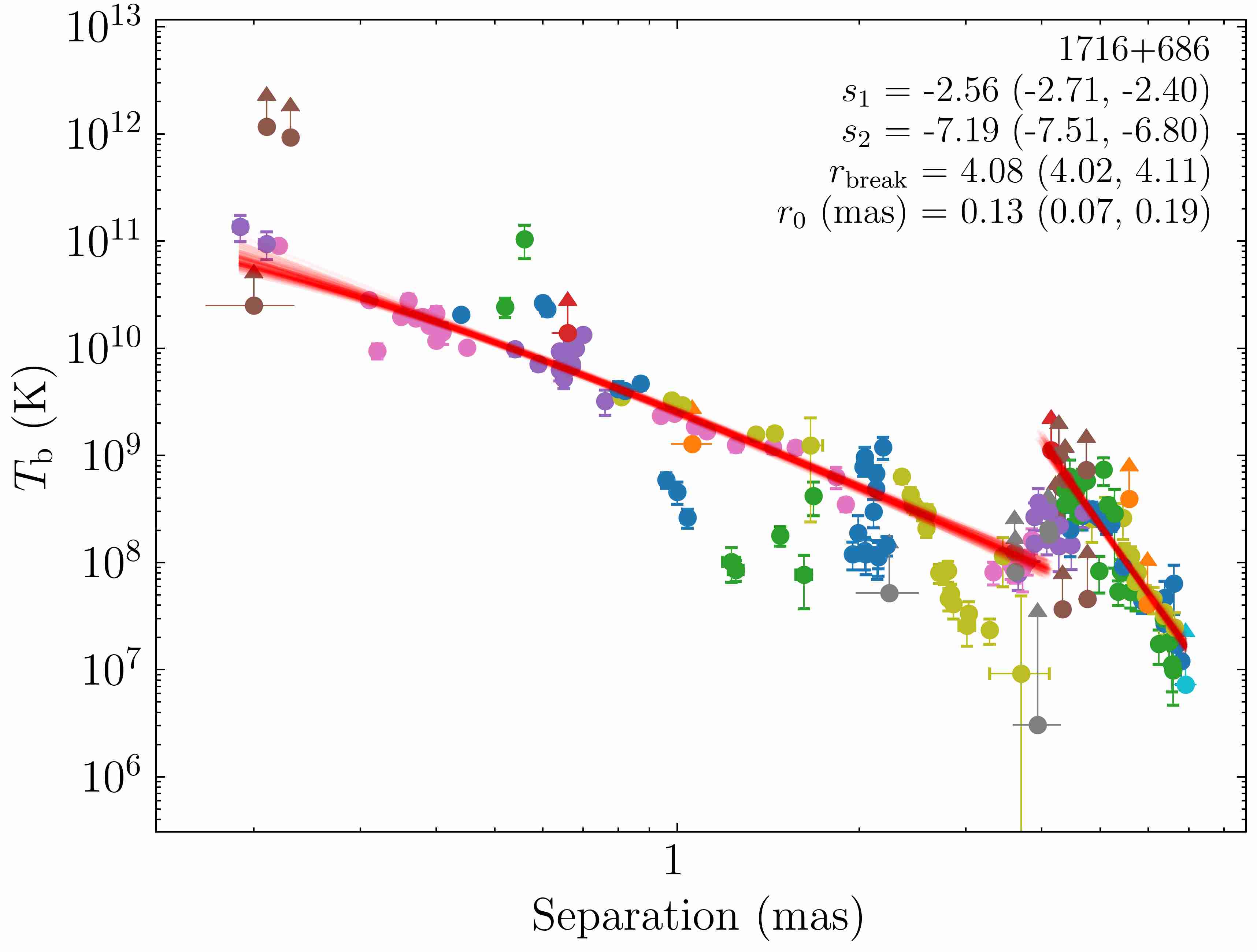}
\includegraphics[width=0.67\columnwidth]{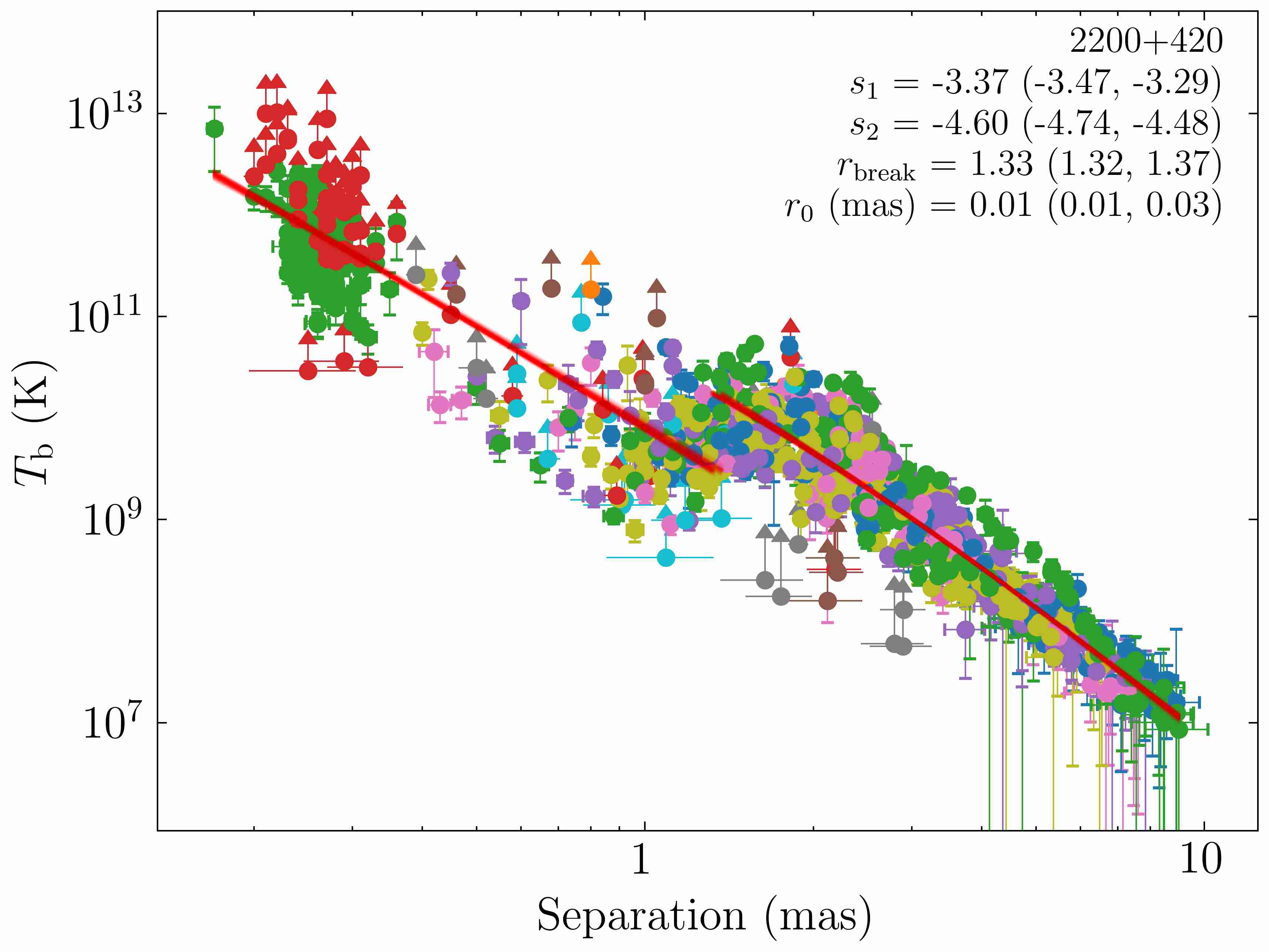}
\caption{Plots of brightness temperature versus radial distance from the 15\,GHz core for selected jets with strong evidence of a break or complex behaviour. Different colors denote individual jet components. The arrows indicate the lower limits on $T_{\rm b}$, obtained from the upper limits on $R$ with equation~\ref{eq: lobanov_limit}. Solid lines indicate results of the automated fit by a double power-law model following Eq.~\ref{eq:tbgr}. Plots for all sources are available on-line from the journal as supplementary material.}
\label{fig:tbgrad}
\end{figure*}

We identified 233 cases whose brightness temperature profiles are better fit by a double power-law model.
The $T_\mathrm{b}(r)$ distributions for these jets are shown in Fig.~\ref{fig:tbgrad}, and the fitted parameters are provided in Table~\ref{tab:results}.
The corresponding distribution of the power-law indices $s_1$ and $s_2$ is given in Appendix~\ref{sec:apndx_distr_break}.
For the majority of the known sources which have a change in the jet geometry from a parabolic to conical streamline (the jets where the transition zone is observable with the VLBA at 15\,GHz, Table~\ref{tab:transjets}), we detect either a significant or slight enhancement in the $T_{\rm b}(r)$ profile at the position of the jet break. 
The same behavior has  been reported for NGC~1052 \citep{2004AA...426..481K, 2019AA...623A..27B}, 3C~111 and BL~Lac \citep{2022A&A...660A...1B}.

\subsection{Relation between the broken profiles of $R(r)$ and $T_\mathrm{b}(r)$} 
\label{sec:br_distrs}

For the majority of cases, $R(r)$ and $T_{\rm b}(r)$ change coherently, implying that if there is a break or variation in one dependence, then it is likely presented in another dependence, even if the fit with a double power-law is insignificant. 

Out of 117 jets with the break in $R(r)$ and 233 cases in $T_{\rm b}(r)$, 97 sources strongly favour the model with the break in both distributions.
For these sources, the median difference between $r_{\rm br,T}$ and $r_{\rm br,R}$ amounts to 0.4\,mas. The non-zero discrepancy can be explained by the fact that the automated routine provided poorly constrained position of the break due to complex behaviour of $R(r)$ and $T_{\rm b}(r)$. 

\input{Sect_Rbr_dist}


\input{Sect_Compar}

\section{Peculiar jet properties and complex distributions}
\label{sec:pecular_and_complex}
\subsection{Complex cases}
\label{sec:complex}

For some sources, the distributions of $R(r)$ and $T_{\rm b}(r)$ resemble a repeating zigzag pattern (e.g. 1716$+$686, Fig.~\ref{fig:srgrad}) or a lighting bolt shape (e.g. 0738$+$313, Fig.~\ref{fig:tbgrad}). 
These data clearly require a more complex model to be fitted, which is beyond the scope of this paper.
By visual inspection, we identified at least 40 most pronounced cases of complex distributions, and label them in Table~\ref{tab:results}.
In section~\ref{sec:individual}, we discuss some of the interesting cases.
Probably, sources that show breaks in their distributions will exhibit more complex behavior with new observations. In this case, as jet components will travel along the jet, they will highlight its different parts, therefore filling the transition zone around the break (e.g. 1147$+$245).

In some of these complex cases, the variations of $R(r)$ and $T_{\rm b}(r)$ at the same jet locations are coherent for different jet components. i.e. different components follow the same zigzag profiles (e.g. BL~Lac, Fig.~\ref{fig:tbgrad}; 3C~111, Fig.~\ref{fig:pa}).
Meanwhile, in a handful of sources, these variations are incoherent, such that each component draws their own  $R(r)$ and $T_{\rm b}(r)$ zigzag patterns, which results in smearing and broadening the total distributions.
The best examples of such behaviour are 3C~273 (see Appendix ~\ref{sec:3c273}) and 3C~279..
In Fig.~\ref{fig:pa}, we plot the radial dependence of the position angle (PA) of individual components measured relative to the core for the selected sources.
In the case of 3C~111, the PA exhibits small but coherent variations at the same jet position for different components.
Whereas for 3C~273 and 3C~279, there is a notable variety of individual feature trajectories on the sky.
For instance, the $T_{\rm b}(r)$ profiles of 3C\,273 jet components 2, 9, 26 and 29 vary within some range, showing local maxima and minima.
This may suggest that each maximum corresponds to a minimum in the angle between the local jet direction and line-of-sight, and, therefore, Doppler-boosted \citep{2002PASA...19..486P} and vice versa. 
Thus, at the same jet location (e.g. 10\,mas), components 26 and 29 were in opposite extreme orientations relative to the jet axis.
We suggest that there is a motion along helically twisted pressure maxima within the jet, evidence of which is, for example, observed in jets of 3C~273 \citep{2001Sci...294..128L}, 0836$+$710 \citep{2012ApJ...749...55P} and 3C~345 \citep{2024A&A...684A.211R}, or the rotation of the jet around its axis which leads to such profiles. Therefore, individual components move along that radial streamline directed towards the observer in different locations, and their Doppler factors are different \citep{2002PASA...19..486P}. 
Meanwhile, for the sources having persistent zigzag $R(r)$ and $T_{\rm b}(r)$ profiles, it is reasonable to suggest the existence of a permanent disturbance (like a jet shape transition or a bend). Thus, different parameters of individual features travelling along the jet will not effect this disturbance leading to a break in $R(r)$ and $T_{\rm b}(r)$ at the same position.

\subsection{Non-radial motion, bent and straight jets. Line-of-sight scenario}
\label{sec:nonradial}

The analysis of 259 MOJAVE sources indicates \citep{2013AJ....146..120L} that nearly all of the 60 most heavily observed jets show significant changes in their innermost position angle over time (20\degr{}--150\degr{}). 
The epoch-stacking analysis  \citep{2017MNRAS.468.4992P, 2020MNRAS.495.3576K} suggests that different features occupy only a portion of the full jet cross section, and that the latter appears only after multi-epoch observations are summed together.
In this case, the emission pattern located closer to the observer relative to the jet axis, i.e. the near side, will make a smaller angle to the LOS than the far side, so their Doppler factors will be different. In the same manner, due to the projection effects, the LOS changes will induce an apparent change clearly in the jet aperture.
At the same time, an extreme case of quasar 0858$-$279 with a true jet bent occurs due to interaction between the relativistic plasma and the surrounding dense medium. A jet feature at the bent is observed brighter than the core due to a re-acceleration in a shock and dominates the overall emission \citep{2022MNRAS.510.1480K}.

Recent analysis of our sample \citep{2021ApJ...923...30L} showed that the majority of 173 jets  observed over a 10 yr period have inner PAs which vary on decadal time-scales over a range of 10\degr{} to 50\degr{}. 
For the majority of sources, PA variations show non-trivial but rather smooth behaviour, and some jets display very large changes in PA, up to 200\degr{}. 
Since our per-source analysis shows strict evidence for the association of the breaks in distributions with apparent jet bends, the analysis of $T_{\rm b}(r)$ and $R(r)$ for such cases requires a more complex model to be considered.

Also, we consider the components having significant non-radial motions, i.e. whose trajectories do not extrapolate back to the jet base. These non-radial components indicate directional changes of the jet features \citep{2009AJ....138.1874L, 2015ApJ...798..134H}. 
In total, 227 components in 226 sources were selected. Notably, 162 jets out of these 226 sources (70 per cent) show a break in our analysis in any $T_{\rm b}(r)$ or $R(r)$ dependence. The median difference between the location of the break and the median position of a non-radial component is of $0.31\pm0.28$, providing convincing evidence that the broken profiles can be explained by the change in the LOS of the emitting regions.

\begin{figure}
    \centering
    \includegraphics[width=0.9\columnwidth]{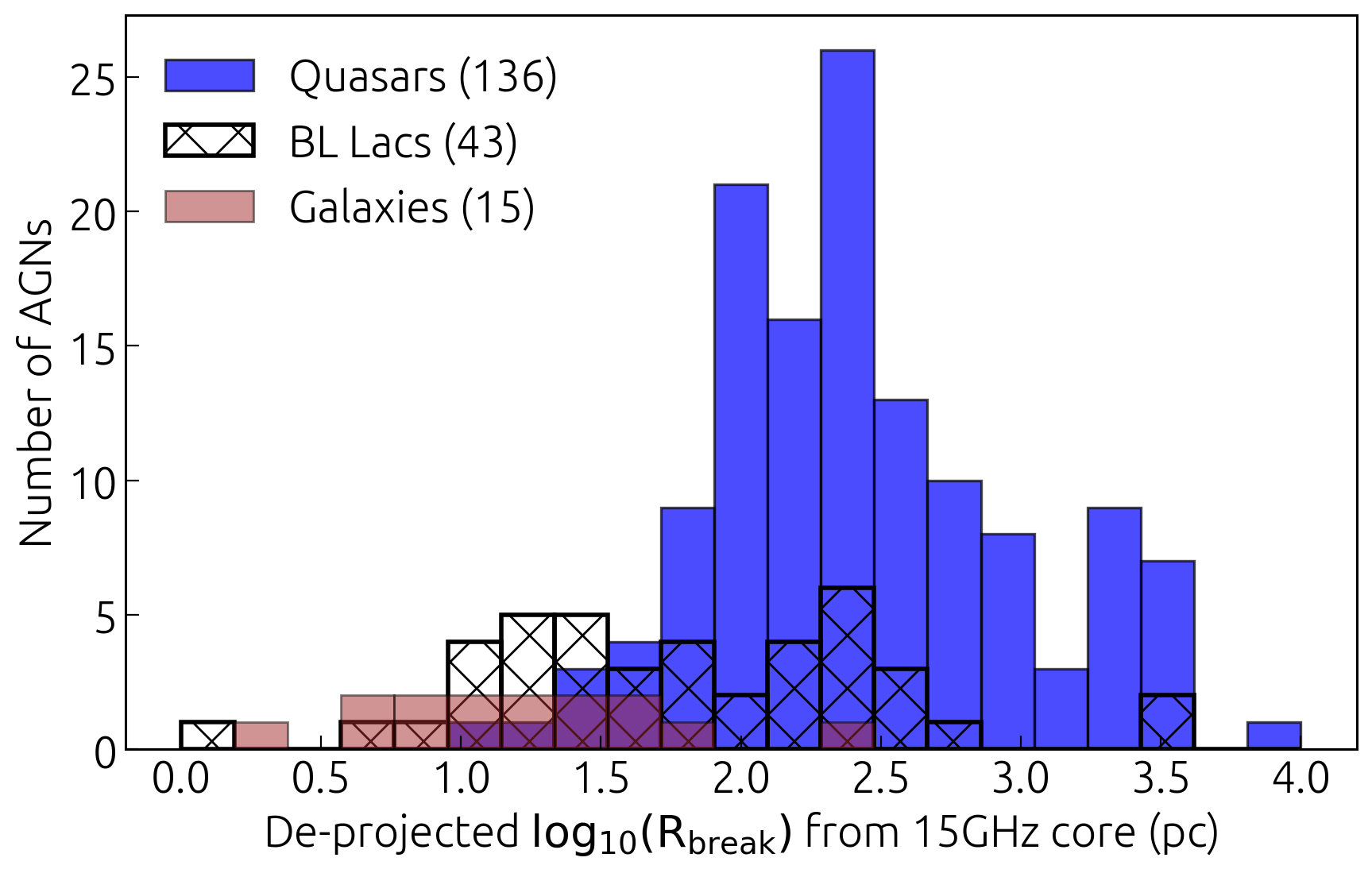}\\
    \includegraphics[width=0.89\columnwidth]{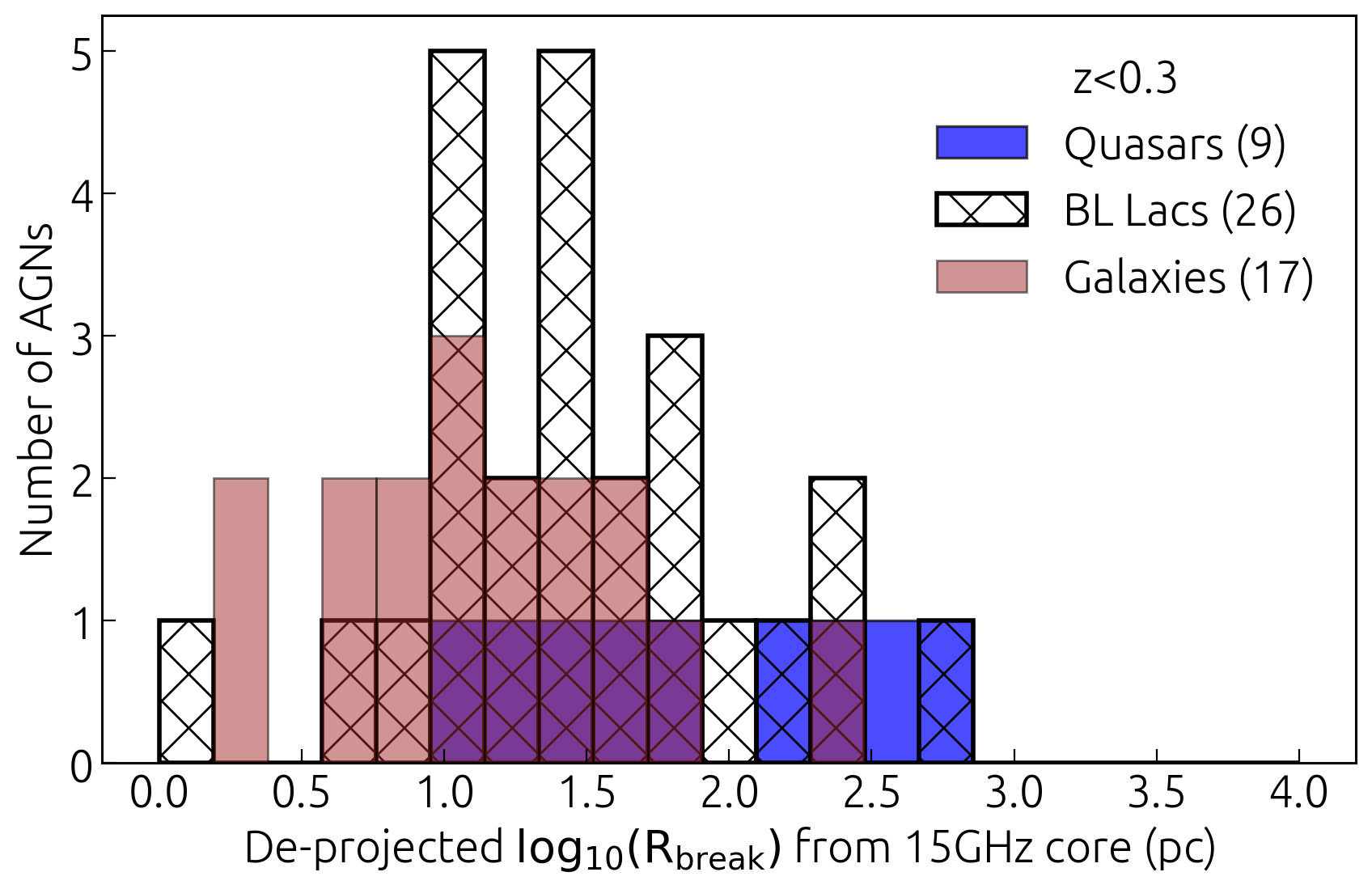}
    \caption{Distribution of the distances of the break from the apparent 15\,GHz core for the jets with complex $T_{\rm b}(r)$ gradients (top) and for the sources at redshift $z<0.3$ (bottom).}
    \label{fig:rbr_dist}
\end{figure}


\begin{figure*}
    \includegraphics[width=0.67\columnwidth]{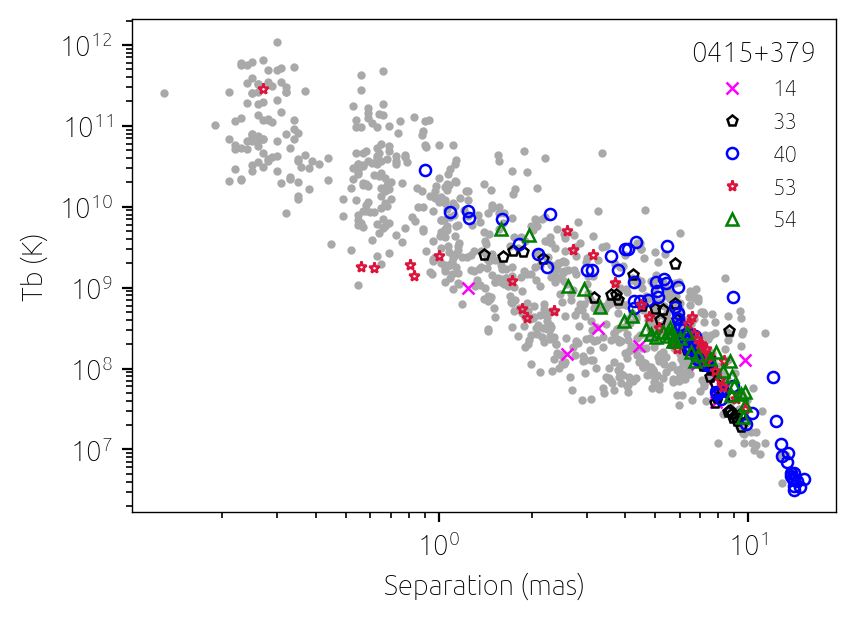}
    \includegraphics[width=0.67\columnwidth]{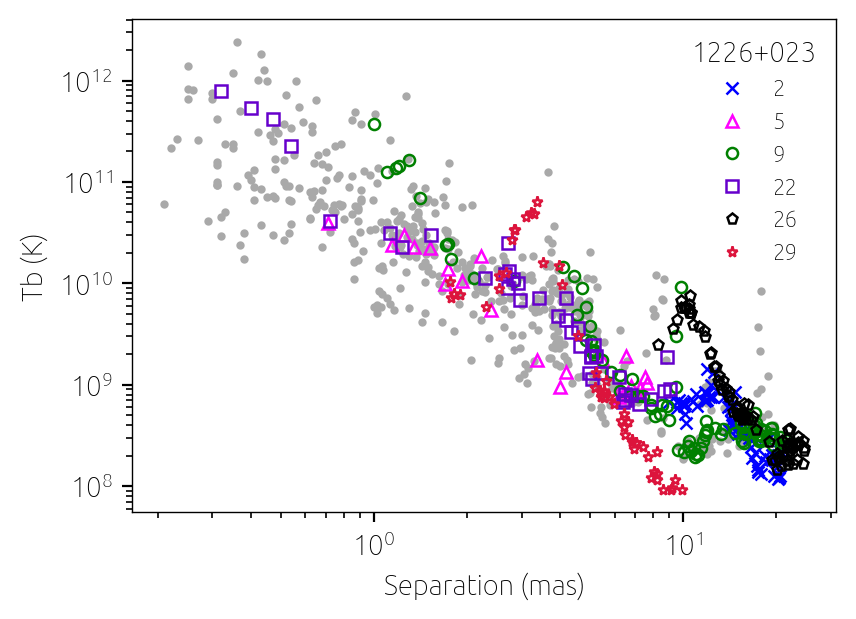}
    \includegraphics[width=0.67\columnwidth]{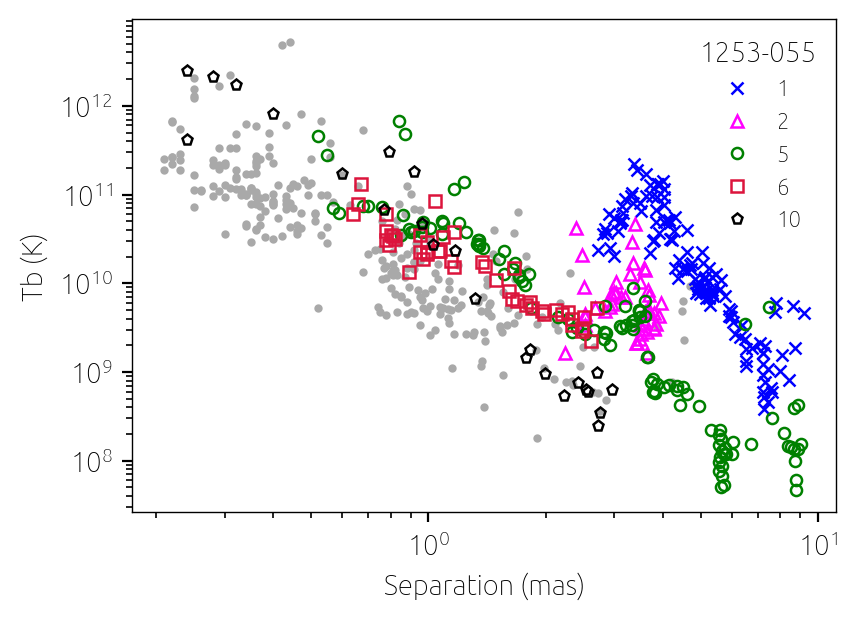}\\
    \includegraphics[width=0.67\columnwidth]{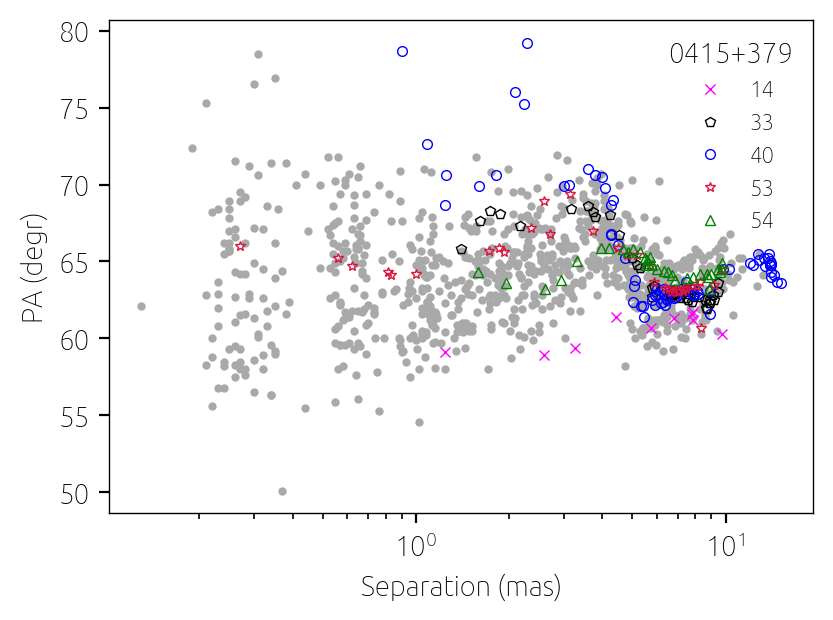}
    \includegraphics[width=0.67\columnwidth]{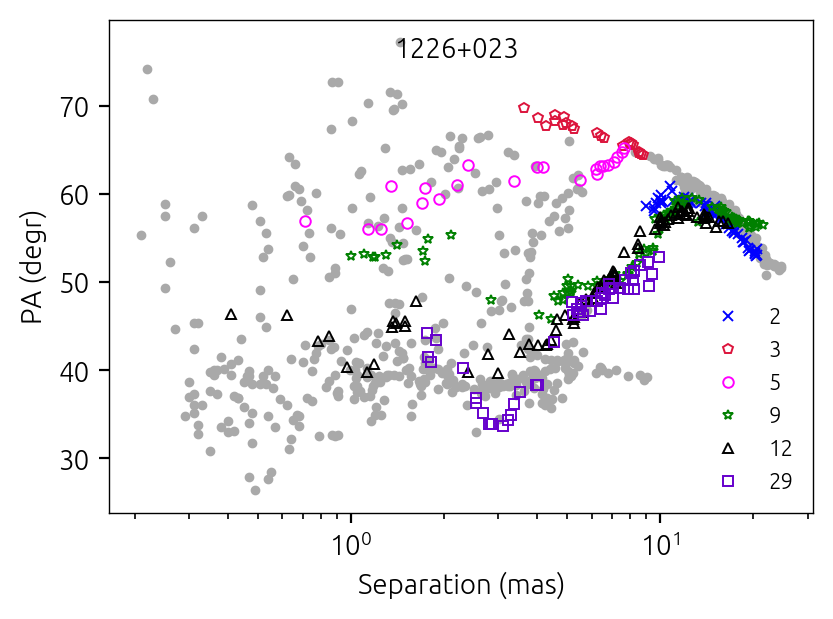}
    \includegraphics[width=0.67\columnwidth]{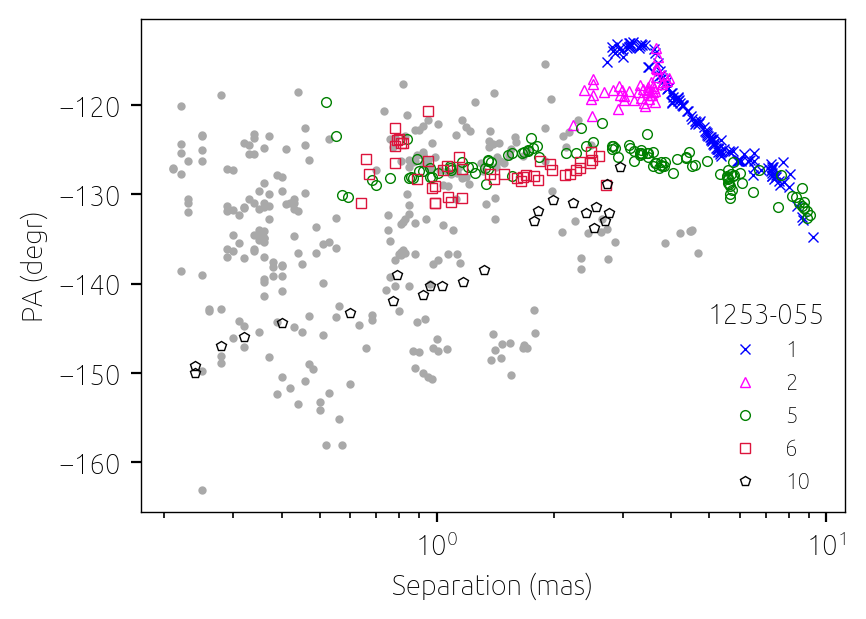}
    \caption{(Top row) Brightness temperature versus radial distance  measured relative to the 15\,GHz core (from right to left) for 3C~111, 3C~273 and 3C~279. The individual jet components whose motion along the jet was observed at many epochs are highlighted by colour, the rest jet features are shown by grey dots. Errors and limits are not shown. (Bottom row) The corresponding position angles of the same components.}
    \label{fig:pa}
\end{figure*}

\subsection{Helical jets}

One possible explanation  for the zigzag $T_{\rm b} (r)$ profile is helical jet structures \citep[e.g.][]{2003ApJ...597..798H, 2018ARep...62..116B}. One of the best examples is quasar 1716$+$686 (Fig.~\ref{fig:srgrad}) which exhibits a clear helical jet seen on the stacked image \citep{2023MNRAS.520.6053P}.
Helical patterns can arise as a result of variation in the flow direction, like precession; random perturbations to the jet, jet-surrounding medium interaction and bends; jet stratification; and developing instabilities.
For example, observed helical structure of the jets of M~87 \citep{2023MNRAS.526.5949N} and 3C~279 \citep{2023NatAs...7.1359F} has been interpret as being driven by plasma Kelvin-Helmholtz instabilities, which can be generated by the activity of the supermassive black hole and accretion processes, e.g. due to Lense-Thirring precession \citep{2023Natur.621..711C}.

To explain the edge-brightened asymmetry of total intensity transverse to the jet direction in 3C\,273, \citet{2012IJMPS...8..265G} suggested that it could be due to the dependence of the synchrotron radiation on the angle between the helical magnetic field and the line of sight in the plasma frame (see also our detailed discussion of 3C\,273 in Appendix ~\ref{sec:3c273}).
Sources showing complex behavior of their $T_{\rm b}(r)$ and $R(r)$, e.g. 0106+678, 0430+052, 0836+710, 1222+216, and 1828+487, indeed exhibit the same asymmetric intensity profiles as 3C\,273 (see their stacked images shown in Figure~2 of \citet{2017MNRAS.468.4992P}).

Other observational evidence of helical jets is helical trajectories formed by individual jet components (e.g. 3C\,345 in \citealt{1995A&A...302..335S}; NRAO\,150 in \citealt{2014A&A...566A..26M}; 0836+710 in \citealt{2012ApJ...749...55P}; 2136+141 in \citealt{2006ApJ...647..172S}).
Or this can be a helical magnetic field in a sheath or boundary layer surrounding the jet. It can be traced through its toroidal component from the Faraday rotation measure (RM) gradient across the jet \citep[see e.g.][]{2003ApJ...589..126Z}.
A number of sources with complex profiles (see Figs.~\ref{fig:srgrad},~\ref{fig:tbgrad} and ~\ref{fig:pa}) indeed exhibit or show hints of RM gradients: 0333+321 \citep{2008ApJ...682..798A}, 0836+710 \citep{2010ApJ...720...41A}, 1156+295 \citep{2009MNRAS.393..429O}, 1219+285 \citep{2017MNRAS.467...83K}, 1226+023 \citep{2002PASJ...54L..39A}, 1458+718 \citep{2009MNRAS.393..429O, 2017MNRAS.467...83K}, 1611+343 \citep{2010ApJ...722L.183T}, 1633+382 \citep{2013MNRAS.429.3551A}, 1641+399 \citep{2012AJ....144..105H}, 1803+784 \citep{2009MNRAS.400....2M}, 1828+487 \citep{2014MNRAS.438L...1G}, 2201+315 \citep{2017MNRAS.467...83K}, 2230+114 \citep{2012AJ....144..105H}, 2200+420 \citep{2016ApJ...817...96G}, 2251+158 \citep{2012AJ....144..105H}.

\subsection{Shocks}
\label{sec:shocks}

The position of the geometry transition in the M87 jet \citep{2012ApJ...745L..28A} coincides with the stationary feature HST-1 \citep{1999ApJ...520..621B}, which can be associated with a recollimation shock formed due to change in the ambient pressure.
Recently, the existence of a stationary component at the jet shape transition location was shown in other nearby AGNs \citep{2018ApJ...860..141H, 2020MNRAS.495.3576K}.
Besides, the studies of individual AGN jets show that the break in the $T_{\rm b}(r)$ distribution can be associated with standing shocks \citep{2005AJ....130.1418J, 2010ApJ...712L.160R, 2013AA...551A..32F, 2018AA...610A..32B}. 
Additionally, detailed studies of individual sources suggests the existence of recollimation shocks in their jets: e.g. 3C~66A \citep[0333$+$321,][]{2005ApJ...631..169B}, 3C~111 \citep[0415+379,][]{2018AA...610A..32B} and CTA~102 \citep[2230+114,][]{2013AA...551A..32F}. Our distribution profiles and break positions correspond well to those estimated in the above-cited works. We discuss these AGNs in detail in Appendix~\ref{sec:individual}.

To examine how many cases with the broken profiles can be associated with the slow pattern speed components, we selected sources showing slow pattern ($\beta_{\rm app}< 0.2c$) jet features based on their kinematics \citep{2021ApJ...923...30L}. 
We identified 64 individual jet components in 47 sources. 
For 24 jet features in 24 sources, $R(r)$ and/or $T_\mathrm{b}(r)$ show a broken profile, and the location of the break coincides with the median position of a stationary component.
One half (30 jets)  of 64 slow pattern components are localized in the inner 0.5~mas from the 15\,GHz core. Thus, there might be more successive associations with the broken profiles we missed, since we do not consider breaks at distances $<1$~mas if they are based only on the components located at $r<0.5$~mas.

\section{Evaluation of physical parameters}
\label{sec:disc}

\input{Sect_Disc}


\section{Summary and Conclusions}

We present a study of the brightness temperature and component size distributions along relativistic jets for the brightest AGNs in the northern sky. The complete flux density limited sample with total VLBA flux density above 1.5~Jy for declination $\delta>-30^\circ$ comprises 447 (analysed; 458 in total) sources. Their observations were made within the MOJAVE programme and its predecessor, the VLBA 2cm Survey from 1995 until 2019 at 15\,GHz. 

Our main findings and conclusions are as follows. 
\begin{enumerate}
    \item The jet feature's size follows a power-law profile as a function of distance to the core $R\propto r^d$ with the median value of $d=1.02\pm0.03$. This result agrees well with the independent previous measures of the deconvolved jet width \citep{2020MNRAS.495.3576K}, therefore, indicating that the jet components nicely trace jet geometry. On the scales probed by the VLBA observations at 15\,GHz, the jets in quasars and BL Lac objects expand freely, while the jets in radio galaxies experience collimation.
    \item Extrapolating the fits of the radial distribution of the component size to the true jet origin, we estimated typical separation of the 15\,GHz VLBI core as 109$\pm$6~$\mu$as. The comparison of this result with the core shift measures between 8 and 15\,GHz yields $r_{\rm core}\propto \nu^{-1/(0.83\pm0.03)}$, indicating that the VLBI cores typically lie within the collimated jet regions. 
    \item We show that the jet regions where a transition from parabolic to conical streamline takes place is characterized by an enhancement of the brightness temperature.    
    \item The brightness temperature gradient along the jets and the dependence of the brightness temperature on the component size follow a power-law profile $T_{\rm b}\propto r^s$ and $T_{\rm b}\propto R^{\hat{s}}$ with a median index of $s=-2.82$ and $\hat{s}=-2.87$, respectively. Using theoretical relations between the estimated power-law indices, we derived a parameter range of physical conditions in the jets. The median values agree well with a simple model where the components are optically thin blobs moving down the conical outflows ($d=1$) with a constant speed, the power-law evolution of the emitting particle density ($n=-2$), dominant toroidal magnetic field ($b=-1$) and optically thin spectra ($\alpha=-0.8$). The distributions of the power-law index are also consistent with the shock-in-jet model which incorporates adiabatic expansion under an assumption that the Doppler factor increases along the jet.
    \item We measured a variety of the brightness temperature profiles down the jet.  Half of the sources (233 jets) are poorly described by the single power law, implying the presence of different inhomogeneities in their jets. A number of broken profiles can be associated with slow moving components formed by a shocked region. We found a significant correlation between the observed broken $T_{\rm b}(r)$ and $R(r)$ profiles with the observed non-radial motions in the jets. Convincing evidence for the association of complex distributions is shown with the asymmetric intensity profiles, jets bends and helically twisted magnetic fields. We conclude that the complex profiles are determined by curvature effects, such that a change in the position angle of the emitting region relative to the line of sight causes Doppler factor variations.
    \item The change of the $T_{\rm b}(r)$ and $T_{\rm b}(R)$ gradients at the position of the jet geometry transition generally depends on the velocity profile and the viewing angle. For radio galaxies, it suggests a change of the gradients from steep to flat, being consistent with the observations of several radio galaxies demonstrating a jet transition from a parabolic to conical shape. 
\end{enumerate}

It is not clear why in some sources, we observe complex behavior of $R(r)$, $T_{\rm b}(r)$ and $T_{\rm b}(R)$ and their association with jet peculiarities but not in others.
The presence of different inhomogeneities which have an imprint on the considered profiles could be distinguished in source-by-source analysis, which is currently done for a handful of AGNs because it requires a considerably large amount of observational time. We expect that more sources will exhibit complex profiles once new observations are performed; which will help addressing and inferring physical conditions within the jets of active galaxies.

\section*{Acknowledgements}

We thank the anonymous referee for their constructive comments which helped to improve the manuscript. We thank Ken Kellermann and Andrei Lobanov for their useful comments and Elena Bazanova for language editing.
This study was supported by the Russian Science Foundation, project 20-72-10078\footnote{\url{https://rscf.ru/en/project/20-72-10078/}}.
This work is part of the MuSES project which has received funding from the European Research Council (ERC) under the European Union's Horizon 2020 Research and Innovation Programme (grant agreement No 101142396).
This research made use of the data from the MOJAVE database maintained by the MOJAVE team \citep{2018ApJS..234...12L}. The MOJAVE programme was supported under NASA-Fermi grant 80NSSC19K1579.
The National Radio Astronomy Observatory is a facility of the National Science Foundation operated under a cooperative agreement by Associated Universities, Inc.

\section*{Data Availability}

The fully calibrated visibility and image data at 15~GHz from the MOJAVE programme are publicly available\footnote{\url{https://www.cv.nrao.edu/MOJAVE}}. The results of the calibrated VLBI visibility data model fitting are taken from \citet{2021ApJ...923...30L}.

\bibliographystyle{mnras}
\bibliography{litr}

\appendix
\input{Apndx_TbD}

\input{Apndx_other}

\input{Apndx_IndivS}

\bsp    
\label{lastpage}
\end{document}

%% file: Sect_Intro.tex
\section{Introduction}

Relativistic jets observed in the centers of Active Galactic Nuclei (AGNs) are believed to be driven by accreting supermassive black holes \citep{1982Natur.295...17R}. 
Numerical simulations indicate that understanding the jet launch requires magnetohydrodynamical processes and dynamically important magnetic fields \citep{2001Sci...291...84M, 2006MNRAS.368.1561M, 2014Natur.510..126Z}.
Nevertheless, a complete explanation of how AGN jets accelerate and propagate out to large distances is still needed. Likewise, the physics of individual jet components associated with the propagation of disturbances downstream the jet which appear on radio images as knots of enhanced brightness is still poorly understood.

The standard model of AGNs suggests that their radio emission is explained by synchrotron radiation from relativistic electrons. 
This emission is Doppler boosted due to the relativistic bulk motion aligned close to the line of sight \citep[LOS,][]{2009AA...494..527H, 2007Ap&SS.311..231K}.
This yields extreme brightness temperatures \citep[$T_\mathrm{b}$,][]{2016ApJ...817...96G, 2016ApJ...820L...9K, 2018MNRAS.474.3523P, 2020ApJ...893...68K,2020AdSpR..65..705K,2022ApJ...924..122G}, whose values can exceed the inverse Compton limit \citep[$\thicksim10^{11.5}$~K,][]{1969ApJ...155L..71K,1994ApJ...426...51R} and depart from equipartition of energy between the magnetic field and radiating particles  \citep[$\thicksim5\times10^{10}$~K,][]{1994ApJ...426...51R}.
When a blob of relativistic plasma propagates downstream, it loses energy through synchrotron radiation and adiabatic expansion. These factors lead to a rapid decrease in the brightness temperature along the jet.
Assuming optically thin synchrotron emission with the emissivity:
\begin{equation}
    j_{\nu} \propto N B^{(1 - \alpha)} \nu^{\alpha} 
\end{equation}
a constant speed jet with a power-law distribution of the magnetic field $B(r) \propto r^{b}$, the emitting particle density $N(r) \propto r^{n}$ and the jet width $R(r) \propto r^{d}$ with distance $r$ from the core, the observed brightness temperature then can be parametrized as \citep{2004AA...426..481K}
\begin{equation}
        T_{\rm b}(r) \propto N(r) B(r)^{1-\alpha} R(r) \propto r^{s}\,,
\label{eq:original_Tb_formula}
\end{equation}
and the index $s$ is defined by
\begin{equation}
        s = n + b(1-\alpha) + d\,.
\label{eq:eps_ksi}
\end{equation}
In an assumption of a canonical scenario of equipartition\footnote{This implies the ongoing particle acceleration to compensate for the cooling associated with adiabatic expansion \citep{1979ApJ...232...34B}.} between the magnetic and emitting particles energy densities ($b=-1, n=-2$), conical jet shape ($d=1$) and $\alpha=-1$, equation~\ref{eq:eps_ksi} yields $s=-3$.

One of the first studies of the brightness temperature distribution along a jet outflow was of NGC~1052 at 5--43\,GHz \citep{2004AA...426..481K}, whose brightness temperature gradient along the eastern jet is found to follow a power law with $s\approx-4$.
This was confirmed by \citet{Kadlerphdthesis} in application to 18 sources observed at 2 and 5\,GHz as well the study by \citet{2012A&A...544A..34P}, who have analyzed simultaneous 2 and 8~GHz (Very Long Baseline Interferometry (VLBI) observations of AGNs.
Recently, \citet{2022A&A...660A...1B} conducted similar analysis based on the observations of 28 sources at 15 and 43\,GHz and found similar results.
\citet{Kadlerphdthesis}, \citet{2012A&A...544A..34P} and \citet{2022A&A...660A...1B} obtained the consistent result of $T_\mathrm{b}\propto r^{-2.2}$, because the steeper slope $T_\mathrm{b}\propto r^{-4}$ of NGC~1052 is taken after the region of jet recollimation, which is possibly shocked \citep{2004AA...426..481K}.

\begin{table*}
	\centering
	\caption{Source properties. Columns are as follows: (1) IAU B1950 name; (2) Alias; (3) Redshift; (4) Optical class (Q=quasar, B=BL Lac object, G=radio galaxy, N=narrow-line Seyfert 1, U=unknown); (5) Reference for Redshift and Optical Classification that are collected by  \citet{2023MNRAS.520.6053P}. This table is available in its entirety in a machine-readable form as supplementary material.}
	\label{tab:sources}
	\begin{tabular}{llllr} 
	\hline
	 Name & Alias & z & Opt. cl. & Reference\\
	 (1) & (2) & (3) & (4) & (5)\\
	\hline
    0003$-$066 &  NRAO 005   & 0.3467 & B & \citet{2005PASA...22..277J}\\
    0003$+$380 &  S4 0003+38 & 0.229  & Q & \cite{1995AAS..109..267G} \\
    0006$+$061 &  TXS 0006+061 & \ldots & B & \cite{2012AA...538A..26R} \\
    0007$+$106 &  III Zw 2 & 0.0893 & G & \cite{1970ApJ...160..405S} \\
    0010$+$405 &  4C +40.01 & 0.256 & Q & \cite{1992ApJS...81....1T} \\
    0011$+$189 &  RGB J0013+191 & 0.477 & B & \cite{2013ApJ...764..135S} \\
    0012$+$610 &  4C +60.01 & \ldots & U & \ldots \\
    0014$+$813 &  S5 0014+813 & 3.382 & Q & \cite{1987SvA....31..136V} \\
    0015$-$054 &  PMN J0017-0512 & 0.226 & Q & \cite{2012ApJ...748...49S} \\
    0016$+$731 &  S5 0016+73 & 1.781 & Q & \cite{1986AJ.....91..494L} \\
    \hline
    \end{tabular}
\end{table*}

\begin{figure}
\centering
    \includegraphics[width=0.9\columnwidth]{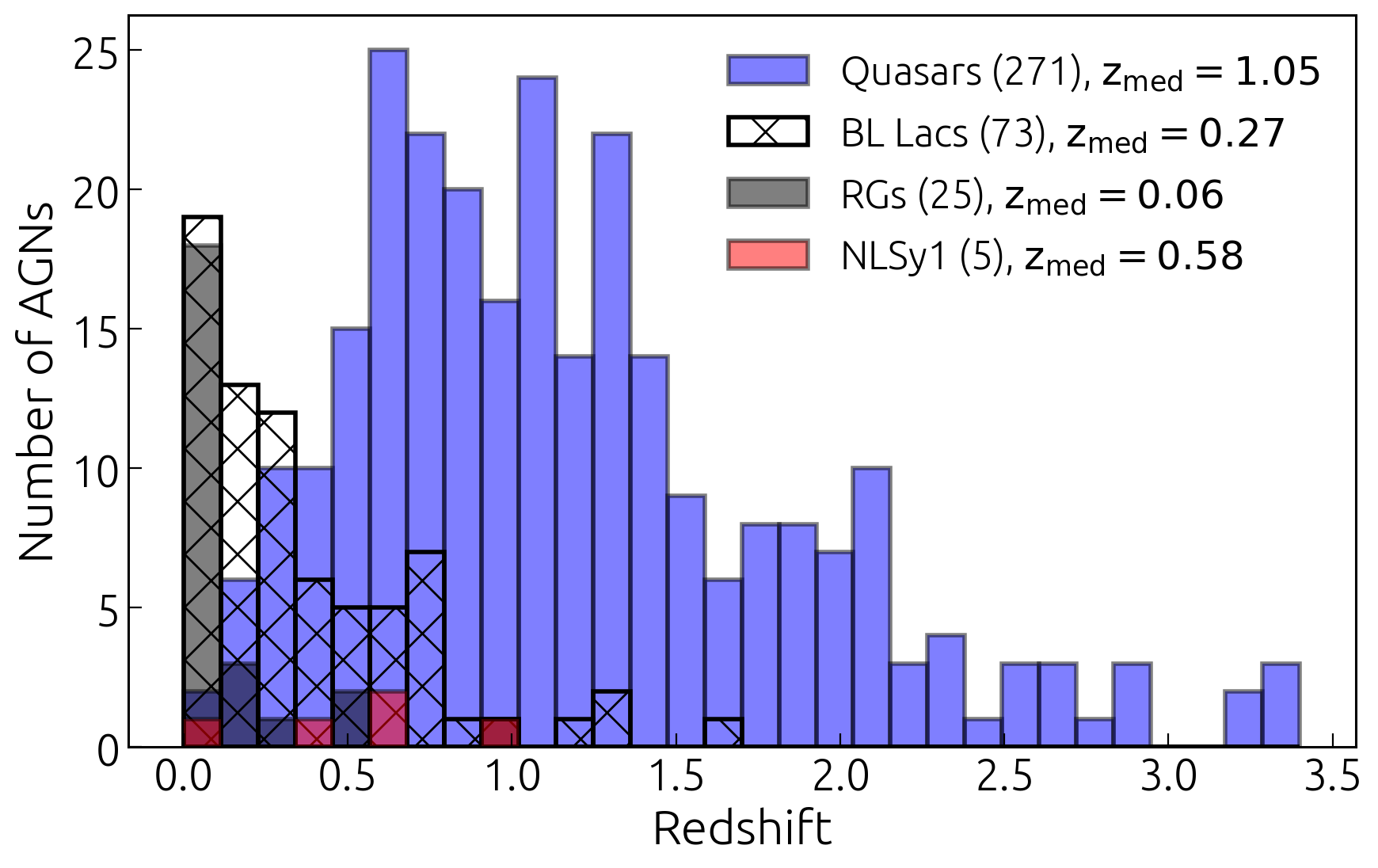}
    \caption{Distribution of the redshifts for the sample. Only sources with the known redshifts are plotted. The median values are given for the different categories.}
    \label{fig:z_distr}
\end{figure}

The detailed studies of individual objects showed departure from a simple power law and significant enhancement of the brightness temperature at the position of a standing jet feature \citep{2005AJ....130.1418J, 2010ApJ...712L.160R, 2013AA...551A..32F, 2018AA...610A..32B}  
and jet bends \citep{2005ApJ...631..169B, 2011MNRAS.415.3049O}.
There is also an indication that jets, such as in M\,87, which show a transition from a parabolically expanding flow (jet acceleration region) to a conical (freely expanding region) geometry \citep{2020MNRAS.495.3576K, 2012ApJ...745L..28A}, are accompanied by a break in the brightness temperature profile along the jet \citep{2004AA...426..481K, 2019AA...623A..27B}.

Therefore, analysis of the brightness temperature can be used not only for an estimation of a parameter range of physical conditions in the AGN jets, but the $T_\mathrm{b}$ profiles provide an excellent tool to trace jet regions with varying or abruptly changing physical conditions, such as different inhomogeneities.
The brightness temperature is most representative and least affected by model fitting peculiarities as compared to the radial evolution of the flux and size of a component. 

To perform the study of the brightness temperature distributions along AGN jets, we analysed a sample of the brightest AGN jets in the northern sky. The sample comprises 447 sources observed within the Monitoring Of Jets in Active galactic nuclei with VLBA Experiments \citep[MOJAVE\footnote{\url{https://www.cv.nrao.edu/MOJAVE}},][]{2021ApJ...923...30L} programme at 15\,GHz. Using the data from almost 24 years of monitoring and almost 40,000 individual jet component measurements, we analysed variations in their brightness temperature and size to constrain the jet geometry, gradients of the magnetic field strength and particle density along the jet, as well as to locate and study discontinuities in the jet.
This study is a large improvement over the previous studies in terms of component size and the use of multi-epoch VLBI data to probe the jet geometry, which is analysed in connection with the brightness temperature radial evolution.

\begin{figure*}
\centering
    \includegraphics[width=0.68\columnwidth]{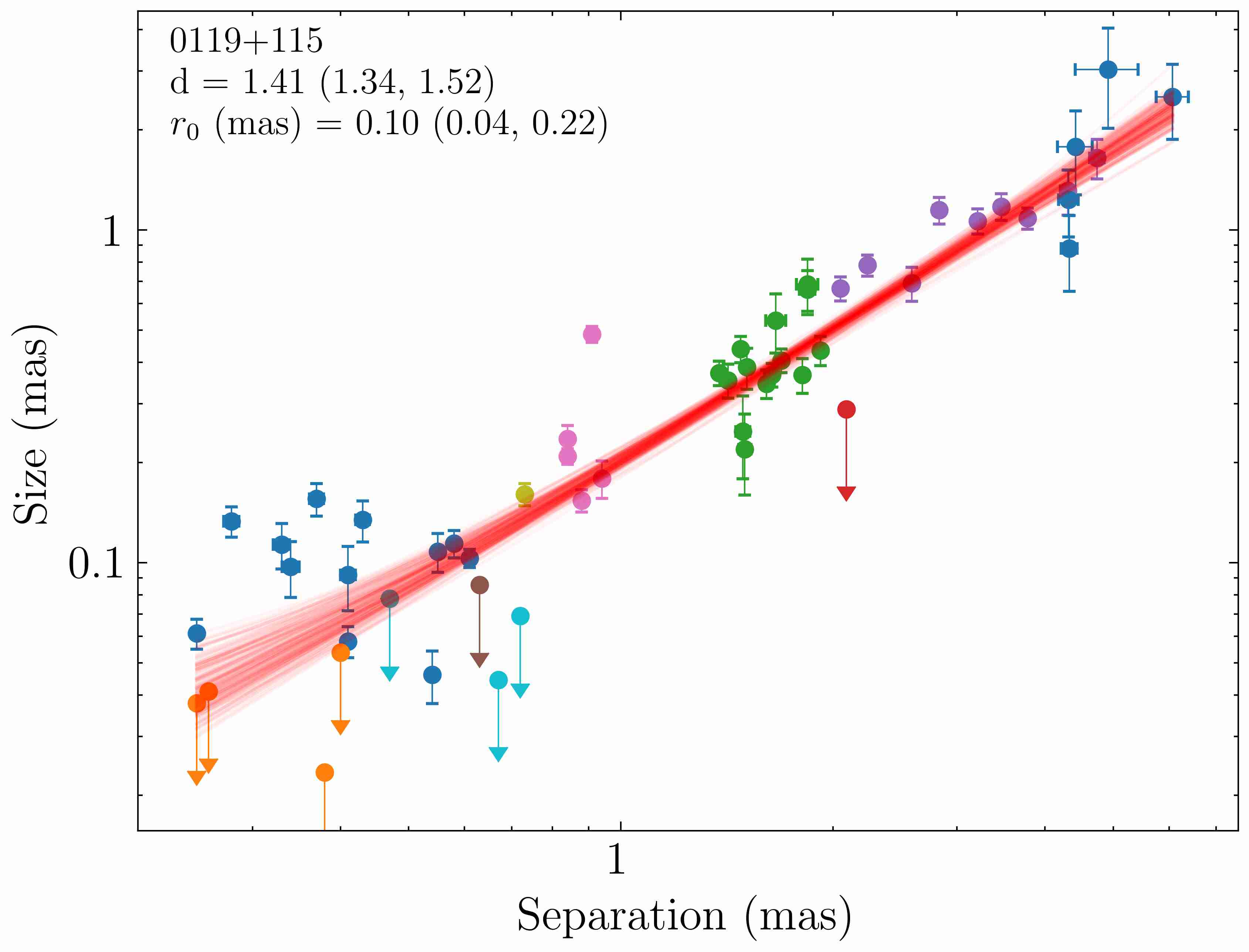}
    \includegraphics[width=0.68\columnwidth]{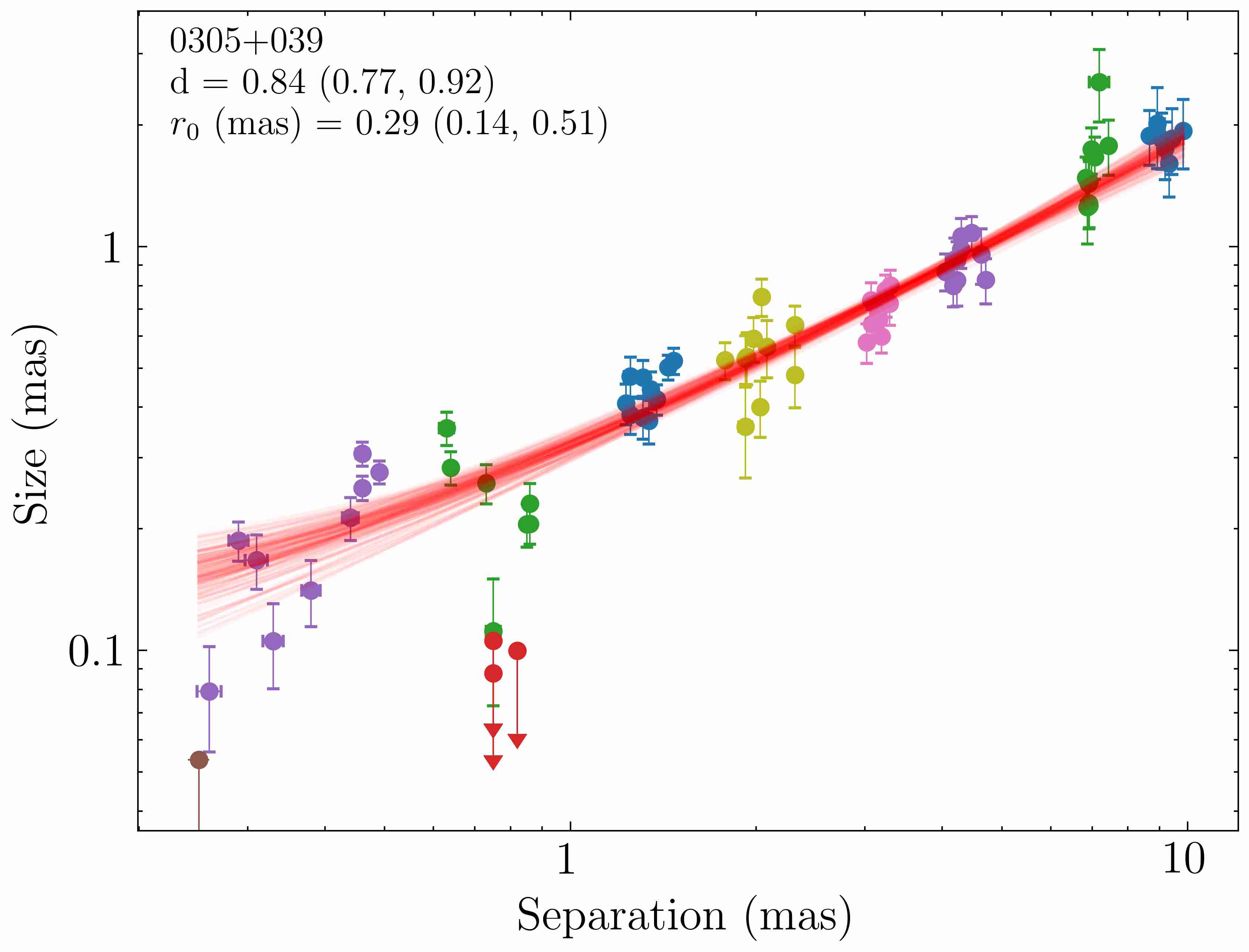}
    \includegraphics[width=0.68\columnwidth]{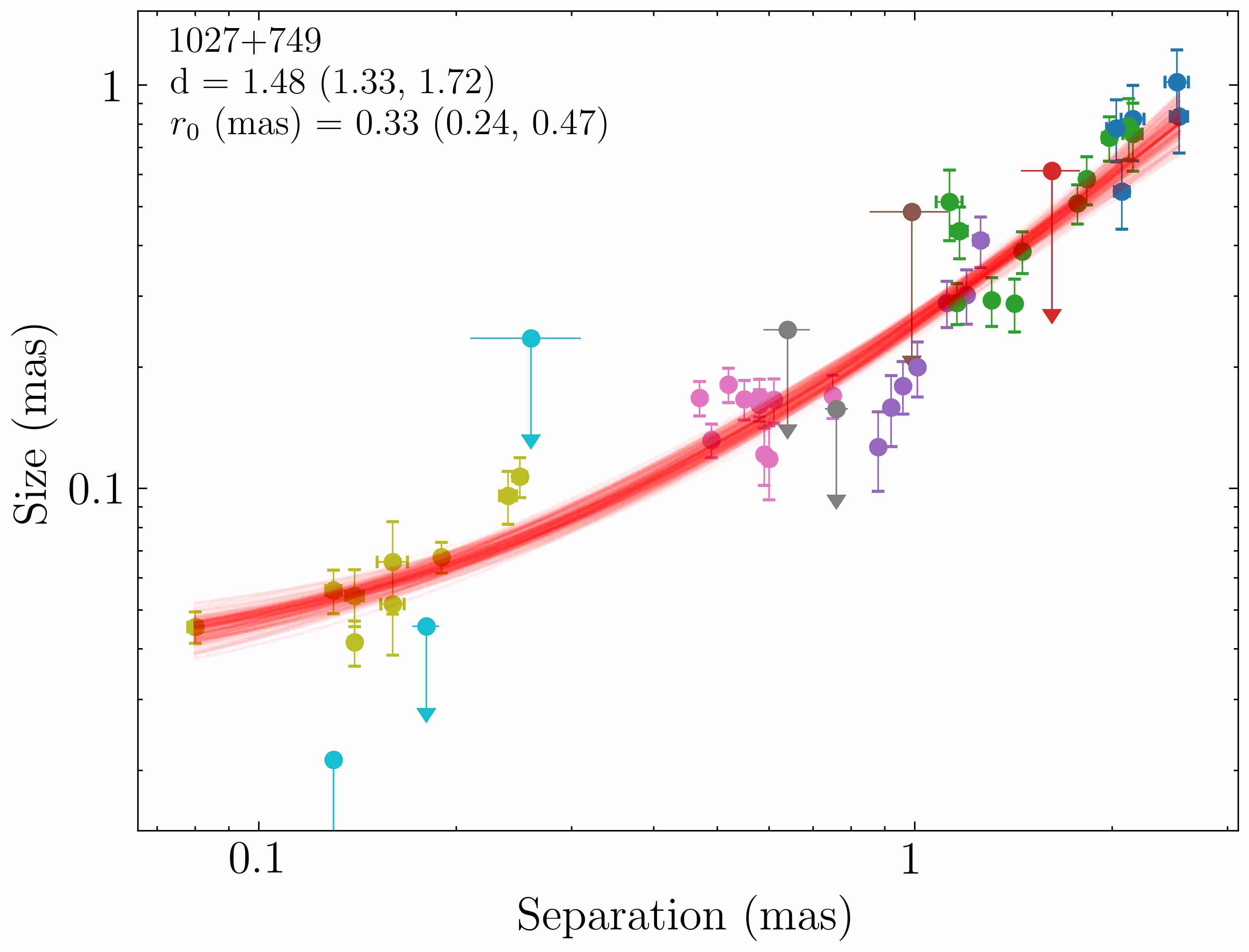}
    \caption{FWHM of the jet component versus radial distance from the 15\,GHz core for 0119$+$115, 0305$+$039 and 1027$+$749. Different colours denote individual components. The arrows indicate the upper limits, Eq.~\ref{eq: lobanov_limit}. The solid red lines indicate the results of the automated fit assuming a single power-law slope $R(r) = a_0 (r+ r_0)^{d}$. The results of the fit are given in a form of a median value and (16, 84) percentile range. The plots for all sources are available online as supplementary material. The dependencies for the sources showing a break in their distributions are shown in Fig.~\ref{fig:srgrad}.}
    \label{fig:s_grad_simple}
\end{figure*}

The paper is structured as follows: in Section~\ref{sec:obs}, we describe the sample and methodology, the results are given in Section~\ref{sec:results} and their discussion in Section~\ref{sec:disc}.
We adopt a cosmology with $\Omega_{\rm m}=0.27$, $\Omega_{\rm \Lambda}=0.73$, and $H_0 = 71~{\rm km~s}^{-1} {\rm Mpc}^{-1}$ \citep{2009ApJS..180..330K}. Spectral index $\alpha$ is defined via the flux density $S_\nu$ observed at frequency $\nu$ as $S_\mathrm{\nu} \propto \nu^{\alpha}$.

\section{Source sample and data analysis}
\label{sec:obs}

\subsection{Sample}

We used 15\,GHz VLBA observations obtained between 1994 August 5 and 2019 August 6 as part of the MOJAVE programme and the data from the NRAO archive.
The sample of the sources with 15\,GHz flux density $\gtrsim0.1$\,Jy is described in detail in \citet{2019ApJ...874...43L}. It includes a complete flux density limited sample with total VLBA flux density above 1.5~Jy for declination $\delta>-30^\circ$.
We excluded the compact symmetric objects 0026$+$346, 0108$+$388, 0646$+$600 and 2021$+$614, whose core location is uncertain.
Quasars 0414$-$189, 0615$+$820, 0640$+$090 and 1739$+$522 have a fine scale structure near the core such that the core location is an uncertain/unstable reference point for kinematic analysis, and, therefore, were not considered.
Quasar 2023$+$335 is subject to anisotropic refractive-dominated scattering \citep{2013A&A...555A..80P}. Its core location is an unstable reference point for kinematics, and it was also dropped.
We also excluded from the study BL Lac object 1515$-$273, represented by the core and quasi-stationary component, and Narrow line Seyfert 1 (NLSY1) galaxy 2115$+$000, has only three jet components visible in five available observation epochs.
The epochs 2017 Nov 18 for 0415$+$379, 2012 Feb 06 for 0912$+$297 and 2006 April 28 for 2043$+$749 were dropped due to poor quality of the data.
Therefore, the final data set comprises 447 AGNs listed in Table~\ref{tab:sources} and consists of 271 flat spectrum radio quasars, 135 BL Lacertae objects, 25 radio galaxies, 5 radio-loud NLSY1 and 11 optically unidentified sources. 
The redshift distribution for the sample is given in Fig.~\ref{fig:z_distr}.

\begin{figure}
\centering
    \includegraphics[width=0.9\columnwidth]{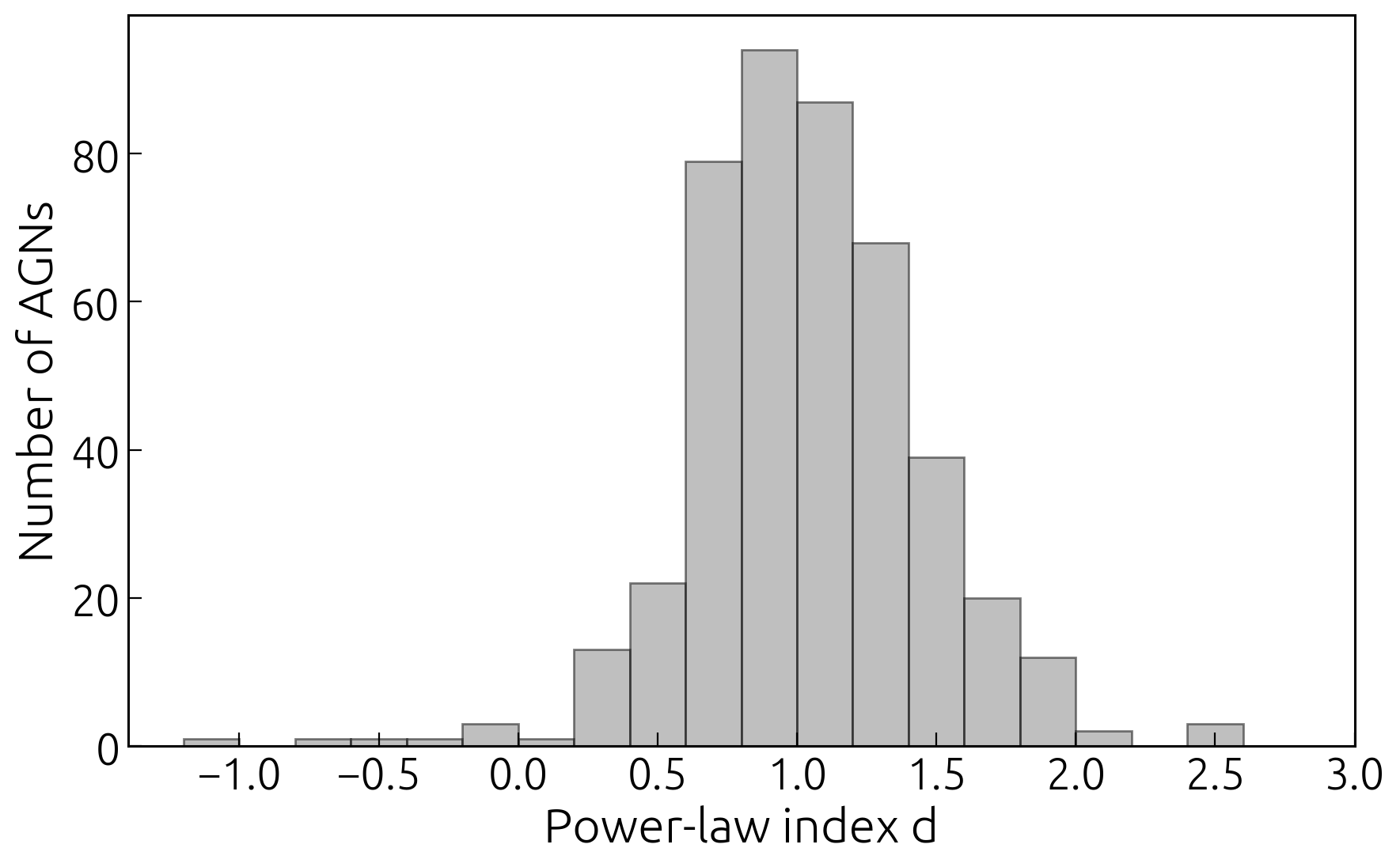}
    \caption{Histogram of the power-law indices $d$ in radial dependence of the component size $R\propto r^d$ of all sources.}
    \label{fig:fit_R_r}
\end{figure}

For the analysis, we consider the kinematics of the components from \citet{2019ApJ...874...43L} and \citet{2021ApJ...923...30L} and spectral indices from \citet{2014AJ....147..143H}.

\subsection{Estimating parameters}
\label{sec:est_pars}

Parsec-scale jets typically appear on the VLBI radio images as a series of individual bright knots which move downstream along the jet with an apparently superluminal speed \citep{2019ApJ...874...43L}. 
There is a much weaker diffuse inter-knot emission, which is difficult to image with the  VLBA, especially for z > 0.1 jets.
The jet emission structure can be parameterized by a number of two-dimensional Gaussian or delta-function intensity profiles using the \textit{modelfit} task in the \texttt{Difmap} software package \citep{1997ASPC..125...77S}.
The model description fit technique and the jet models are presented in \citet{2021ApJ...923...30L}.
Here, we focus on the jet components only, while the analysis of the core properties is presented in \cite{2005AJ....130.2473K,2009ApJ...696L..17K}, \citet{2006ApJ...642L.115H,2021ApJ...923...67H} and \citet{2018ApJ...862..151H}.
By the ‘core’ we mean the apparent base of the jet, which commonly
appears as the most compact and brightest feature in the VLBI images of AGN jets.
It is associated with the region where synchrotron self-absorbed emission becomes visible \citep{1979ApJ...232...34B} with optical depth $\tau\approx1$ and flat or inverted radio spectrum \citep{2012A&A...544A..34P,2012A&A...545A.113P}.
The measured distances of jets components, obtained at different epochs, are all referred to the position of the core.
For the sources showing a two-sided outflow morphology (e.g. 0238$-$084), we analysed only the approaching jet.
For the components represented by an elliptical Gaussian in the model fitting, we considered the equivalent-area circular FWHM size for the analysis of jet component sizes. While, when considering the brightness temperature distributions, we used both major and minor axes of the elliptical Gaussian.

An estimate of the minimum resolvable size in the image was calculated from the following relation by \citet{lobanov_05} assuming a Gaussian emission profile:
\begin{equation}
     \theta_{\rm lim} = 2 \biggr[ \frac{{\rm FWHM}\; {\rm ln}2}{\pi} {\rm ln}\Big( \frac{{\rm SNR}}{{\rm SNR}-1}\Big)\biggr]^{1/2}\,,
     \label{eq: lobanov_limit}
\end{equation}
where FWHM is the full width at half maximum of the restoring beam;  SNR denotes the ratio of the component peak amplitude to the noise rms in the naturally weighted residual image obtained after the source structure was subtracted.

We found that most widely used estimates of the component angular size uncertainties \citep{1999ASPC..180..301F,2008AJ....136..159L,2012A&A...537A..70S} provide either too big or too small errors, depending on the SNR of the component.
This is clearly apparent from the scatter of the $T_\mathrm{b}(R)$ dependence for some sources, which is much narrower than the errors. 
Thus, we employed the error estimates of \cite{2008AJ....136..159L}, but calibrated them considering $T_\mathrm{b}(R)$ dependence for ten sources, which showed a small scatter. We introduced a size error scale factor and inferred it from fitting the single power law to the $T_\mathrm{b}(R)$ dependence for these sources (see Appendix~\ref{sec:calibration_sources_pics}).
This is similar to the method used by \cite{2001ApJ...549..840H} for estimating the positional accuracy of the model components from the fit of the multi-epoch kinematic data. We found a typical scale factor $\approx 0.35$ with the sources showing its values with a range from 0.2 to 0.5 (see Appendix~ \ref{sec:calibration_sources_pics}). This is consistent with \citet{2011A&A...532A..38S}, who found that the method of \cite{2008AJ....136..159L} significantly (3--40 times) overestimated uncertainties for their multifrequency VLBA data set; see also \citet{2012A&A...545A.113P,2022MNRAS.515.1736K}. The uncertainties obtained by this method are conservative (an upper limits) because some dispersion in the $T_\mathrm{b}(R)$ dependence could have intrinsic origin even for the chosen sources with the narrowest dependence. Thus, we decided to employ a single size error scale factor value (0.35) for all subsequent inferences.

As the coordinates of the components at a given epoch are referenced through the position of the core at this epoch, one has to account not only for the uncertainty in the components positions, but also for the ‘core shuttle’ effect \citep[][and references therein]{2017MNRAS.468.4478L,2019MNRAS.485.1822P}. Thus, we introduced the $N_{\rm epoch}$ per-epoch shifts of the components relative to the mean reference position. We put a Gaussian prior with a 0.05~mas width on each of the per-epoch shift parameter and treated all shifts as independent parameters. 

To obtain the error of the brightness temperature $T_\mathrm{b}$, one should account for the residual amplitude scale uncertainty left after self-calibration. However, as with a positional error due to the core shuttle effect, adding this error equally to all components will be incorrect, as this error scales the flux of all components at any given epoch equally. Thus, we introduced the $N_{\rm epoch}$ parameters and put a Gaussian prior around 1 with a 0.05 width, assuming a 5 per cent amplitude scale error \citep{2014AJ....147..143H}. To obtain the error of $T_{\rm b}$, we used both flux density and size errors and propagated the uncertainty assuming their independence \citep[see also][]{2021ApJ...923...67H}.


We compute the brightness temperature of the model fitted VLBI components in the source rest frame as \citep{2005AJ....130.2473K}
\begin{equation}
T_\mathrm{b} = 1.22\times10^{12}\frac{S_{\rm comp}(1+z)}{\nu^2 R_\mathrm{maj} R_\mathrm{min}}~[\mathrm{K}],
\label{eq:tbgr}
\end{equation}
where $R_\mathrm{maj}$ and $R_\mathrm{min}$ are the major and minor axes of the Gaussian component (or corresponding limits, equation~\ref{eq: lobanov_limit}, whichever is larger) in milliarcseconds, $S_{\rm comp}$ is the flux density of the component in Jy and the observing frequency, $\nu$, is given in GHz.

\begin{figure*}
\includegraphics[width=0.67\columnwidth]{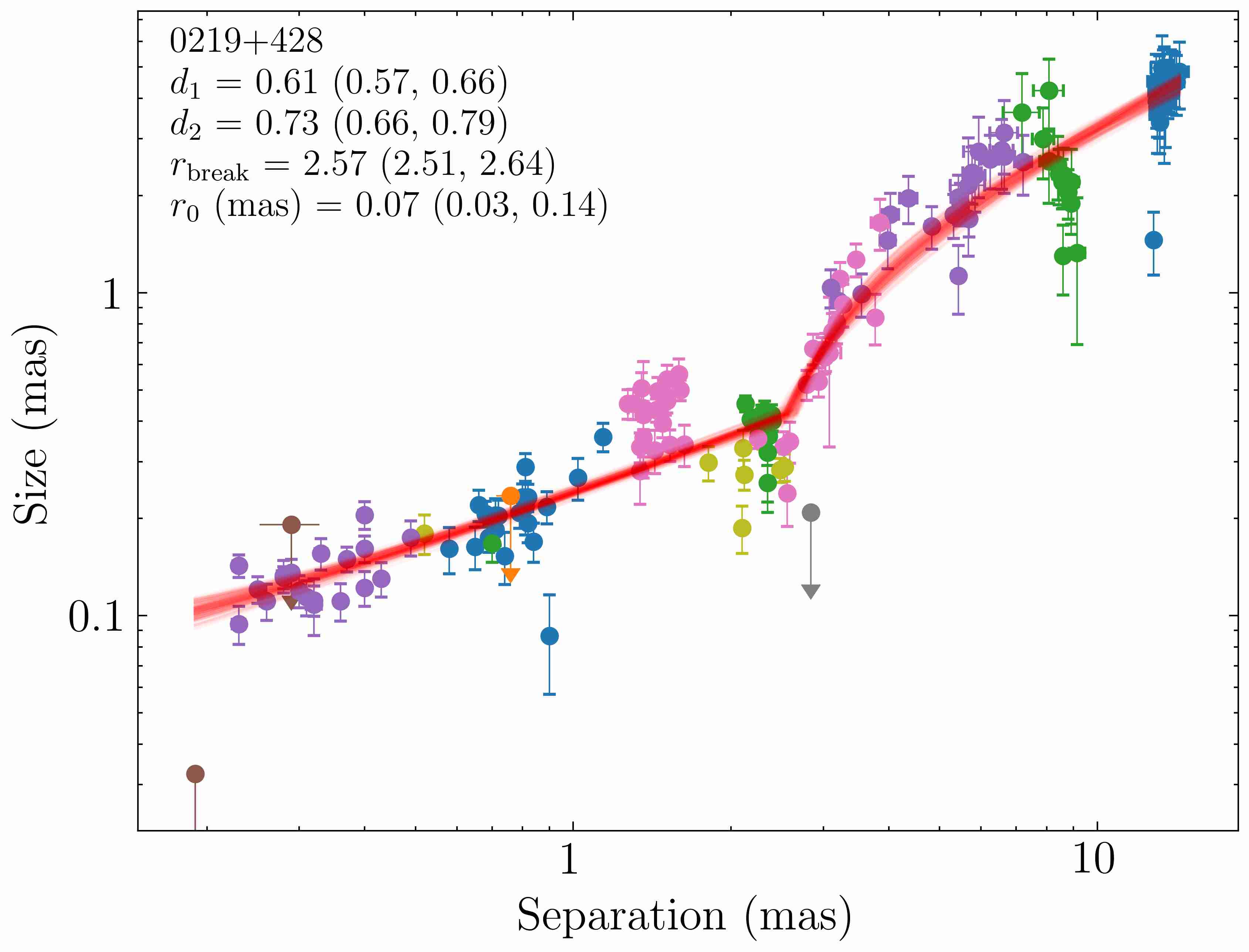}
\includegraphics[width=0.67\columnwidth]{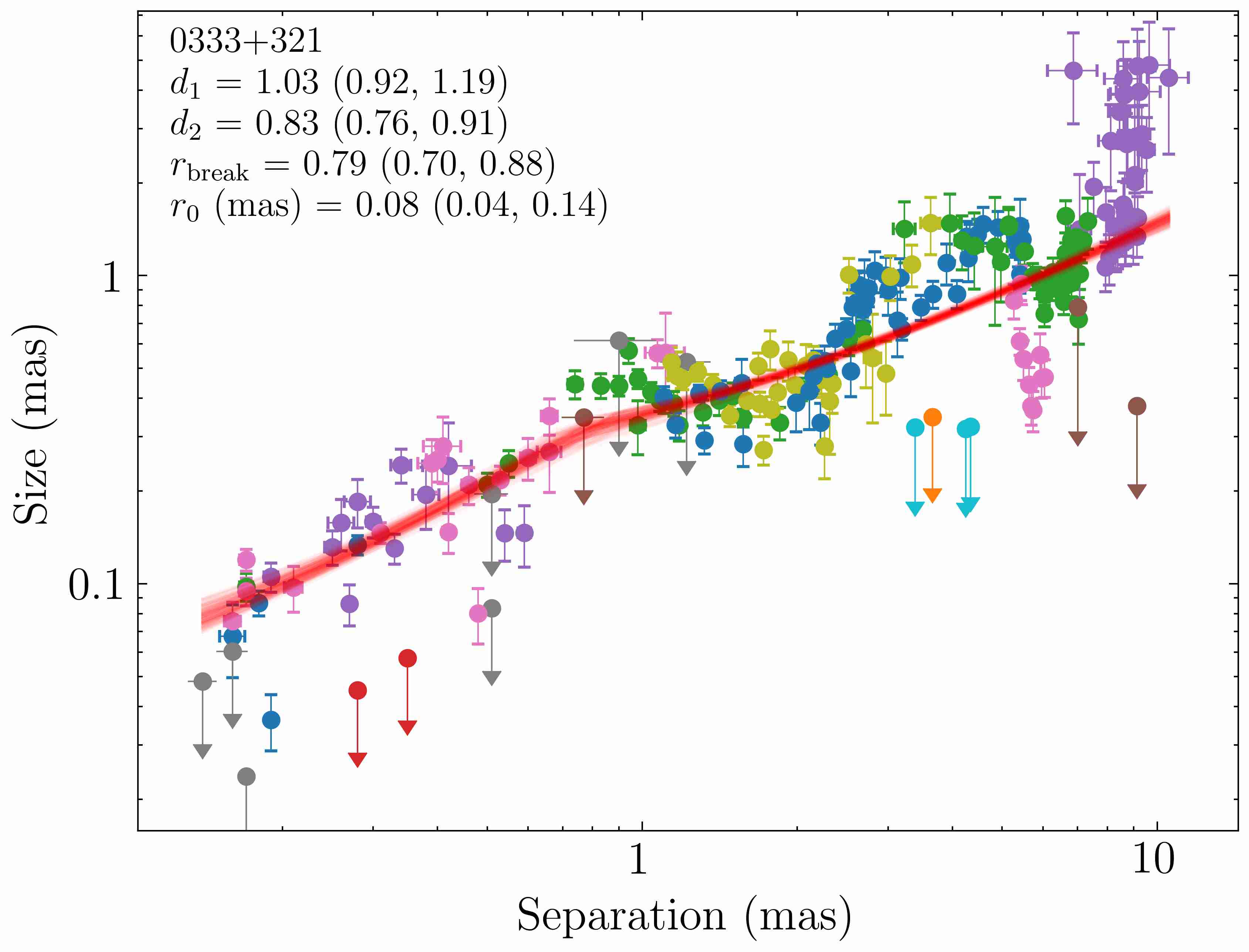}
\includegraphics[width=0.67\columnwidth]{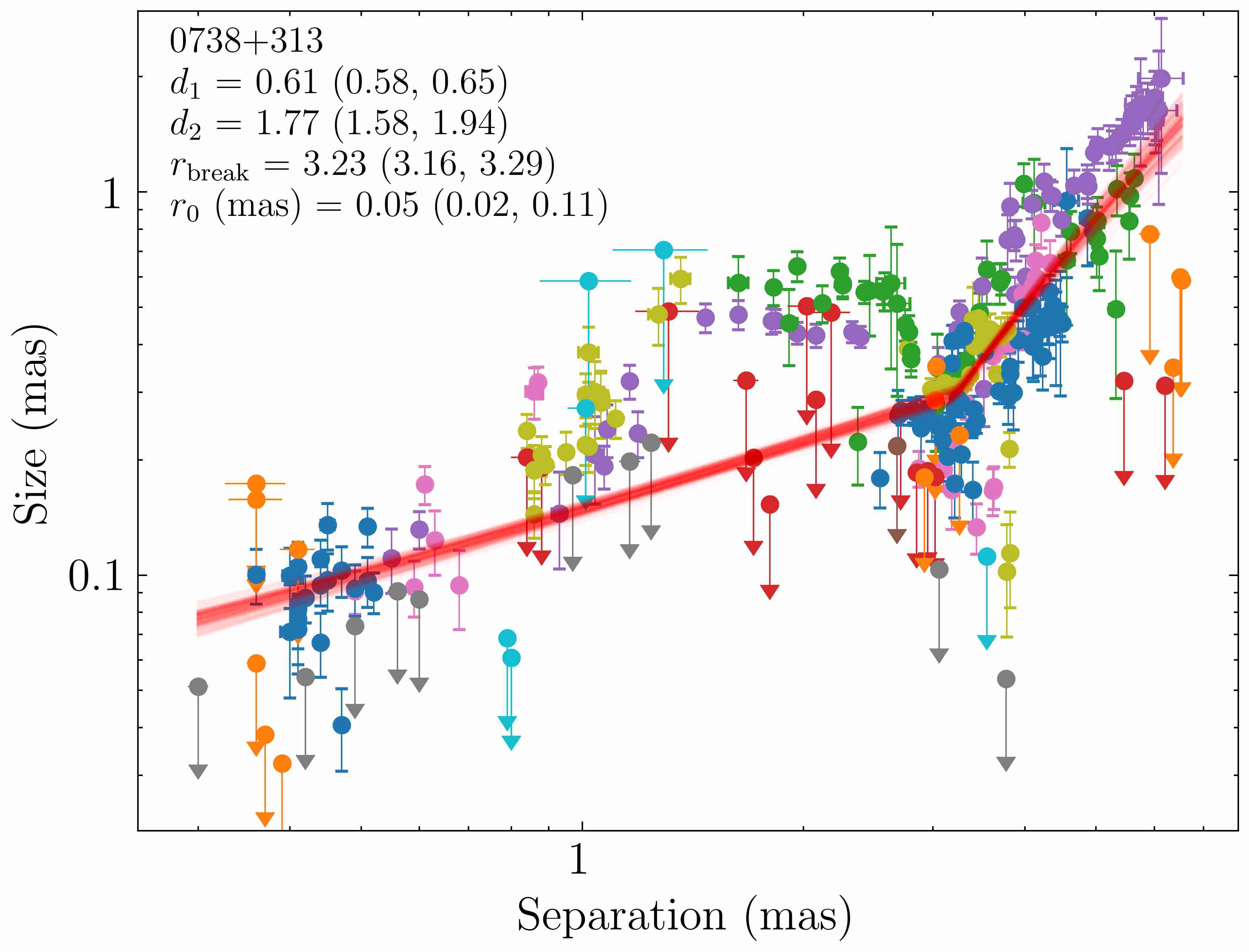}\\
\includegraphics[width=0.67\columnwidth]{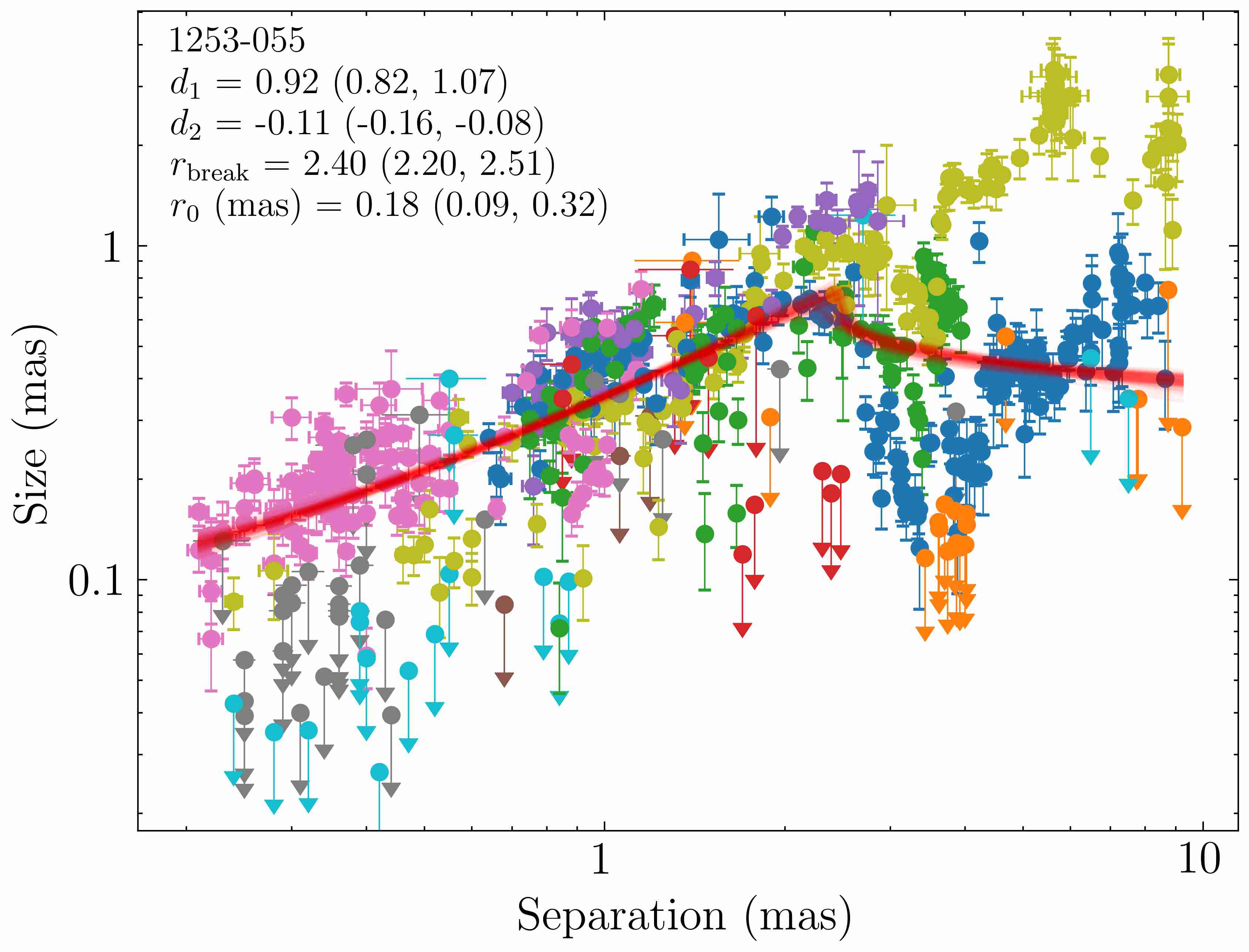}
\includegraphics[width=0.67\columnwidth]{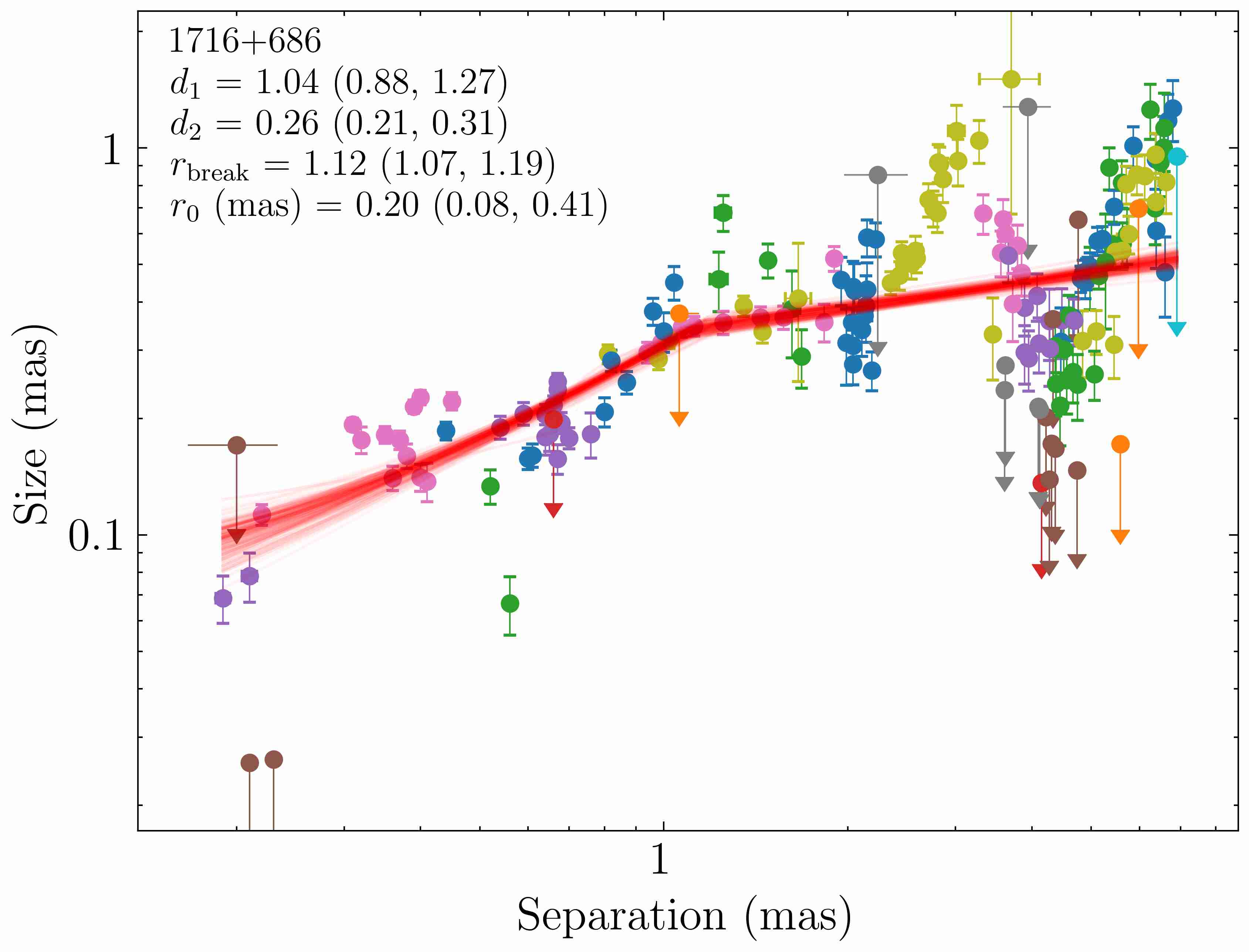}
\includegraphics[width=0.67\columnwidth]{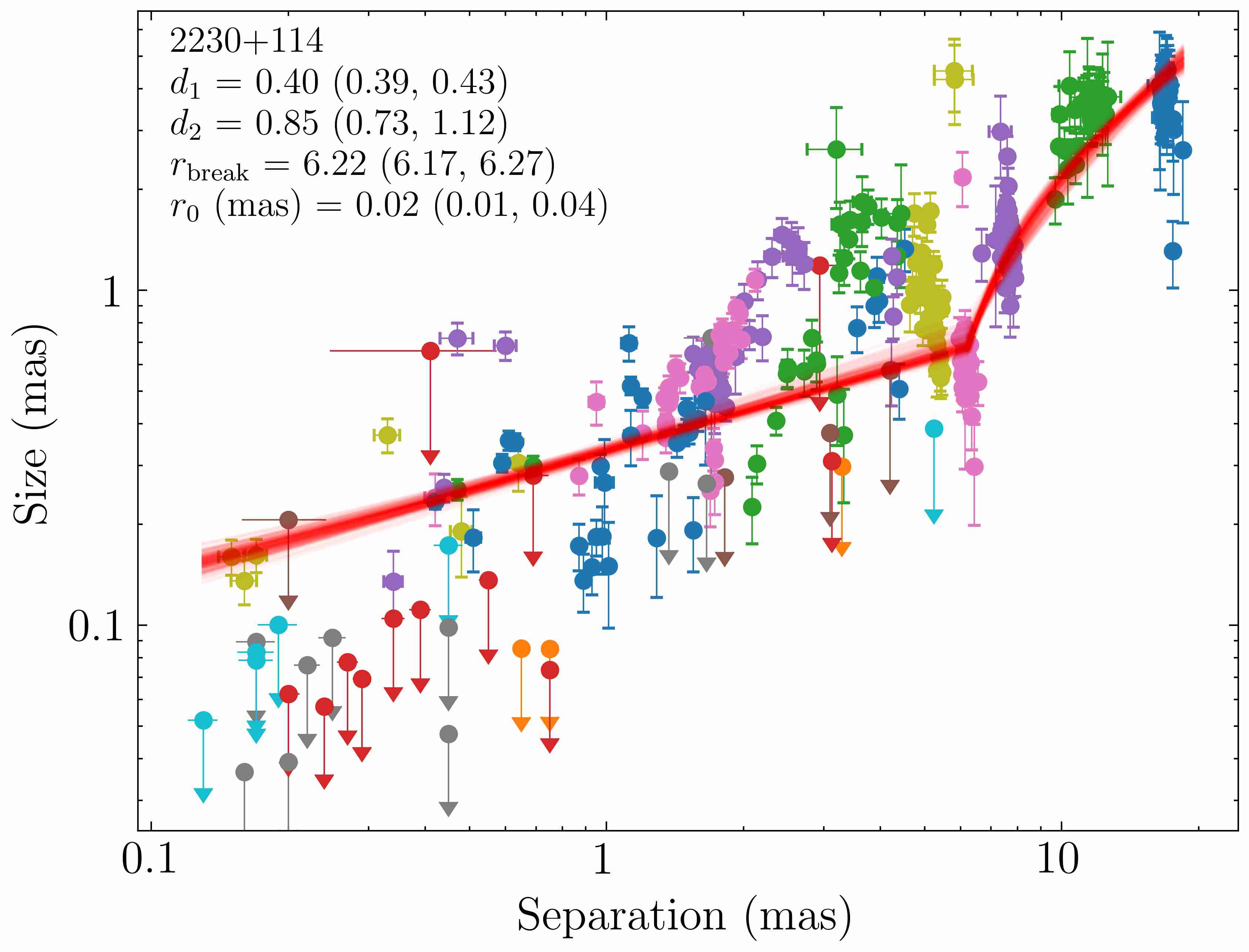}
\caption{Plots of the jet component FWHM versus the radial distance from the 15\,GHz core for the selected jets with strong evidence of a break. Different colours denote individual jet components. The arrows indicate the upper limits, Eq.~\ref{eq: lobanov_limit}. The solid lines indicate the results of the automated fit using a double power-law model, Eq.~\ref{eq:Rr}. The results for the other sources are available online as supplementary material.}
\label{fig:srgrad}
\end{figure*}

\subsection{Parameterization}
\label{sec:parametrization}

We fit the brightness temperature distributions and the size of components as a function of radial distance from the core using  single power-law models: $R(r) = a_0 (r + r_0)^d$ and $T_{\rm b}(r) = a_0 (r+r_0)^s$. Here, a prior assumption was made that $r_0$ takes on positive values because $R(r=0)$ and $T_{\rm b}(r=0)$ correspond to the size and the brightness temperature of the radio core, respectively.

Next, we performed an automated search for possible breaks in the evolution of the jet width and $T_{\rm b}$ by fitting a double (broken) power-law model to the data. To account for a geometry transition, we considered the following relations between the size of the component $R$ and its radial distance from the core $r$ \citep{2020MNRAS.495.3576K}:
\begin{equation}
\begin{split}
    R(r) = a_1(r+r_0)^{d_1},~{\rm if}~r<r_{\rm b,R}\,,\\
    R(r) = a_2(r+r_1)^{d_2},~{\rm if}~r>r_{\rm b,R}\,,
\end{split} \label{eq:Rr}
\end{equation}
where $r_0$ corresponds to the separation of the 15\,GHz VLBI core from true jet origin due to synchrotron opacity, and $r_1$ represents how much one underestimates the length of the jet if it is derived from data after the break \citep{2020MNRAS.495.3576K}. The parameter $a_2$ is chosen such that the two power-laws join each other at the break location $r_{\rm b,R}$. 

For the brightness temperature dependence on the radial distance, we use:
\begin{equation}
\begin{split}
    T_{\rm b}(r) = a_1(r+r_0)^{s_1}, ~{\rm if}~r<r_{\rm b, T}\,,\\
    T_{\rm b}(r) = a_2(r+r_1)^{s_2}, ~{\rm if}~r>r_{\rm b, T}\,.\\
\end{split} \label{eq:tbbreak}
\end{equation}
Here, parameter $r_0$ corresponds to the brightness temperature at the 15\,GHz core.

The same analysis but for the brightness temperature dependence versus the size of components $T_\mathrm{b}(R)$ is given in Appendix~\ref{sec:tbvssize}.
Throughout the text, we use a notation $r_{\rm break}$ which refers to the location of the break in any of the $R(r)$, $T_{\rm b}(r)$ and $T_{\rm b}(R)$ dependencies.

In this paper, we consider the radial separation of components from the core. In the case of the bent jets, this separation is smaller than the total path components travel along the jet. In Appendix~\ref{sec:test-realpath}, we analyse integrated displacement of the components from the core to estimate the magnitude of this effect.

\subsection{Fitting and model selection}
\label{sec:bestfit}

To fit these dependencies, we employed the \textit{Diffusive Nested Sampling} algorithm \citep{brewer2011} implemented in the \texttt{DNest4} package \citep{JSSv086i07} for sampling the posterior distribution of the model parameters. For fitting $R(r)$ and $T(r)$ dependencies, we employed the robust Student-$t$ likelihood \citep{lange1989robust} that helps in mitigation the effect of the outliers, e.g. due to the local substructures (\ref{sec:pecular_and_complex}) or single erroneous component cross-identification. At the same time for fitting $T(R)$ dependence we used the conventional Gaussian likelihood as most of the sources demonstrates the modest scatter in $T(R)$ \ref{sec:est_pars}. For the every profile, 300 fits were performed.

We made use of the Bayesian approach\footnote{We made use of the Bayesian factors as a merit of the relative model fitness \citep{2008ConPh..49...71T}. This is the difference of marginalised likelihoods $Z$ (also known as model likelihoods or evidences, which is average of the likelihood under a prior for the specific model choice) which updates the ratio of the probabilities of the competitive models from their a-priori (before collecting the data) value.} to infer the parameters space for the two models with and without a break (single and double power-law fits) and then used a Bayesian model comparison to prefer one model with respect to the other\footnote{We adopt the empirical scale from \citet{2008ConPh..49...71T} to evaluate the strength of evidence when comparing the models with and without a break (see their Table~1). Therefore, the difference of logarithms of the marginalised likelihoods ${\rm log}Z$ indicates: 
$<1$ --- inconclusive, 
[1,2) --- weak, 
[2,2.5) --- moderate, 
[2.5,5) --- strong 
and $>5$ --- decisive evidence.}.
The following additional criteria was used to reject unreliable cases: (i) $r_{\rm break}\lesssim1$\,mas, and the first slope is based only on the components located at $r<0.5$\,mas; the region before or after the break cannot consist of the measurements (ii) made over less than five individual epochs or (iii) of the only quasi-stationary component. The preferred situation is when every region is covered by more than one component observed at multiple epochs.


%% file: Sect_Rbr_dist.tex
\subsection{Distance from the core to the break position}
\label{sec:drbreak}

We examine how the break distances are distributed over different AGN classes. For this, we consider sources which show significantly broken fits in $T_{\rm b}(r)$ and $R(r)$ and divide them according to their optical classification. 

To convert the observed angular distances to a deprojected linear distances in the jet frame, we estimate the viewing angles as 
\begin{equation}
    \theta = {\rm arctan}\frac{2\beta_\mathrm{app}}{\beta^2_\mathrm{app}+\delta^2_{\rm var} -1},
\end{equation}
where $\beta_\mathrm{app}$ is the maximum observed apparent speed obtained in \citet{2021ApJ...923...30L}, and $\delta_\mathrm{var}$ is the VLBI Doppler factor estimate from \citet{2021ApJ...923...67H}. 
For the sources with unknown $\delta_\mathrm{var}$, we calculate the critical viewing angle as $\theta_\mathrm{cr} = (1+\beta^2_{\rm app})^{-0.5}$, which maximizes the Lorentz factor \citep{2021ApJ...923...67H}. 
These requirements reduced the number of considered sources to 207. 

The resultant distribution of the deprojected distances is presented in Fig.~\ref{fig:rbr_dist}.
Transitions occur at smaller distances for the jets in radio galaxies ($<$100\,pc), and all the jets in quasars show a break on scales larger than 10\,pc, with the median value being about 300\,pc. At the same time, the BL~Lac objects are localized in between, with the median of 56~pc.
The distributions of the deprojected distances for the sources with a comparable range of redshifts ($z<0.3$, 52 sources) is shown in Fig.~\ref{fig:rbr_dist}. From the AD-test, quasars and radio galaxies are drawn from a different population, $p_{\rm AD}=0.0075$. 
This result is consistent with the findings of \citet{2015MNRAS.453.4070P}, \citet{2017A&A...606A.103H}, \citet{2021AA...647A..67B}, \citet{2021AA...649A.153C} that the jet recollimation zone in quasars with more powerful jets is located at larger distances from the central engine than in BL~Lacs and radio galaxies.
Thus, acceleration and collimation occur over a more extended region in quasars with respect to jets in BL~Lacs and radio galaxies.
This scenario is in agreement with the kinematic study of \citet{2019ApJ...874...43L}, who found a strong correlation between apparent jet speed and synchrotron peak frequency, with the highest jet speeds being found only in AGNs with low synchrotron peak frequency values, i.e.\ in powerful jets.

%% file: Sect_Compar.tex
\section{Comparison with other studies}

The jet morphology in a large sample of AGN jets was studied by \citet{2017MNRAS.468.4992P}, who put together from 5 up to 137 single-epoch total intensity images obtained at 15\,GHz at time intervals from 1.3 to 21 years.
Analysing the stacked images of 362 sources, they found a typical jet geometry close to conical at scales from hundreds to thousands parsecs.
Later, \citet{2020MNRAS.495.3576K} performed a similar analysis looking for the geometry transition signs from a parabolic to conical shape. They used stacked images of 319 jets at 15~GHz supplemented by singe-epoch 1.4~GHz maps for 95 of them and found median $d=1.02$. Our median value over all sources of $d=1.02$ is consistent with the result of the aforementioned works and, therefore, indicates that the component size is a good tracer of jet geometry.
Whereas some components having size smaller than the jet width may represent local substructures.

An early study \citep{Kadlerphdthesis} of the jet width profiles formed by individual jet components for 19 sources using single-epoch observations at 1.7 and 5\,GHz obtained median $d_{\rm 1.7~GHz}=0.8\pm0.1$ and $d_{\rm 5~GHz}=0.8\pm0.1$. For the same sources, our median value is $d=1.1\pm0.2$.
Recently, \citet{2022A&A...660A...1B} studied collimation profiles of 28 AGN jets at 15 and 43\,GHz, selected such that they are contained in both\footnote{Except for complex-structure radio galaxy 0316+413 (3C84).} the MOJAVE data archive and the Boston University blazar group sample archive \citep{2016Galax...4...47J}.
\citet{2022A&A...660A...1B} also used modelfit components in the analysis and obtained a median value of $d=0.8\pm0.1$. 
This sample of AGNs is included in our data set, and from the source-by-source comparison, we obtained median $d=0.9\pm0.1$, thus our results are in good agreement with these two studies.

The brightness temperature gradients were also analysed by \citet{2012A&A...544A..34P} in application to 30 AGNs having a rich jet structure consisting of at least three model fitted jet components at 2 and 8\,GHz, selected from a sample of 370 bright, flat-spectrum, compact extragalactic radio sources.
\citet{2012A&A...544A..34P} obtained the mean value of the power-law index $s_{\rm 2~GHz}=-2.2\pm0.1$, $s_{\rm 8~GHz} = -2.1 \pm 0.1$. 
Our median value calculated for the common 22 sources at 15~GHz gave $s=-2.8\pm0.2$.
For the sample of 19 sources observed at 1.7 and 5\,GHz, \citet{Kadlerphdthesis} estimated an average $s_{\rm 1.7~GHz}=-2.2\pm0.1$ and $s_{\rm 5~GHz}=-2.3\pm0.2$.
Considering the 11 common sources with \citet{Kadlerphdthesis}, we obtained the average value $s=-3.2\pm0.2$.
The resultant average value of the power-law index over a sample of 28 sources at 15 and 43\,GHz \citep{2022A&A...660A...1B} is $s=-2.2\pm0.1$, compared to our value $s=-2.5\pm0.1$ for the same sources.

The analysis of the brightness temperature versus component size for the observations of 30 sources at 2 and 8\,GHz \citep{2012A&A...544A..34P} resulted in the mean $\hat{s}_{\rm 2~GHz} = -3.0 \pm 0.4$, $\hat{s}_{\rm 8~GHz} =-3.1\pm0.3$. Considering the common 22 sources, we obtained $\hat{s}=-3.0\pm0.1$.
For the data set of 19 sources at 1.7 and 5\,GHz, \citet{Kadlerphdthesis} estimated average $\hat{s}_{\rm 1.7~GHz} = \hat{s}_{\rm 5~GHz} = -2.6$. Considering the same sources, we got $\hat{s}=-2.9\pm0.1$.

\citet{2020MNRAS.495.3576K} noted that even single-epoch low-frequency VLBI imaging observations with a good $uv$-coverage are sensitive enough to detect the jet morphology at large scales. 
In section~\ref{sec:nonradial} we discussed strong variations of the component position angles in the inner jet, thus at any given time, traveling jet components do not fill out the entire jet width.
Therefore, we suggest that the discrepancy between our study and the studies which used single-epoch observations could be due to a sparse coverage of the latter, i.e. they do not sample enough of the jet structure. In turn, this work contains too many small components which could represent local sub-structures which can also introduce some bias. 
Another reason for this discrepancy could be the effect of missing short baselines at 15\,GHz, which results in resolving out structures visible in the lower frequency data. Also, a steeper spectral index of the components in a high-frequency range can cause a smaller values of the brightness temperature gradients (equation~\ref{eq:eps_ksi} and equation~\ref{eq:ss_index}).
For example, flow density stratification across the jet width seen in 3C~273 \citep{2021AA...654A..27B}.

Finally, the discrepancy could be due to the fitting procedure. Unlike other studies, our fitting procedure employs robust Student-$t$ likelihood (section~\ref{sec:bestfit}) and advanced error model (section~\ref{sec:est_pars}), including uncertainties in the position of components, core shuttle effect and amplitude uncertainty left after the self-calibration. Accounting for various uncertainties in the position eliminates possible attenuation bias \citep{carroll2006measurement}, that could shift slope estimates toward zero. To investigate the influence of the likelihood type and positional uncertainties on the fit, we selected 11 sources common to \citet{Kadlerphdthesis} and fitted the $T(r)$ dependence using various combinations of the likelihoods (Gaussian or Student-$t$) and error contribution (with and without accounting for the aforementioned uncertainties). It turns out that using the robust likelihood results in a steeper fitted dependence $T(r)$. 
Interestingly, our results for $R(r)$ obtained with the same methods are consistent with those from other studies \citep{Kadlerphdthesis,2020MNRAS.495.3576K,2022A&A...660A...1B}. Moreover, as shown in \autoref{sec:tbvssize}, our estimates of slopes $d$, $s$ and $\hat{s}$ are self-consistent in the sense that $s$/$\hat{s}$ = $d$. This implies that the underlying $T(r)$ dependence is heavily influenced by the various effects (section~\ref{sec:pecular_and_complex}) that could bias the estimate of its slope.

%% file: Sect_Disc.tex
\subsection{Particle density gradient}
\label{sec:collimation_and_brightness_temerature}

\begin{figure}
    \centering
    \includegraphics[width=0.9\columnwidth]{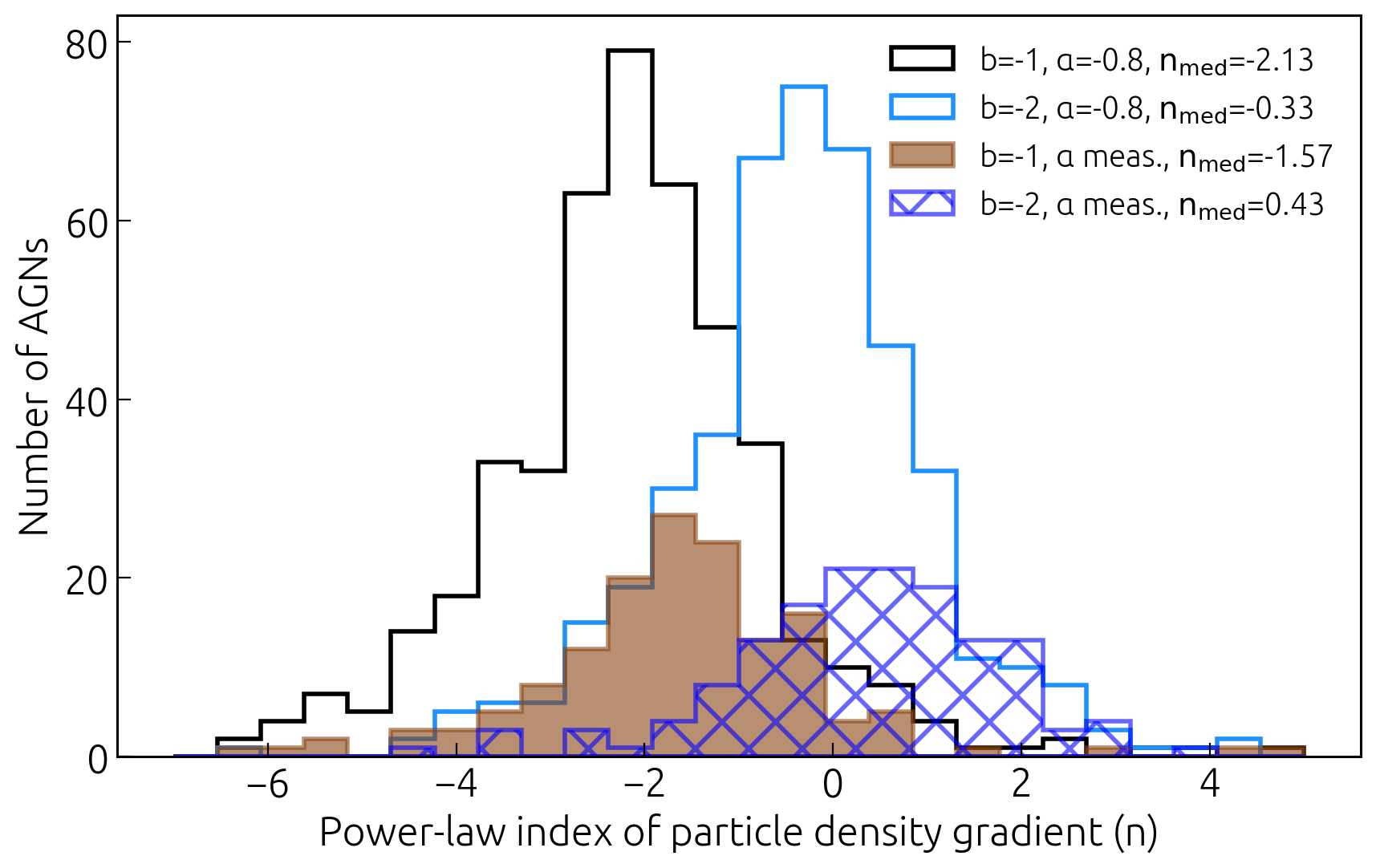}
    \caption{Histogram of the power-law indices $n$ in a particle density distribution along the jet $N(r)\propto r^{n}$, assuming toroidal ($b=-1$) and poloidal ($b=-2$) scaling of the magnetic field strength, $B(r)\propto r^{b}$. Median values of the resultant power law indices $n$ are given. The unfilled distributions are calculated for 447 sources and $\alpha=-0.8$ using fitted values for indices $s$ and $d$, following Eq.~\ref{eq:eps_ksi}. For the filled and dashed distributions we considered the average spectral indices of jet components measured between 8 and 15\,GHz \citep{2014AJ....147..143H}, which reduces the number of sources to 148. }
    \label{fig:bn}
\end{figure}

The estimated power-law indices $s$ and $d$ give a possibility for deriving a parameter range of physical conditions along the jets at pc-scales probed by 15\,GHz VLBA observations.
We considered two different configurations of the magnetic field, $b=-1$ (toroidal) and $b=-2$ (poloidal) in application to the observed $T_{\rm b}(r)$ and $R(r)$ gradients and calculated the particle density index $n$ using the equation~\ref{eq:eps_ksi}.
We assumed constant bulk motion speed (i.e. $\delta \propto r^{p}$ and $p = 0$) 
and $\alpha=-0.8$, which is the mean of the jet component spectral indices distribution over 190 MOJAVE sources observed between 8 and 15\,GHz in 2006 \citep[$\alpha=-0.81\pm0.02$;][]{2014AJ....147..143H}. 
In Fig.~\ref{fig:bn}, we plot the resultant distribution of the $n$-index values, considering $\alpha=-0.8$ for all sources and direct individual measurements of the average spectral indices along ridgelines of 149 jets \citep{2014AJ....147..143H}.

\begin{table}
    \centering
    \caption{Parameter range for the power-law index of particle density gradient $n$, which is estimated from the power-law gradients of $T_{\rm b}(r)$ and $R(r)$  under an assumption of different magnetic field configurations.}
    \label{tab:n_ind}
    \begin{tabular}{lcccc}
        \hline
         Magnetic field & \multicolumn{2}{c}{$n$}\\
         topology& $\alpha_\mathrm{mean}=-0.8$& $\alpha_\mathrm{ind}$\\
         (1) & (2) & (3)\\
        \hline
        toroidal, $b=-1$ & $-2.13$ & $-1.57$\\
        poloidal, $b=-2$ & $-0.33$ & 0.43\\
        \hline
    \end{tabular}
    \flushleft Note: Column (2) presents $n$ values estimated for $\alpha_\mathrm{mean}$ which is the mean of the jet component spectral indices distribution obtained for the 190 MOJAVE sources, observed between 8 and 15\,GHz.
    Column (3) shows $n$ for the average spectral index value $\alpha_\mathrm{ind}$ calculated individually along ridgelines of the 149 MOJAVE sources. Spectral index values are taken from \citet{2014AJ....147..143H}.
\end{table}

Table~\ref{tab:n_ind} summarises the estimated median values of the power-law index of particle density gradient.
The poloidal magnetic field configuration ($b=-2$) leads to implausibly small $n$-values. Meantime, a toroidal magnetic field ($b=-1$) results in median values expected from the scenario of equipartition between the magnetic and emitting particles energy densities ($b=-1$, $n=-2$).
Moreover, a predominantly toroidal magnetic field configuration is expected in the Poynting flux dominated jets and the models where transverse magnetic field component is enhanced by a series of shocks.
Besides, the resultant value of $n$-index agrees well with the scenario which assumes mass conservation and a particle density gradient defined by the geometry of the outflow, $N\propto R^{-2}$ and, therefore, $N\propto r^{-2d}$, $n=-2d$.

\begin{figure}
    \centering
    \includegraphics[trim=0cm 0cm 0.5cm 0cm, width=0.85\columnwidth]{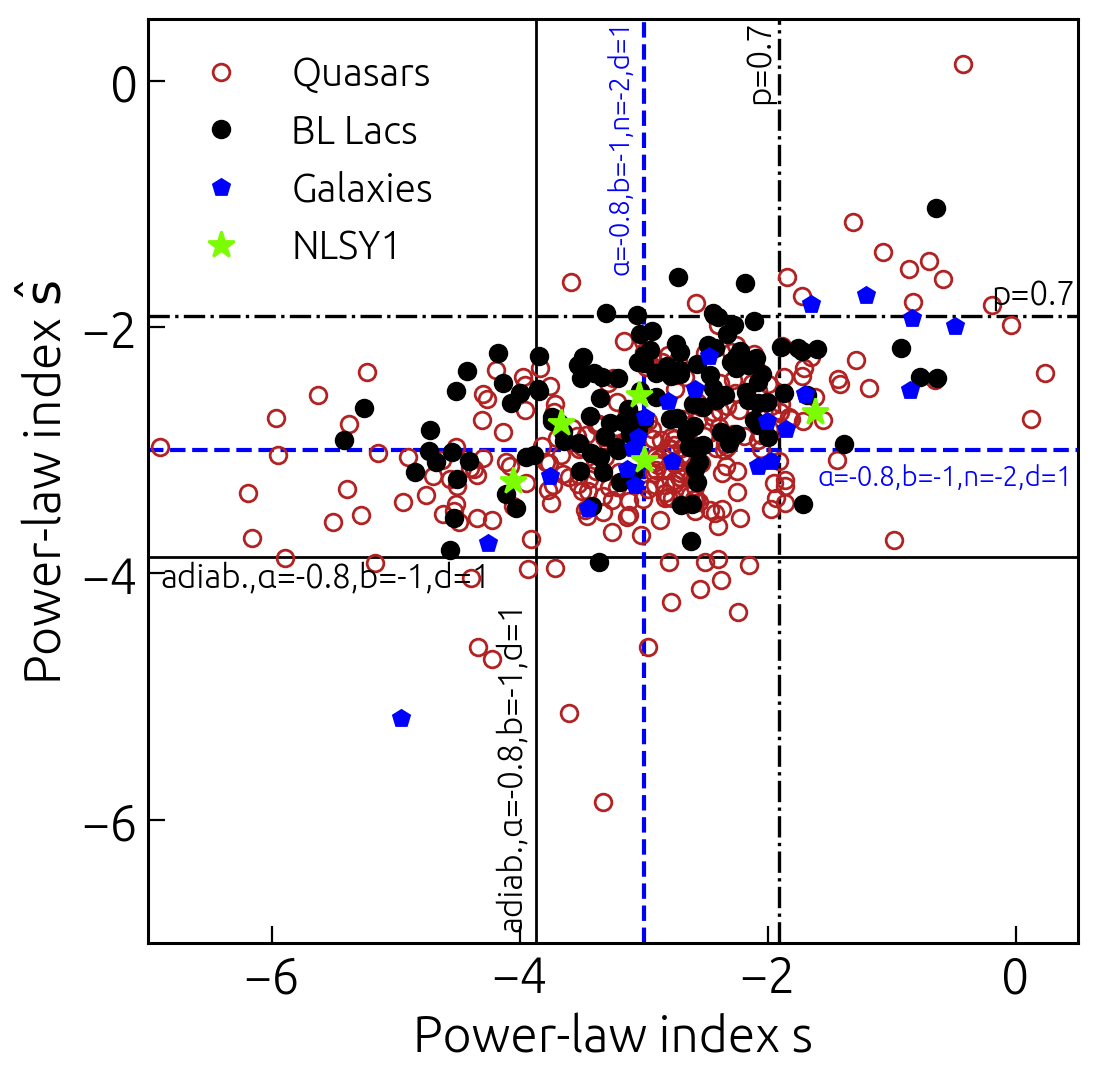}
    \caption{Dependence of $\hat{s}$ versus $s$ in $T_{\rm b}\propto r^{s}$ and $T_{\rm b}\propto R^{\hat{s}}$, respectively, and the parameter space of $n$ and $b$. The dashed lines indicate the region under assumption of moderate variation in Doppler factor, i.e.\ $\delta \propto r^{p}$ and $p = 0.7$. The solid line marks the lower bounds for a shock-in-jet model and assumption of $b=-1$, $\alpha=-0.8$. The dashed-dotted lines indicate values $n=-2$, $b=-1$, $d=1$ and $\alpha=-0.8$.}
    \label{fig:tb_td}
\end{figure}

\subsection{Testing the adiabatic expansion in the shock-in-jet model}

We can place the observed brightness temperature gradient in the context of the shock-in-jet model \citep{1985ApJ...298..114M}, which assumes that each jet component is an independent relativistic shock propagating downstream a jet. We consider that the adiabatic expansion is the dominant energy loss mechanism, while Compton and synchrotron losses can be neglected \citep{2009ApJ...696.1142M}. 
Then the evolution of the brightness temperature can be defined as \citep{1999ApJ...521..509L,2015A&A...578A.123R}:
\begin{equation}
\label{eq:eps_adiab}
    \begin{split}
    & T_{\rm b,jet}(R)\propto R^{\hat{s}_{\rm ad}},\\
    & \hat{s}_{\rm ad} = -[2(2\epsilon+1)-3b(\epsilon+1)-3p(\epsilon+3)]/6\,,\\
    \end{split}
\end{equation}
Here, the energy of electrons scales as $N(\gamma) \propto \gamma^{-\epsilon}$, which results in the spectral index $\alpha = (1-\epsilon)/2$. 
Assuming the mean spectral index value of $\alpha=-0.8$ (see section~\ref{sec:collimation_and_brightness_temerature}), $\epsilon = 2.6$.
In Fig.~\ref{fig:tb_td}, we plot lower bounds assuming $b=-1$, $d=1$ and $\epsilon = 2.6$ by solid lines.
The majority of sources lie above this area, which rules out the assumption of a constant Doppler factor.
\citet{2009ApJ...706.1253H, 2015ApJ...798..134H} found evidence for increasing Lorentz factors down the jet up to de-projected distances of $\thicksim$100 parsecs from the jet core.
Thus, we consider variability in the Doppler factor, i.e. $\delta\propto r^{0.7}$ and show the corresponding values by dash-dotted lines in Fig.~\ref{fig:tb_td}. 
Under this condition, the observed values of $s$ and $\hat{s}$ are consistent with the adiabatic loss phase. 
The same conclusion was obtained, for instance, for BL~Lac object 0716+714 \citep{2015A&A...578A.123R} and quasar CTA~102 \citep{2013AA...551A..32F}.


\subsection{Jet shape transition and acceleration}

In the previous sections we have assumed a conical jet with a constant speed. In this section we derive the relations for the brightness temperature size dependence assuming accelerating jet. The corresponding expressions for the $T_{\rm b}$ gradients can be easily derived under assumption of the particular $R(r)$ dependence. We also discuss the imprint of the jet shape transition due to cease of the bulk plasma acceleration \citep{2020MNRAS.495.3576K} on the observed $T_{\rm b}(R)$ dependence.

In addition to direct observations of jet acceleration in MOJAVE jets \citep{2009ApJ...706.1253H, 2015ApJ...798..134H}, jet geometry transitions have been observed in numerous sources
\citep{2012ApJ...745L..28A,2016ApJ...833..288T,2018ApJ...854..148N,2018ApJ...860..141H,2019ApJ...886...85A,2020MNRAS.495.3576K,2021AA...647A..67B,2021AA...649A.153C}. According to equations~\ref{eq:eps_ksi} and \ref{eq:ss_index}, the change only in the geometry profile $R(r)$ results in a break in the brightness temperature radial profile $T_{\rm b}(r)$ but not in the $T_{\rm b}(R)$. However, if the change of the jet geometry is due to the saturation of MHD acceleration process \citep[e.g][]{2020MNRAS.495.3576K}, it should be accompanied by the change of the both profiles $T_{\rm b}(r)$ and $T_{\rm b}(R)$.

In such jet viewed at the fixed angle $\theta$, the Doppler factor grows along the jet up to its maximal value $\delta_{\rm max} \approx 1/\sin\theta$ at a distance where $\Gamma \approx \delta_{\rm max}$. Then $\delta$ decreases up to the region $r_{\rm break}$ where the saturation takes place with $\Gamma \approx $~const and the jet becomes almost conical \citep[see figs. 8, 9 in ][]{2020MNRAS.495.3576K}. Thus, downstream $r_{\rm break}$, equations~\ref{eq:eps_ksi} and \ref{eq:ss_index} are applicable. However, in the accelerating part of the jet, i.e. upstream the break, the relations for the brightness temperature gradients depend on the geometry of the magnetic field in the jet and the relation between the emitting particles and the magnetic field energy densities. 
For the toroidal field $B \propto (\Gamma R)^{-1}$ and for the poloidal field $B \propto R^{-2}$, assuming velocity $\beta \approx 1$. The number density of emitting particles is defined from the continuity equation as $N \propto \Gamma^{-1} R^{-2}$.
However, in the local equipartition, $N \propto B^{2}$ and for the adiabatic case $N \propto (\Gamma R^x)^{-(s+2)/3}$, where $x = 2$ for 2D-expansion \citep{1997ApJ...483..178B,2004ApJ...604L..81G,2006MNRAS.367.1083K} and $x = 3$ for 3D-expansion \citep{2010RAA....10...47Q}. 

To obtain the expression for the observed $T_{\rm b}(R)$ dependence in the case of the accelerating jet, we consider the brightness temperature at the plasma frame $T_{\rm b}'$ connected with the measured value through:
\begin{equation}
    T_{\rm b} = \delta T_{\rm b}' \propto \delta N B^{(1-\alpha)} R {\nu'}^{(\alpha - 2)}.
    \label{eq:Tbconnections}
\end{equation}
Here, $T_{\rm b}$ is measured at the observed frequency $\nu$ (up to the constant $(1+z)$ factor), while $T_{\rm b}'$ is measured at the frequency $\nu' = \nu/\delta$ which changes along the accelerating jet. Substituting this to equation~\ref{eq:Tbconnections}, we obtain at fixed observing frequency \citep{1994ApJ...426...51R}:
\begin{equation}
    T_{\rm b} \propto N B^{(1-\alpha)} R {\delta}^{(3 - \alpha)}.
    \label{eq:Tbacceleration}
\end{equation}
Substituting the plasma frame fields radius dependencies in equation~\ref{eq:Tbacceleration}, we obtain the following scalings:
\begin{equation}
\begin{split}
    T_{\rm b}^{\rm tor}(R) \propto & \delta(\frac{\delta}{R \Gamma})^{2-\alpha}\,, \\
    T_{\rm b}^{\rm pol}(R) \propto & R^{(2\alpha-3)}\delta^{3-\alpha}/\Gamma\,, \\
    T_{\rm b}^{\rm eq,tor}(R) \propto & (R \Gamma)^{\alpha-3} R \delta^{3-\alpha}\,, \\
    T_{\rm b}^{\rm eq,pol}(R) \propto & R^{-5+2\alpha}\delta^{3-\alpha}\,, \\
    T_{\rm b}^{\rm ad,2D,tor}(R) \propto & R^{-(7\epsilon+5)/6} \Gamma^{-(5\epsilon+7)/6} \delta^{(\epsilon+5)/2}\,, \\
    T_{\rm b}^{\rm ad,2D,pol}(R) \propto & R^{-(5\epsilon+4)/3} \Gamma^{-(\epsilon+2)/3} \delta^{(\epsilon+5)/2}\,,\\
    T_{\rm b}^{\rm ad,3D,tor}(R) \propto & R^{-3(\epsilon+1)/2} \Gamma^{-(5\epsilon+7)/6} \delta^{(\epsilon+5)/2}\,, \\
    T_{\rm b}^{\rm ad,3D,pol}(R) \propto & R^{-2(\epsilon+1)} \Gamma^{-(\epsilon+2)/3} \delta^{(\epsilon+5)/2}\,,
\end{split}
\end{equation}
where `tor' and 'pol' denote the toroidal and poloidal magnetic field case, 'eq' stands for the equipartition and 'ad' for the adiabatic case. The terms 2D and 3D correspond to the expansion type in the adiabatic case.

Thus, to derive the expressions for the brightness temperature gradients in the accelerating jet, one has to specify the velocity profile $\Gamma(r)$ and assume an asymptotic for $\delta(r)$.
For example, assuming the universal acceleration profile $\Gamma \propto R$ \citep{2022MNRAS.509.1899N}, we obtain that $T_{\rm b}(R)$ does not depend on the magnetic field geometry in all four cases considered:
\begin{equation}
    \begin{split}
        T_{\rm b}(R) \propto & R^{(2-\alpha)(\hat{p}-2) + \hat{p}}\,, \\
        T_{\rm b}^{\rm eq}(R) \propto & R^{2\alpha - 5 + \hat{p}(3-\alpha)}\,, \\
        T_{\rm b}^{\rm ad,2D}(R) \propto & R^{-2(\epsilon+1) + \hat{p}(\epsilon+5)/2}\,, \\
        T_{\rm b}^{\rm ad,3D}(R) \propto & R^{-(7\epsilon+8)/3 + \hat{p}(\epsilon+5)/2}\,.
    \end{split}
\end{equation}
Here, $\hat{p}$ is the exponent of the Doppler factor radius dependence $\delta(R) \propto R^{\hat{p}}$. One can assume further that $\delta \propto \Gamma$ for the jet region with $\Gamma < 1/\theta$ \citep{2022MNRAS.509.1899N}, thus $\hat{p}$ = 1. In that case we obtain:
\begin{equation}
    \begin{split}
        T_{\rm b}(R) \propto & R^{\alpha - 1}\,,\\
        T_{\rm b}^{\rm eq}(R) \propto & R^{\alpha - 2}\,,\\
        T_{\rm b}^{\rm ad,2D}(R) \propto & R^{(1 - 3\epsilon)/2} \propto  R^{3\alpha-1}\,,\\
        T_{\rm b}^{\rm ad,3D}(R) \propto & R^{-(11\epsilon+1)/6} \propto R^{(11\alpha-6)/3}\,.
    \end{split}
    \label{eq:DproptoGamma}
\end{equation}

Comparing with equation~\ref{eq:ss_index} and equation~\ref{eq:eps_adiab} we see that with the assumptions of $\Gamma \propto R$ and $\delta \propto \Gamma$, the break in $T_\mathrm{b}(R)$ is not expected with constant $\hat{s} = -2.5$ for canonical ($\hat{n} = -2$, $\hat{b} = -1$ and $\alpha = -0.5$) in the conical domain case. The same holds with constant $\hat{s} = -3.8$ for the poloidal field with the 3D expansion adiabatic case.

However, in the general case, when only $\Gamma \propto R$ holds, the exponent $\hat{s}$ in the jet region upstream of the break $r_{\rm break}$ depends on the break position relative to the region of the maximal Doppler factor $\delta_{\rm max}$ in the jet. The latter strongly depends on the viewing angle \citep{2019MNRAS.486..430K}.
The value of $\hat{p}$ could be negative if the break occurs further downstream of the region with $\delta_{\rm max}$, where $\delta$ decreases. In that case, the exponent $\hat{s}$ upstream the break will be lower than those, presented in equation~\ref{eq:DproptoGamma}. 

For nearby AGN with the detected geometry transition, $r_{\rm break}$ is possibly located in the region of decreasing $\delta$, hence we expect the break in the $T_{\rm b}(R)$ dependence at $r_{\rm break}$ from steep to flat \citep[e.g., see fig.~11 in][]{2019MNRAS.486..430K}.
We observe a notable change in the $T_{\rm b}\propto R^{\hat{s}}$ slopes from steep to flat upstream and downstream the break, correspondingly (see figure~\ref{fig:tbd}).
For example, for the radio galaxy 1637$+$826 with the highest viewing angle in the sample of \citet{2020MNRAS.495.3576K}, $\hat{s}$-index at the break position changes from $-3.02\pm0.14$ to $-1.37\pm0.19$. 
For Cygnus~A, $\hat{s}$-index changes from $-3.9\pm0.3$ to $-1.8\pm0.7$ and for the radio galaxy PKS 1514$+$00 -- from $-3.3\pm0.9$ to $-1.9\pm0.4$.

To summarize, in general case the jet acceleration changes $T_{\rm b}(R)$ relation making it to depend not only on the magnetic field topology and its relation with the emitting particles density, but also on the acceleration profile and the jet viewing angle. However, for nearby sources we expect the break in the $T_{\rm b}(R)$ dependence from steep to flat, that is observed in several radio galaxies.

%% file: Apndx_TbD.tex
\section{Brightness temperature vs component size}
\label{sec:tbvssize}

In the same manner as in section~\ref{sec:est_pars}, one can parameterize the physical parameters as a function of the jet width:
\begin{equation}
\label{eq:rR_dependeces}
    \begin{split}
        B(R) \propto R^{\hat{b}}\,, \\
        N(R) \propto R^{\hat{n}}\,,
    \end{split}
\end{equation}
then the brightness temperature gradient versus the size of components is defined by $T_{\rm b}(R) \propto R^{\hat{s}}$, where 
\begin{equation}
    \label{eq:ss_index}
    \hat{s} = \hat{n} + \hat{b} (1-\alpha) + 1\,.
\end{equation}
Assuming relation $R\propto r^{d}$, the indices $\hat{s}$ and $s$ (equation~\ref{eq:eps_ksi}) are, therefore, connected through the geometry of the outflow \citep{2012A&A...544A..34P}: 
\begin{equation}
    \label{eq:s_s_d}
    \hat{s} = 1+ \frac{n+b(1-\alpha)}{d} = \frac{s}{d}\,.
\end{equation}

The evolution of the brightness temperature with the component size $T_{\rm b}(R)$ was fitted by $T_{\rm b}(R) = a_0 R^{\hat{s}}$ and to account for the break by
\begin{equation}
\begin{split}
    T_{\rm b}(R) = a_1 R^{\hat{s}_1}, ~{\rm if}~r<r_{\rm b, TR};\\
    T_{\rm b}(R) = a_2 R^{\hat{s}_2}, ~{\rm if}~r>r_{\rm b, TR}.\\
\end{split}
\label{eq:tbrfit}
\end{equation}

\begin{figure}
    \centering
    \includegraphics[width=0.9\columnwidth]{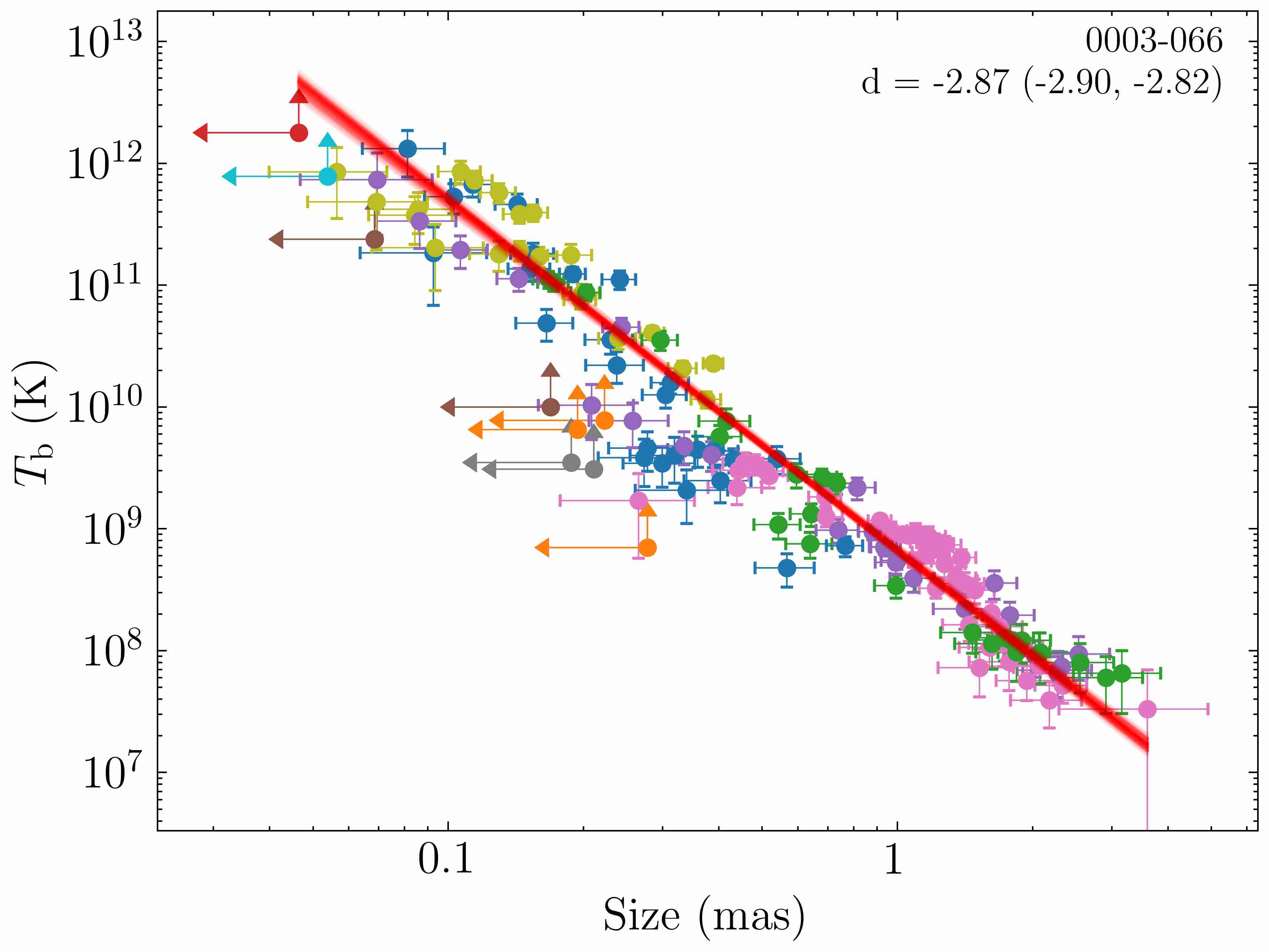}
    \caption{Brightness temperature vs the component FWHM for 0003$-$066. Different colours denote individual components. The arrows denote the upper limits, equation~\ref{eq: lobanov_limit}. The solid lines indicate results of the automated fit by a single power-law model $T_{\rm b}(R) = a_0 R^{\hat{s}}$. The plots for all sources are available online as supplementary material.}
    \label{fig:tbd}
\end{figure}

\begin{figure}
    \centering
	\includegraphics[width=0.9\columnwidth]{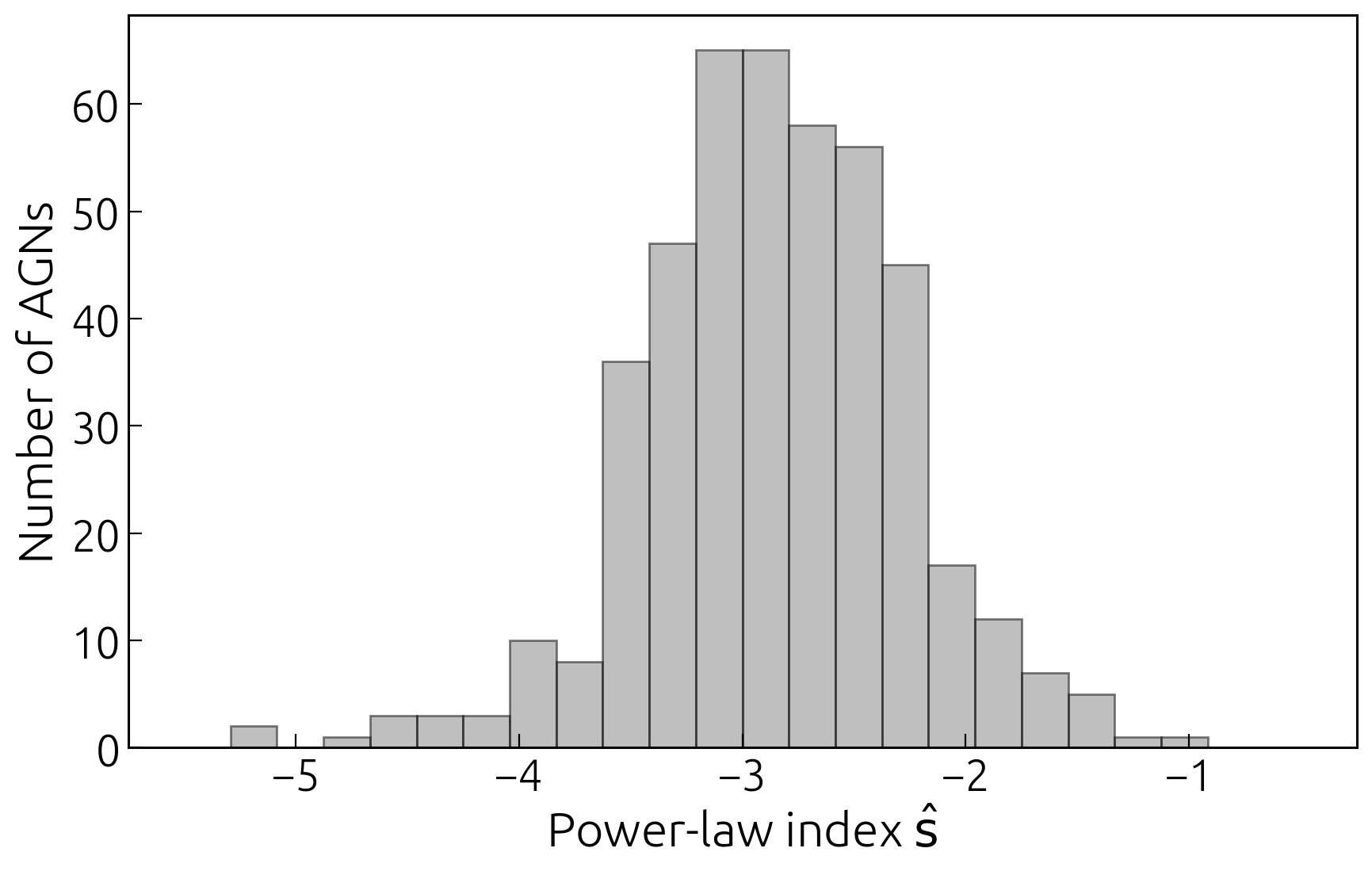}
    \caption{Distribution of the power-law indices $\hat{s}$ in $T_{\rm b}\propto R^{\hat{s}}$ of all sources.} 
    \label{fig:tbdindx}
\end{figure}

Figure~\ref{fig:tbd} shows the radial distributions of the brightness temperature versus component size with the resultant fit by a single power-law model. Figure~\ref{fig:tbdindx} and Table~\ref{tab:fit_results} summarise the power-law index $s'$ distribution with the median of $-2.87$ over all sources.
There is a small spread of median values of $\hat{s}$-indices between quasars, radio galaxies and NLSY1, while BL~Lacs are characterised by a flatter profile.
The Anderson-Darling rejects the null hypothesis that for quasars and blazars $\hat{s}$ values are drawn from the same distribution, at the 0.001 significance level. 
For a range $z<0.1$, the AD-test shows a marginal evidence for difference of populations of RGs and BL~Lacs $\hat{s}$; the corresponding p-value is 0.026.

\begin{figure}
    \centering
    \includegraphics[width=0.9\columnwidth]{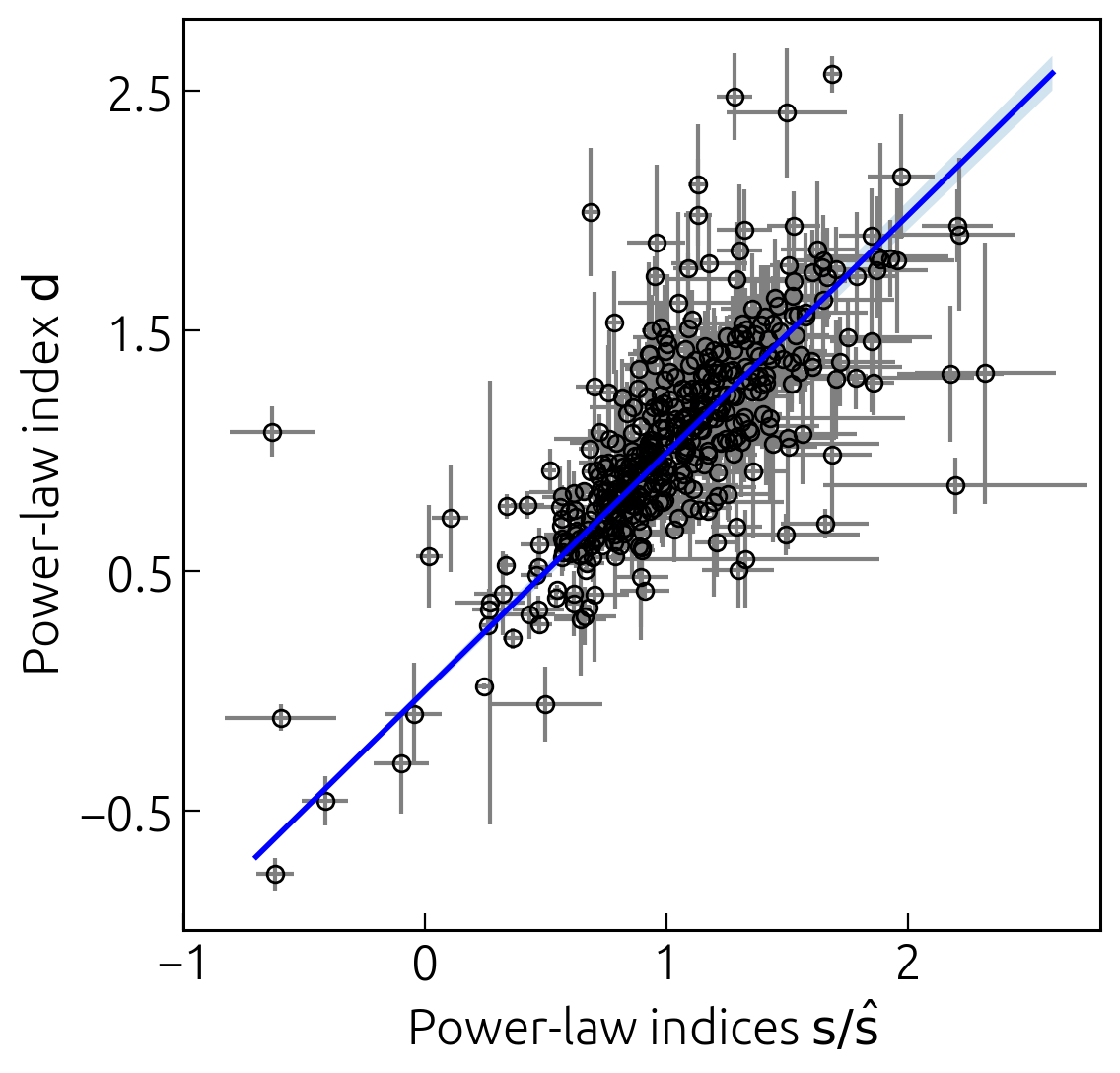}
    \caption{Correspondence between the estimated values of indices $d$ and $s/\hat{s}$. The solid line is a linear regression $(0.99\pm0.02) x + (0.0\pm0.02)$.} 
    \label{fig:d_sss}
\end{figure}

In Fig.~\ref{fig:d_sss}, we plot the dependence of $d$-index against the ratio of $s$ to $\hat{s}$-index (assuming single power-law fits). To account for possible outliers, we employed robust linear regression with Student-$t$ likelihood and obtained the slope $0.99\pm0.02$ and intercept $0.00\pm0.02$ (95 per cent credible intervals). The median value of $s/\hat{s}=0.99\pm0.02$ is consistent with the median $d=1.02\pm0.03$. This is an independent check that the size of the jet features reflects the jet geometry.

\begin{figure*}
    \includegraphics[width=0.67\columnwidth]{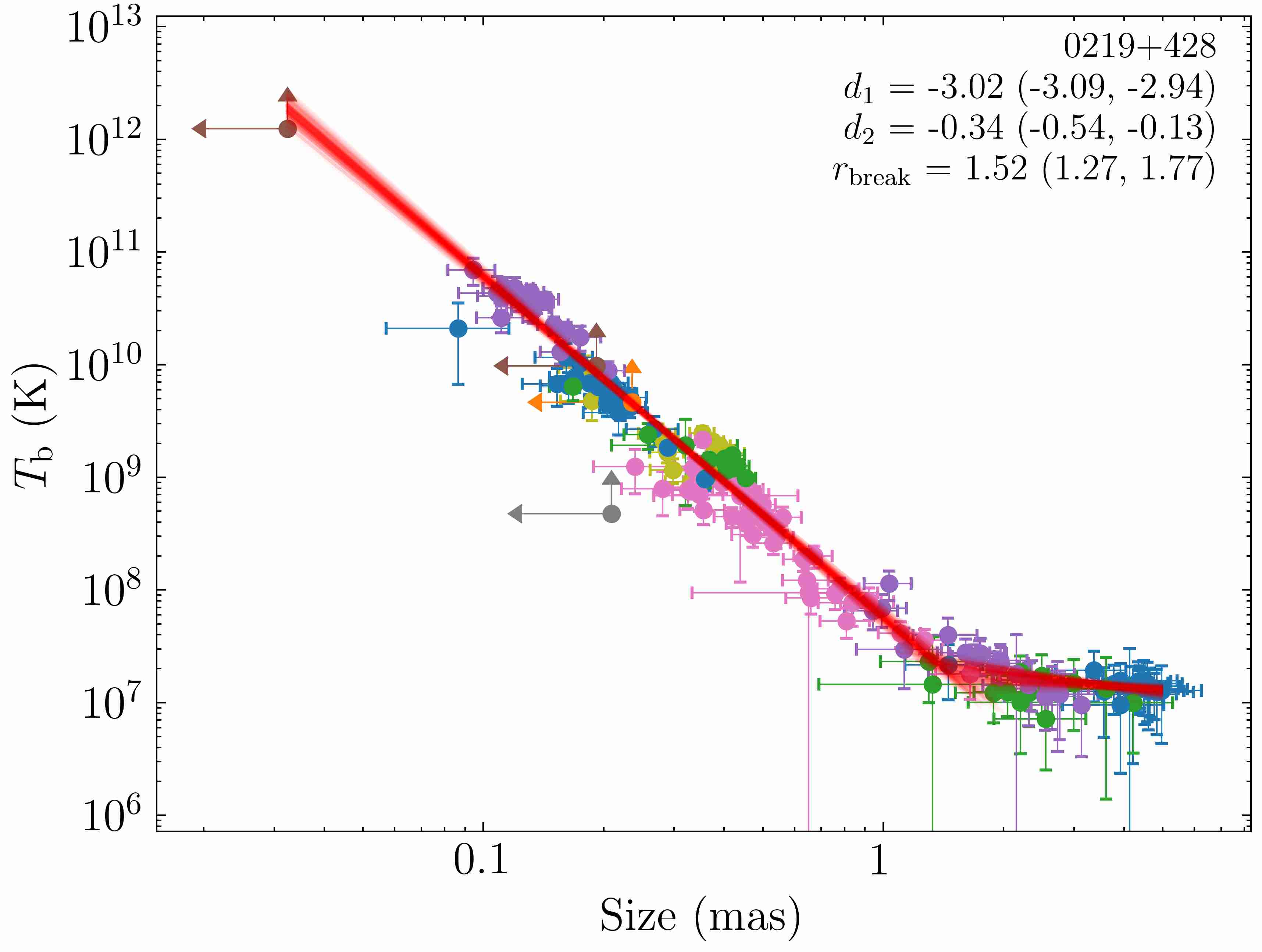}
    \includegraphics[width=0.67\columnwidth]{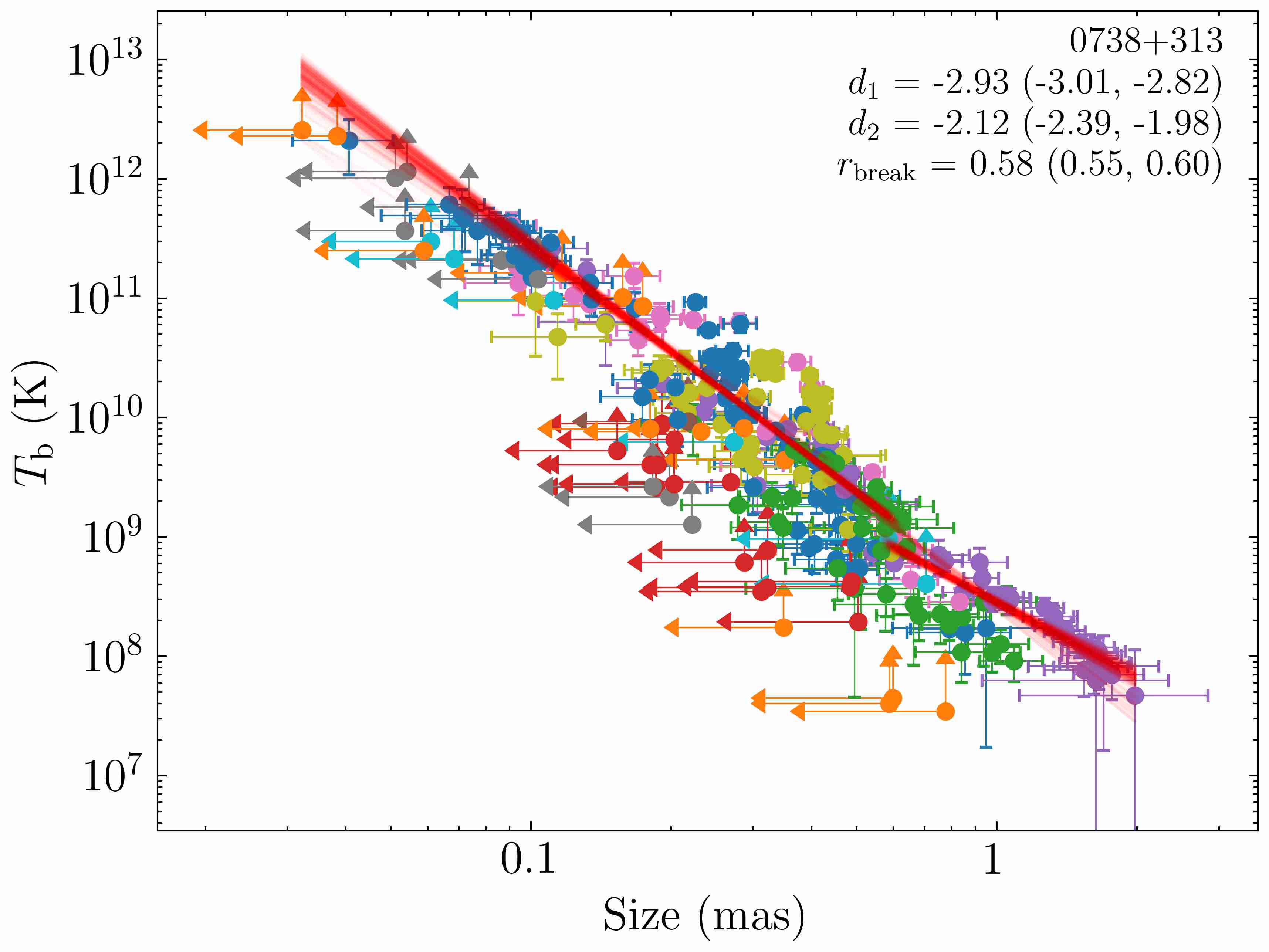}
    \includegraphics[width=0.67\columnwidth]{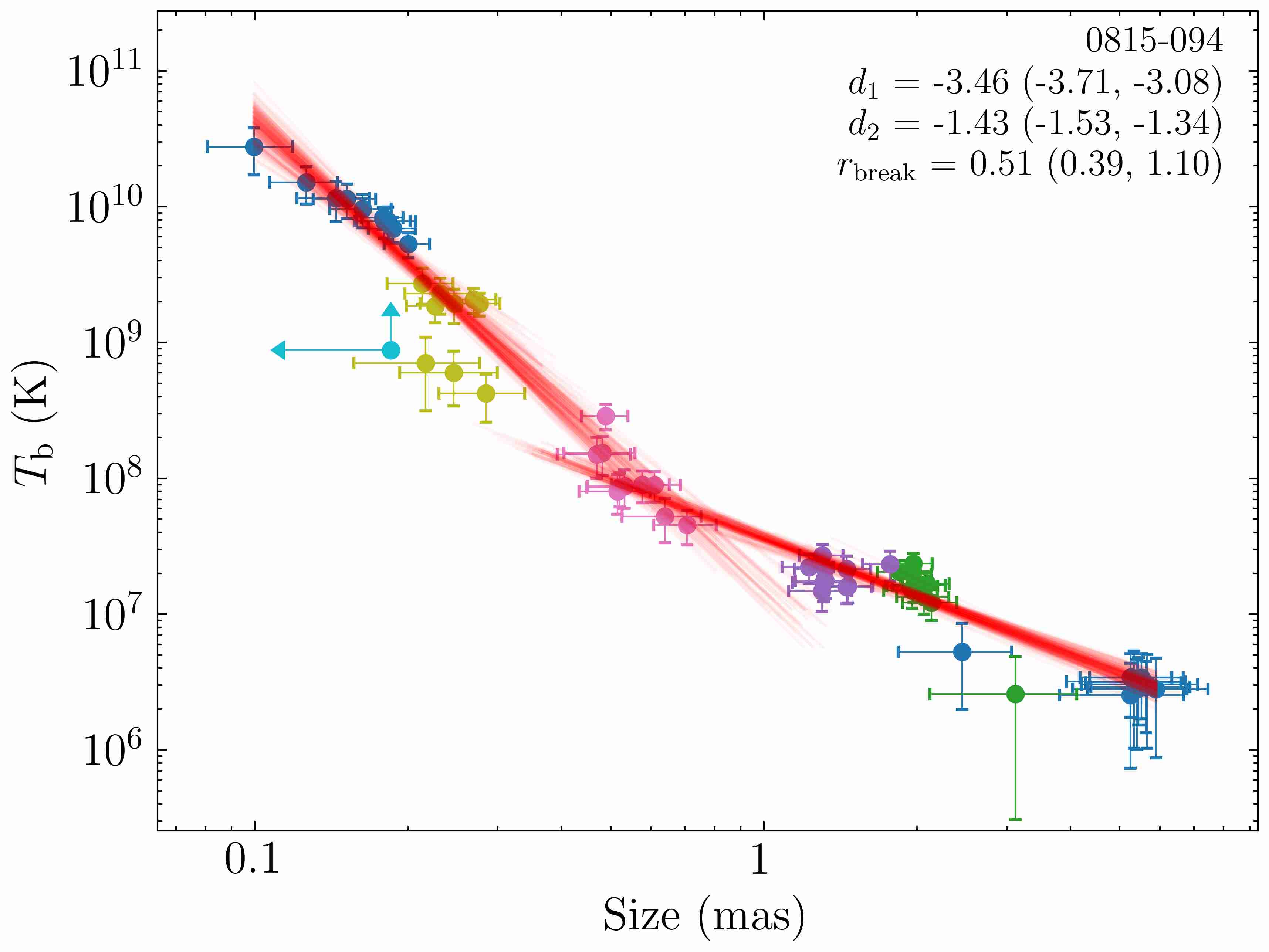}\\
    \includegraphics[width=0.67\columnwidth]{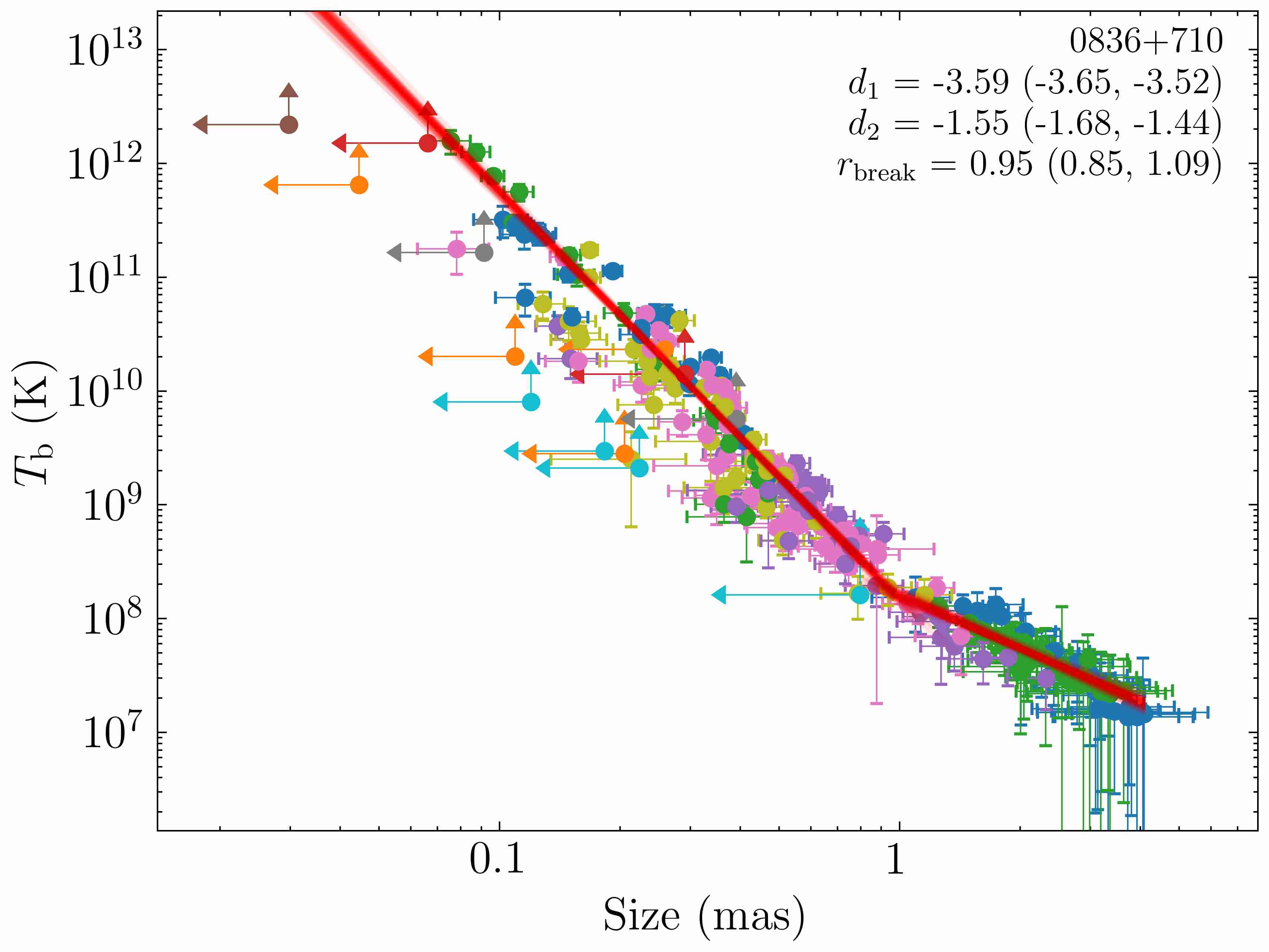}
    \includegraphics[width=0.67\columnwidth]{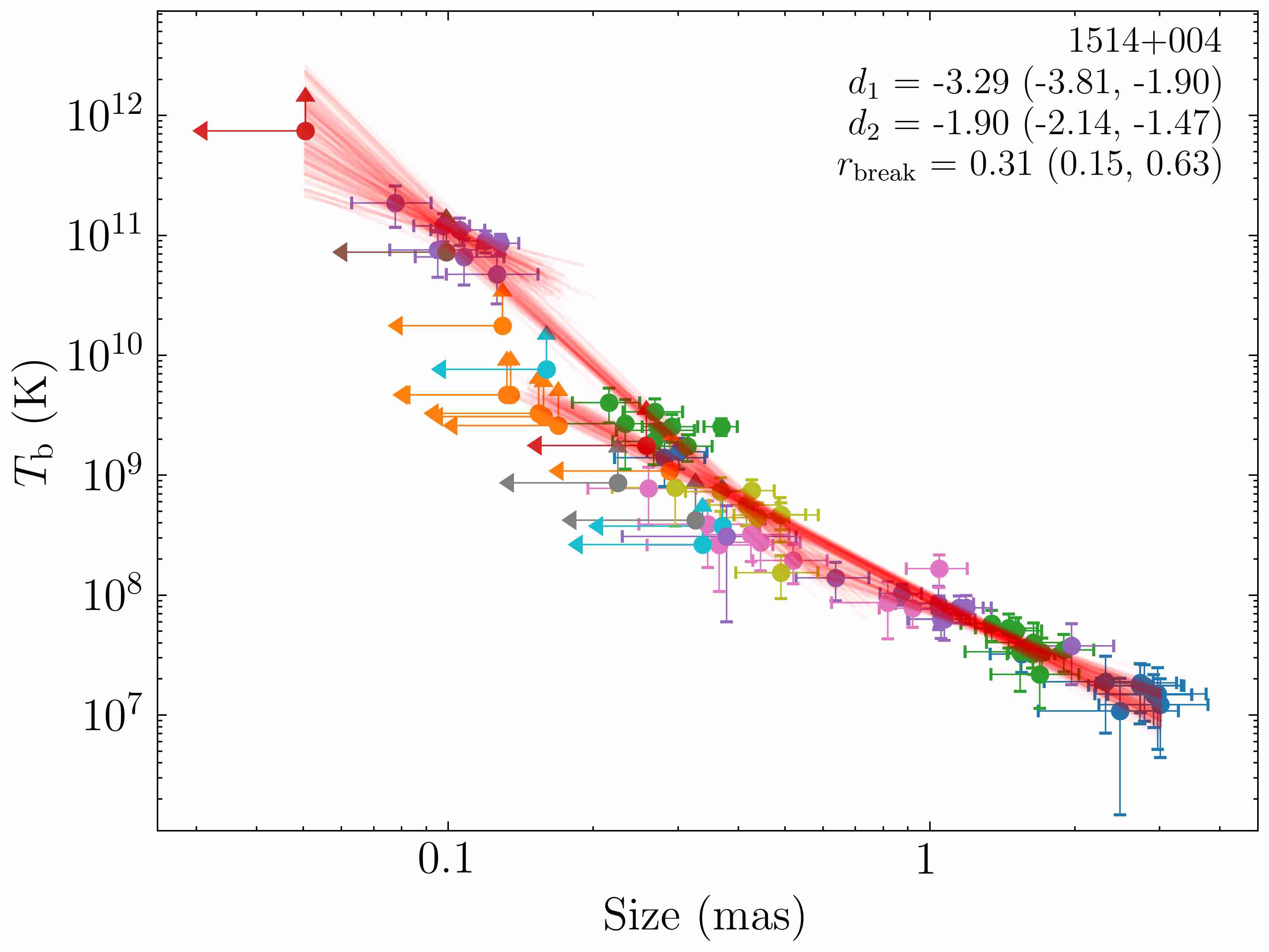}
    \includegraphics[width=0.67\columnwidth]{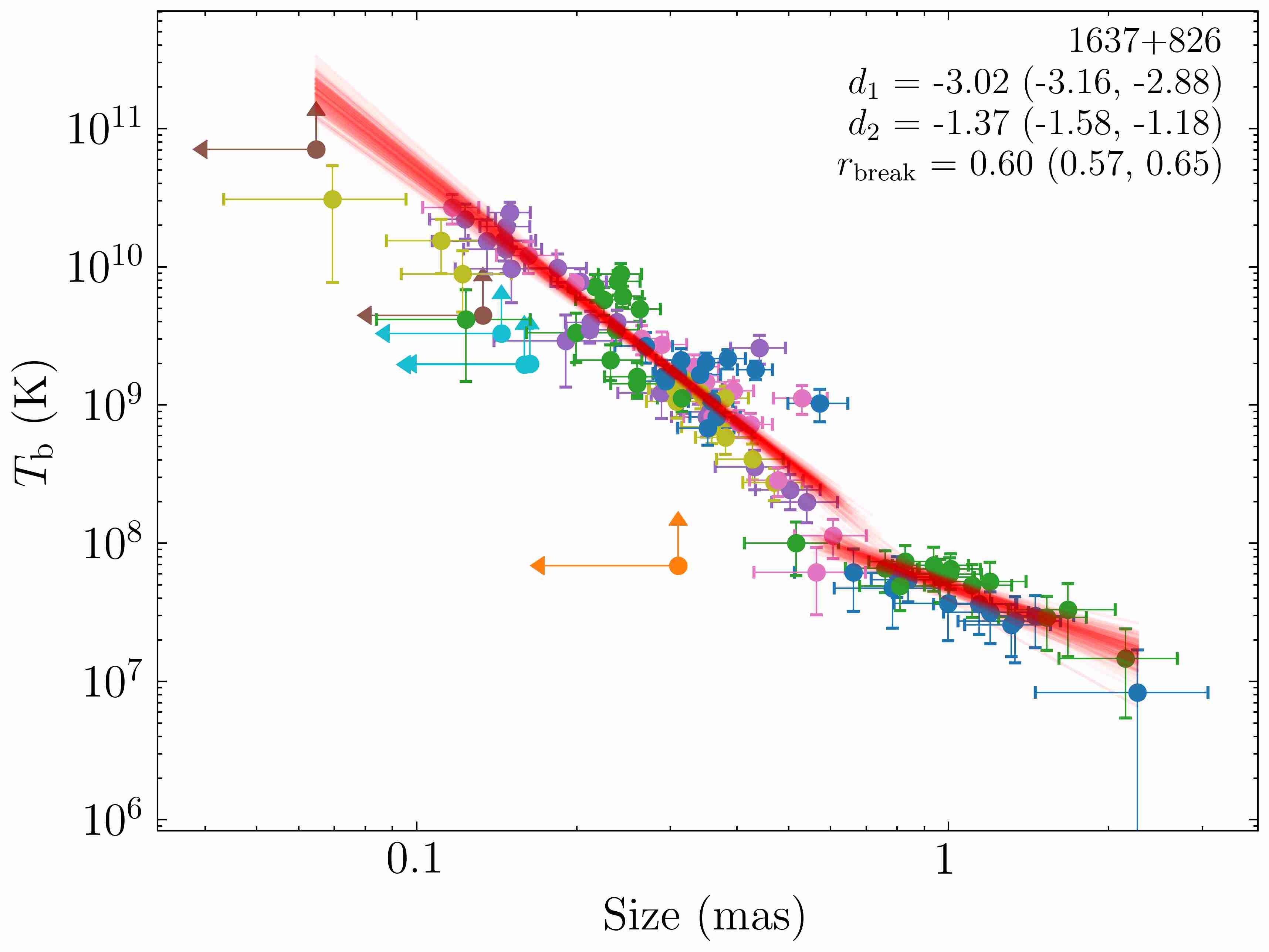}
    \caption{Brightness temperature vs the component FWHM for selected jets with strong evidence of a break. Different colours denote individual components. The solid lines indicate results of the automated fit by a double power-law model following equation~\ref{eq:tbrfit}. The arrows indicate the lower limits on $T_{\rm b}$, obtained from the upper limits on $R$ with equation~\ref{eq: lobanov_limit}. The plots for all sources are available online as supplementary material.}
    \label{fig:tbd_break}
\end{figure*}

The $T_{\rm b}(R)$ profiles of the 172 sources that are better fit by a double-power-law model, are shown in Fig.~\ref{fig:tbd_break}.
The corresponding distribution of the power-law indices $\hat{s}_1$ and $\hat{s}_2$ is given in Appendix~\ref{sec:apndx_distr_break}.
In the majority of known sources with a change in the jet geometry, we detect a break in the distribution of the brightness temperature versus component size.

The location and appearance of the break in $T_{\rm b}(R)$ is not so explicit due to complex behaviour of the brightness temperature and component size near the break.
The $T_{\rm b}(R)$ profiles are more straight and narrower compared to $T_{\rm b}(r)$ and $R(r)$ (e.g. 0333$+$321, 2351$+$456). We suggest that the brightness temperature variations can be partially driven by the spread in component sizes $R$ at the same radial distance $r$. 
Out of 172 sources with the break in $T_{\rm b}(R)$, 132 jets also showed a break in $T_{\rm b}(r)$, and 70 sources are fit by a double power-law model in all three distributions.

\section{Brightness temperature -- size dependencies for the sources used for the uncertainty calibration.}
\label{sec:calibration_sources_pics}
\begin{figure*}
\includegraphics[width=0.67\columnwidth]{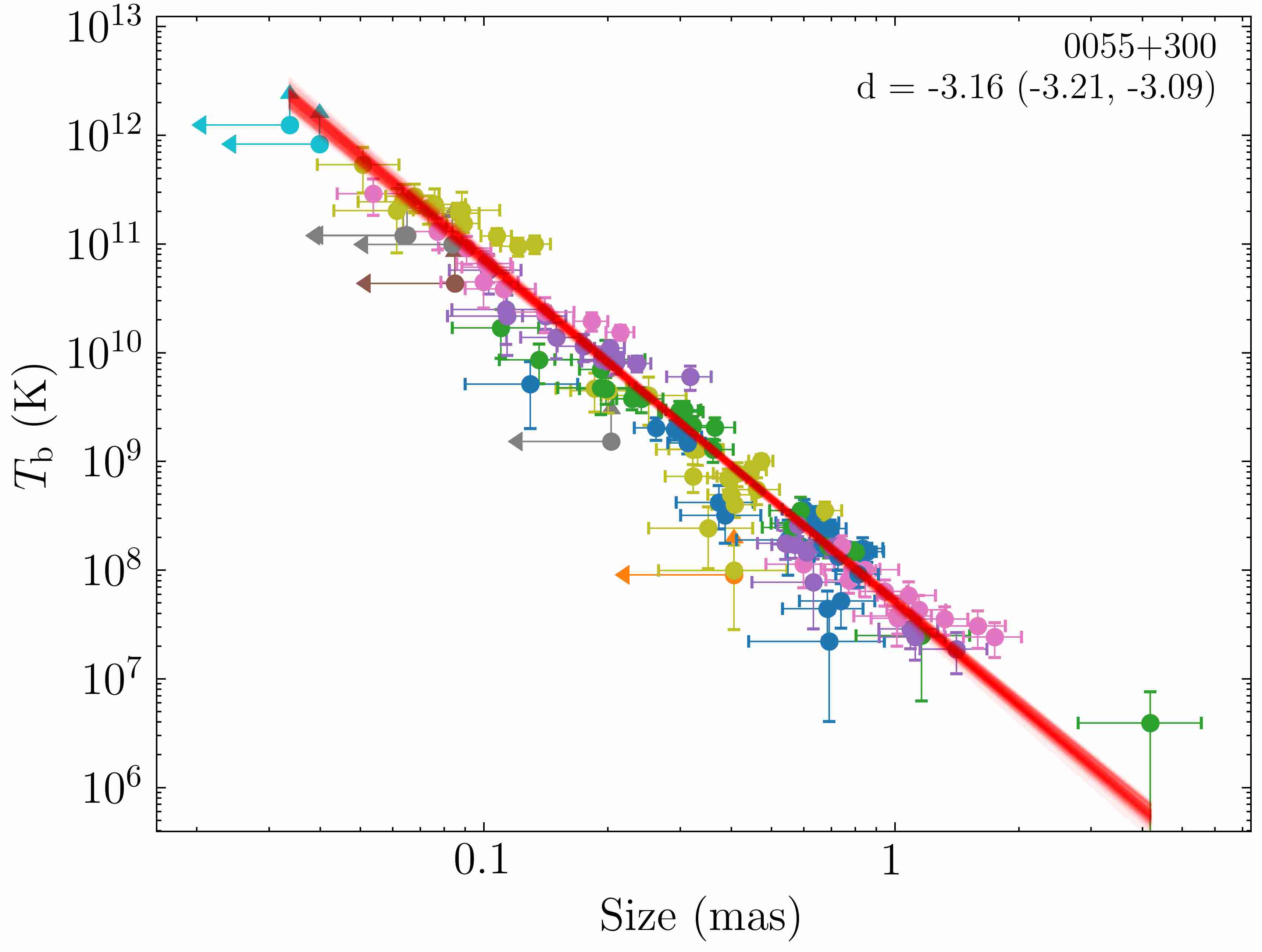}
\includegraphics[width=0.67\columnwidth]{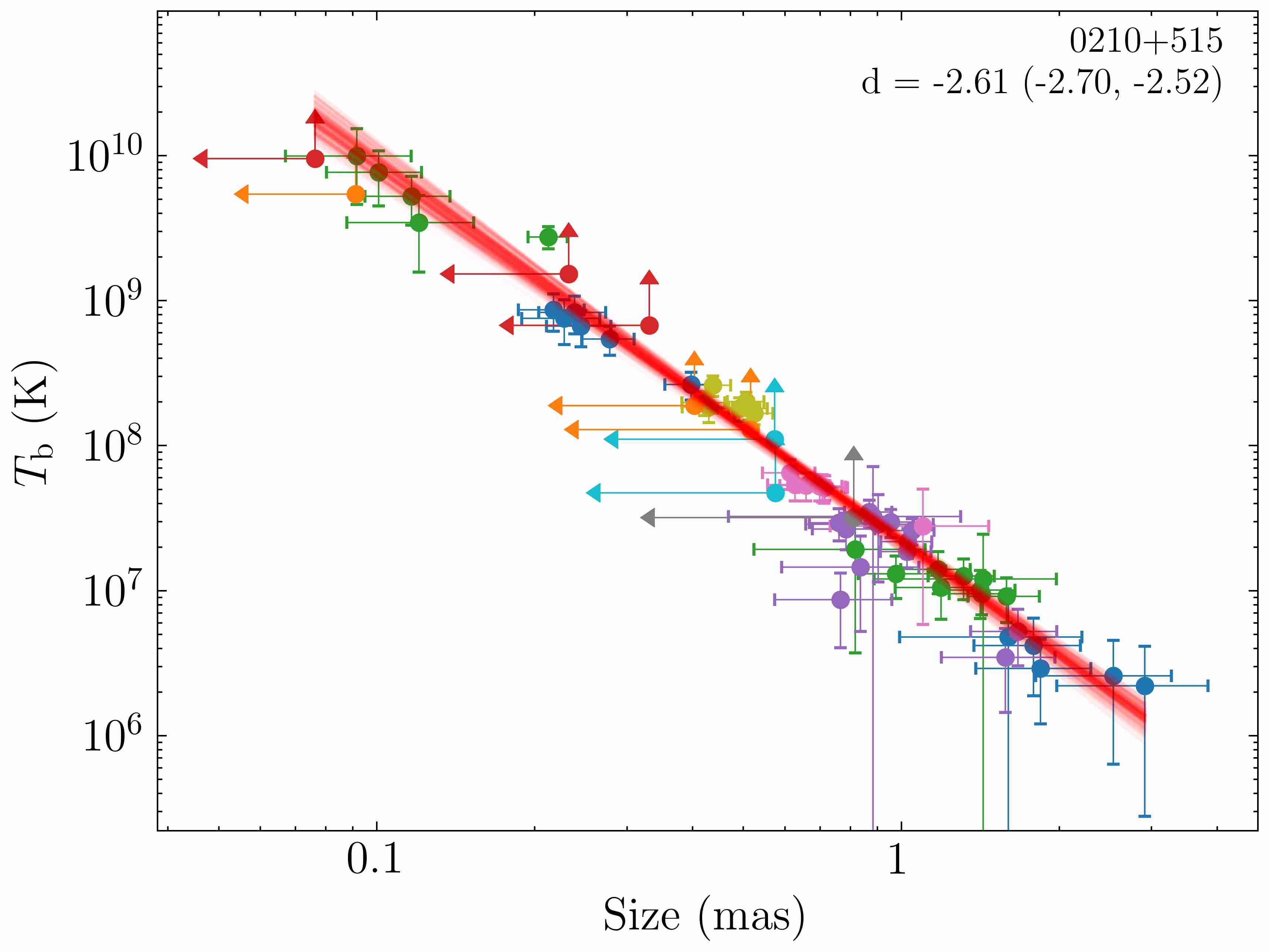}
\includegraphics[width=0.67\columnwidth]{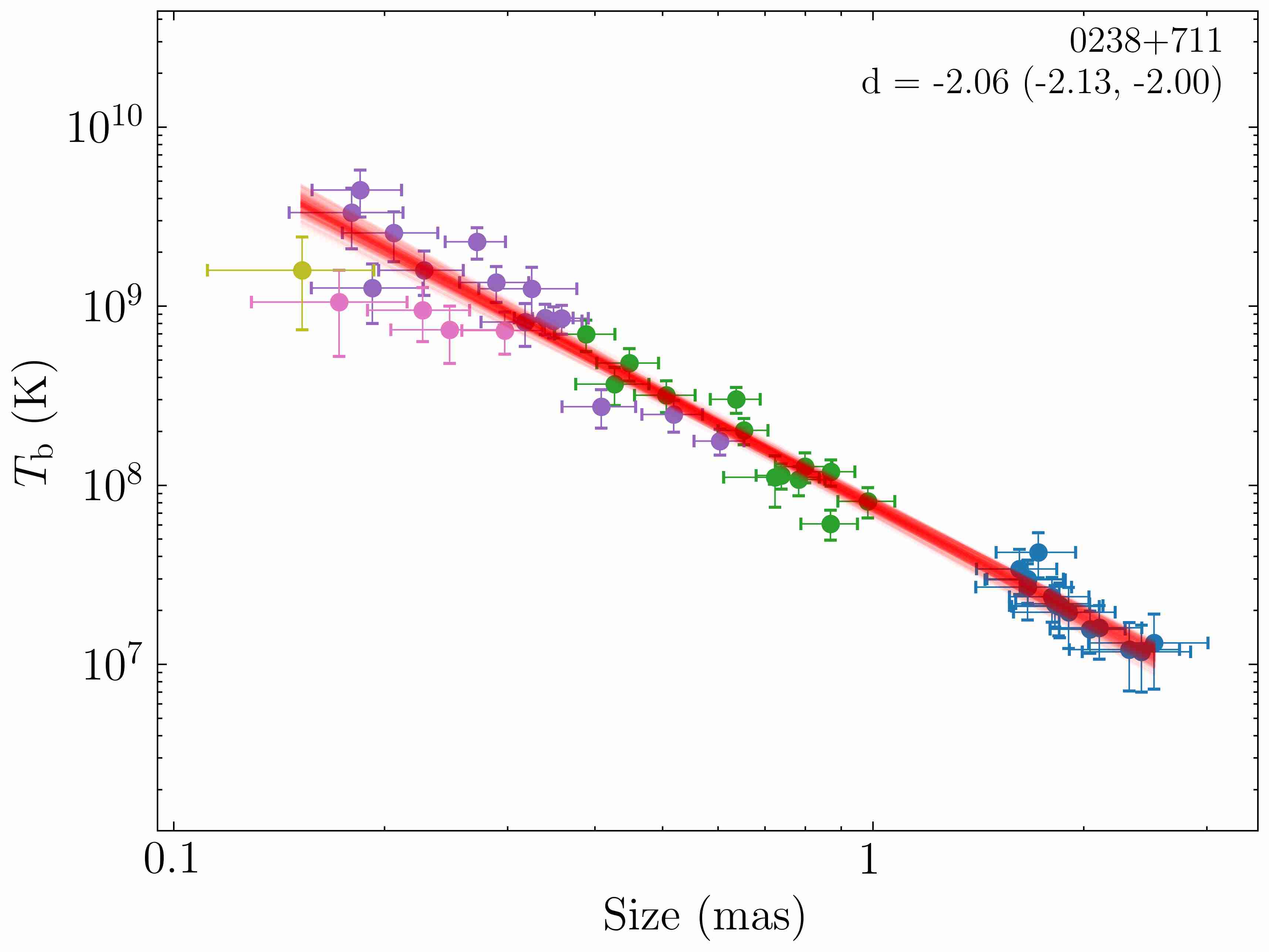}\\
\includegraphics[width=0.67\columnwidth]{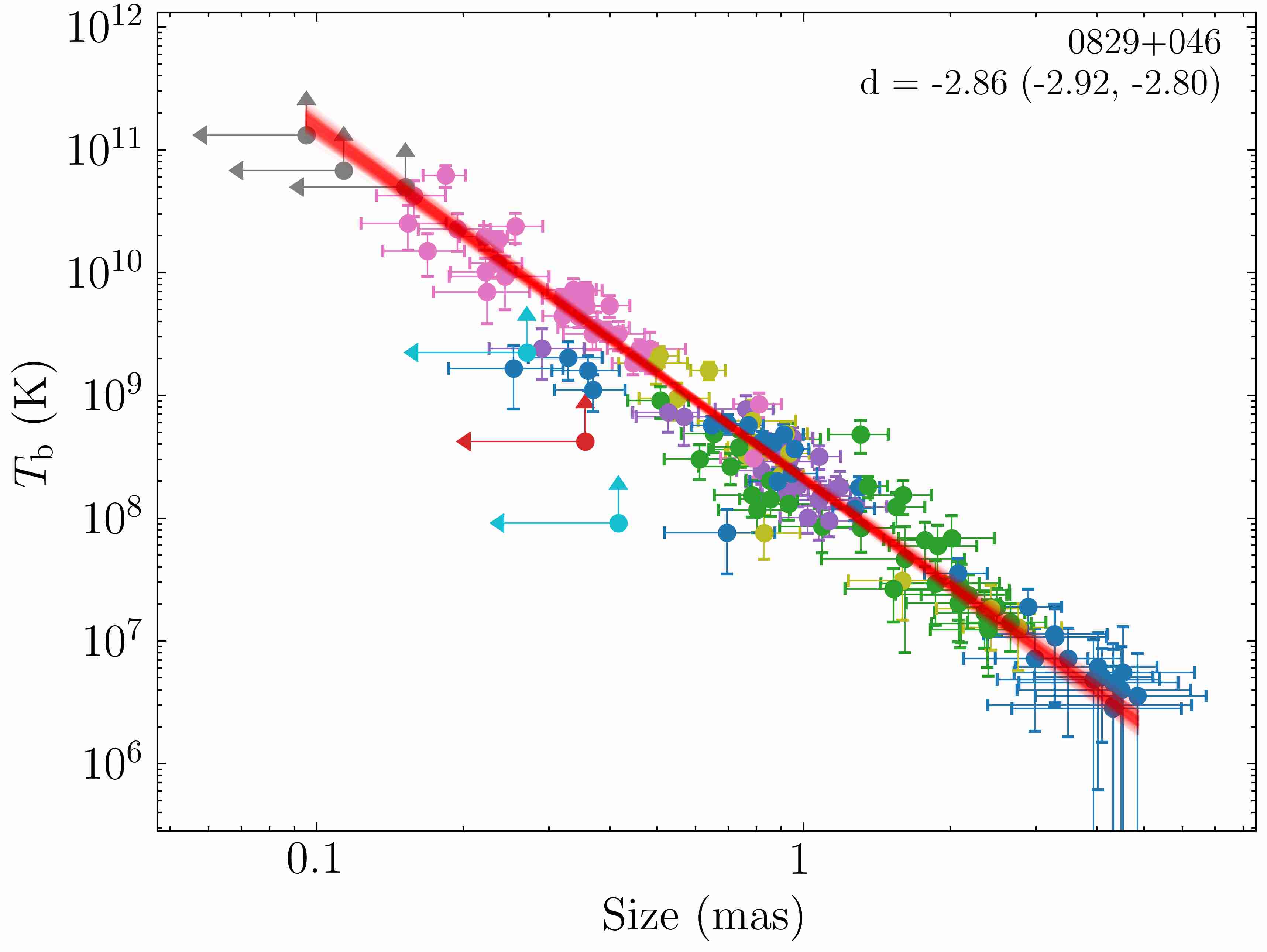}
\includegraphics[width=0.67\columnwidth]{figs/errors/0829+046_dt_G_fit.jpeg}
\includegraphics[width=0.67\columnwidth]{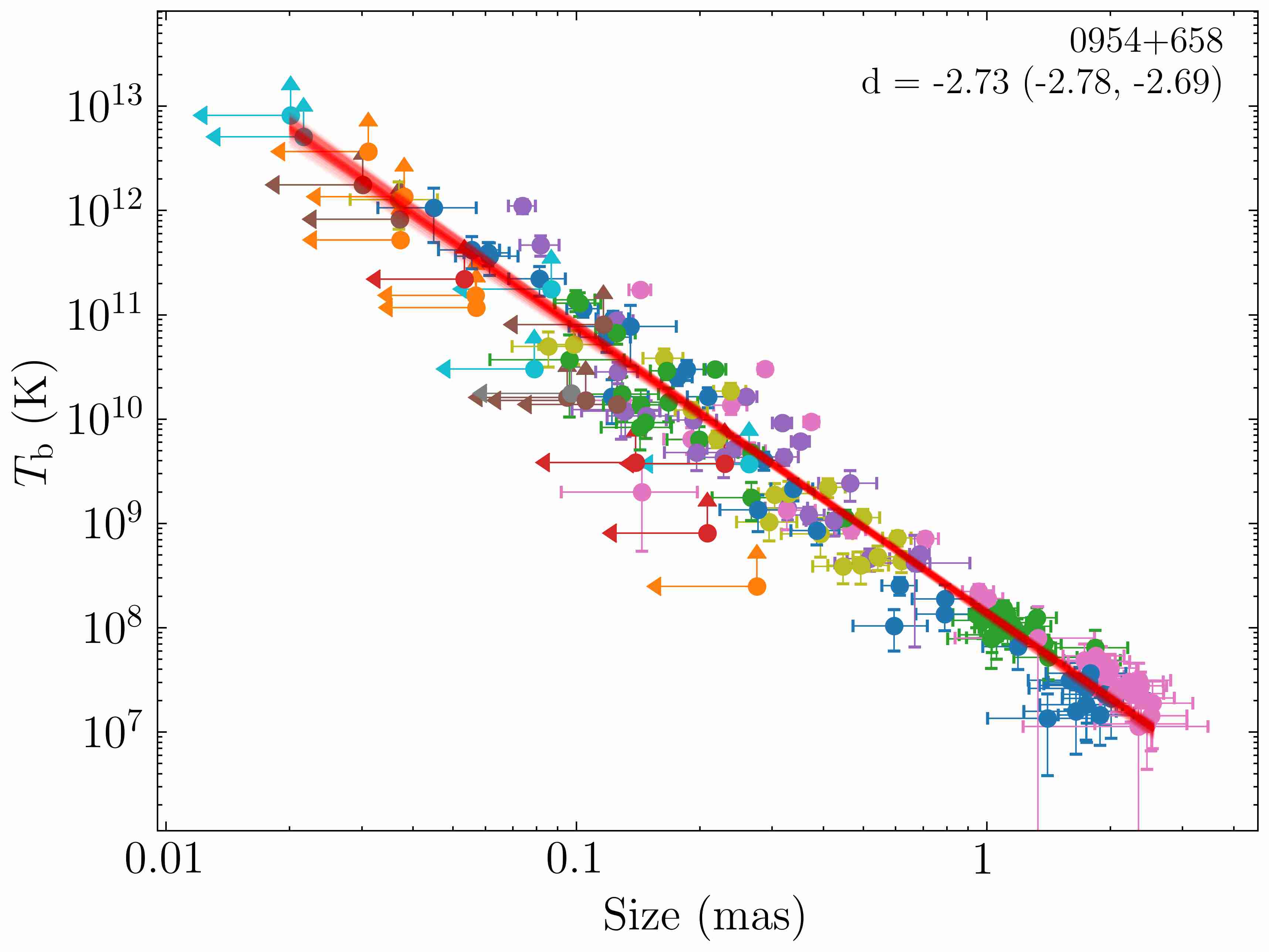}\\
\includegraphics[width=0.67\columnwidth]{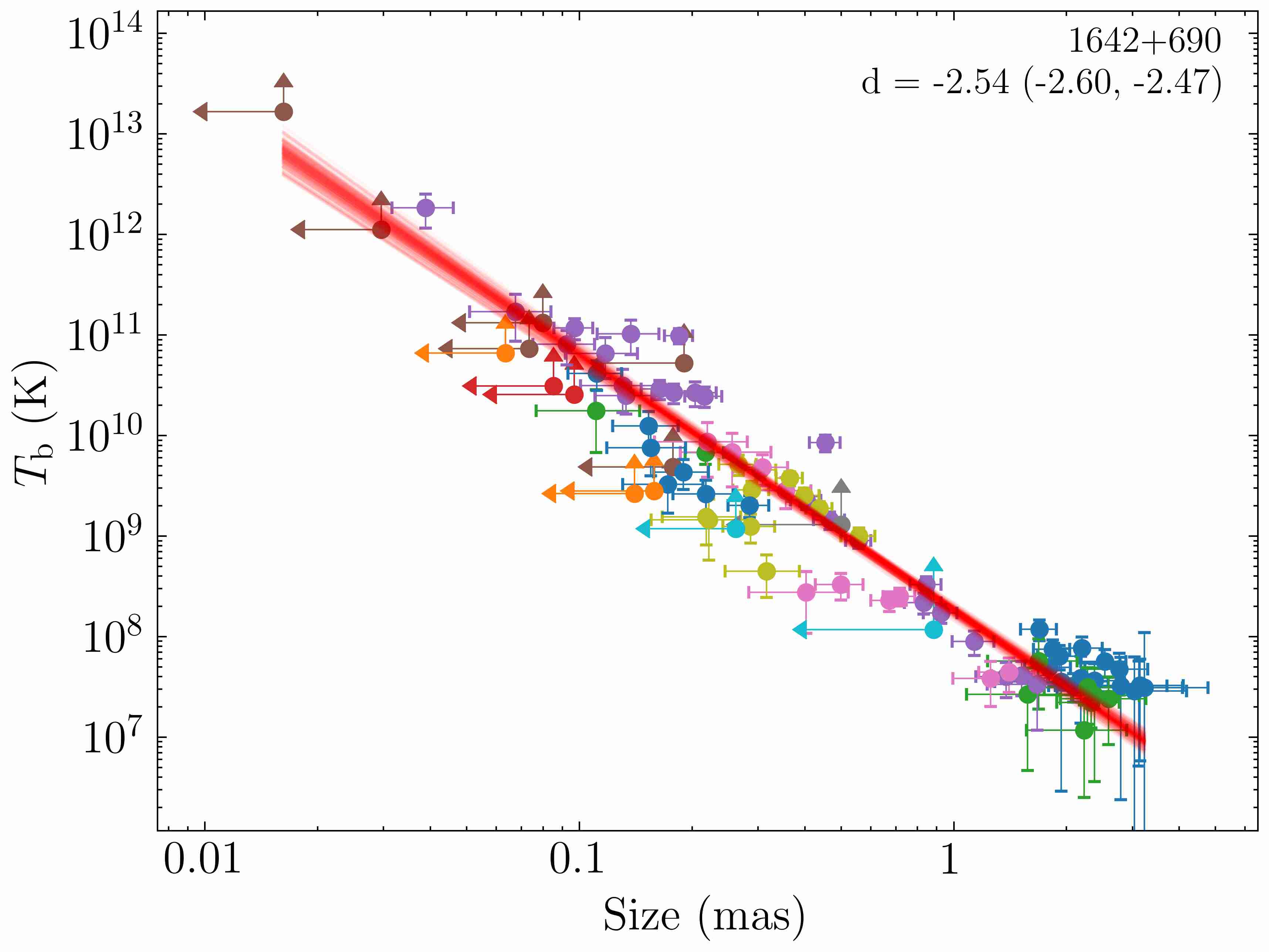}
\includegraphics[width=0.67\columnwidth]{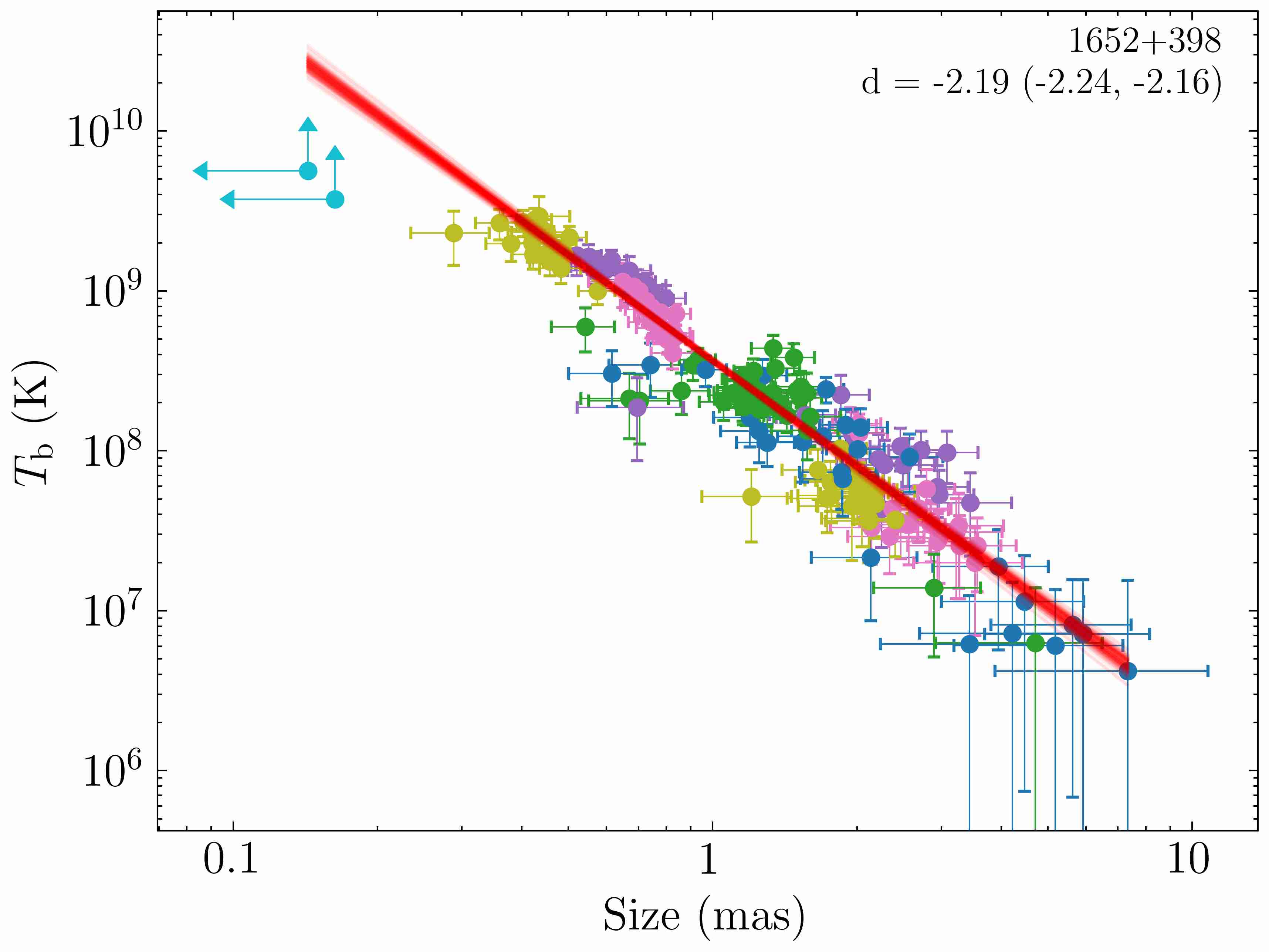}
\includegraphics[width=0.67\columnwidth]{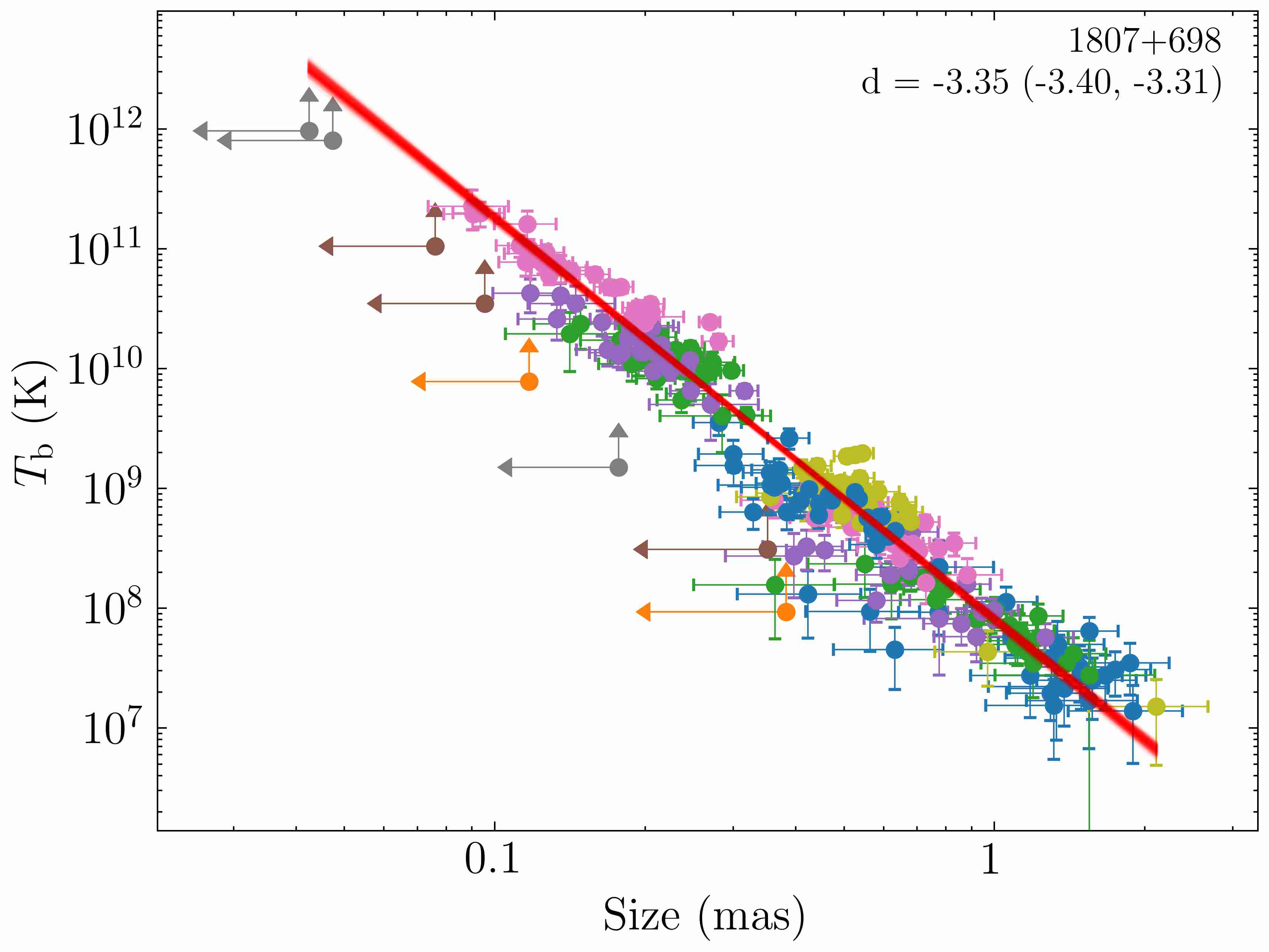}\\
\includegraphics[width=0.67\columnwidth]{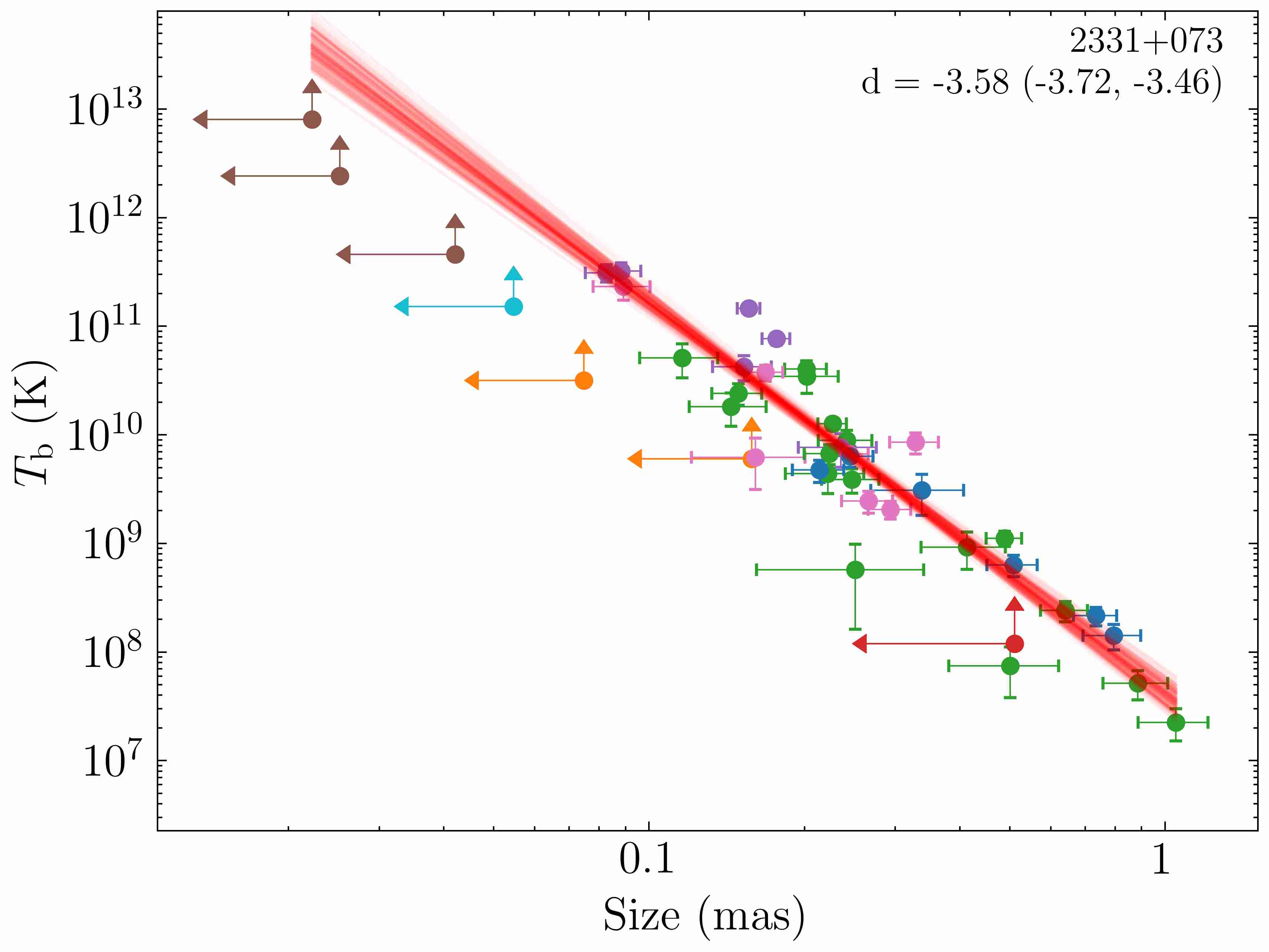}
\includegraphics[width=0.67\columnwidth]{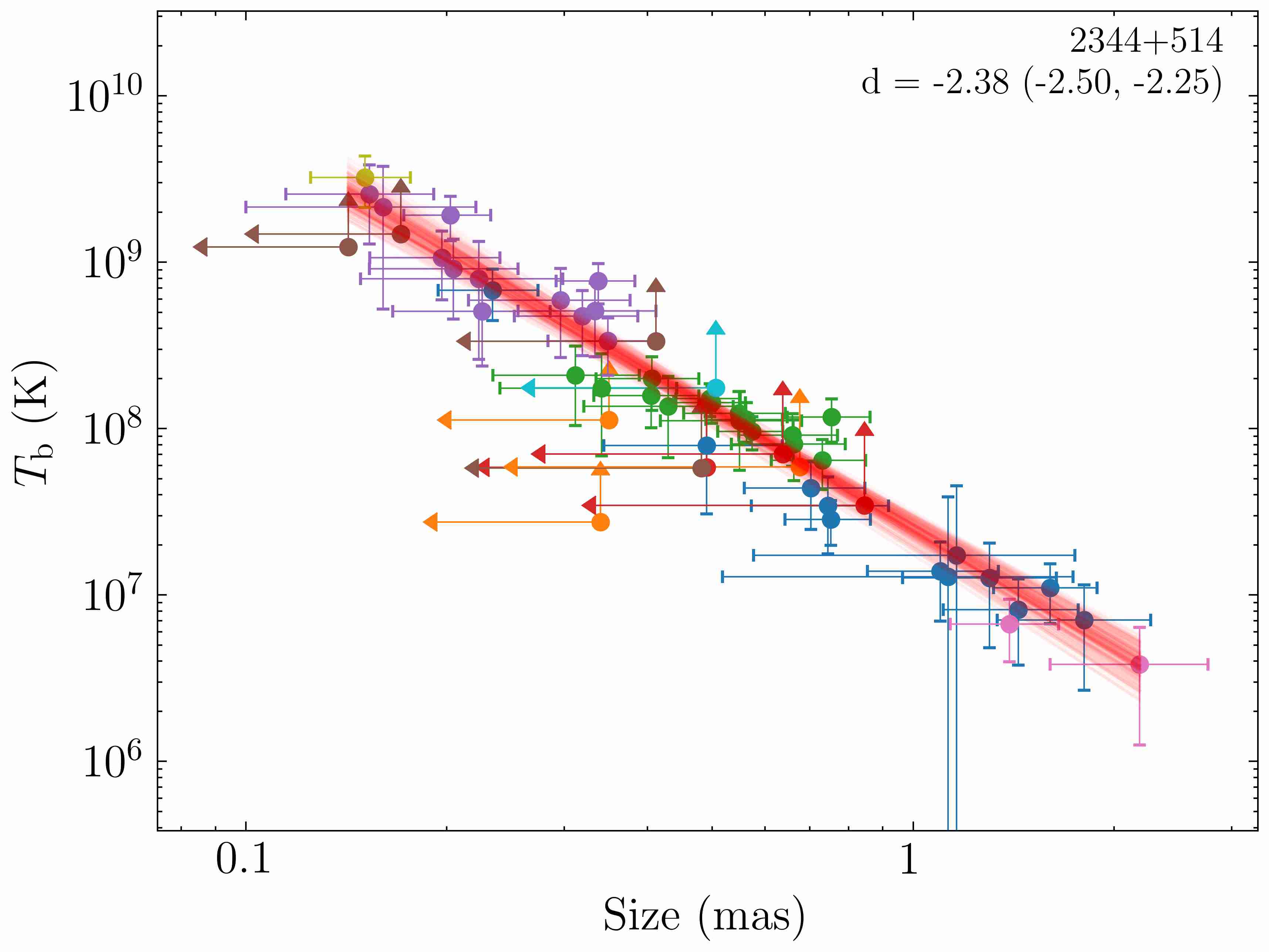}
\caption{The brightness temperature vs. component FWHM dependence and its power-law fits for sources used to calibrate the uncertainty estimates from \citet{2008AJ....136..159L}, Section~\ref{sec:tbvssize}. The uncertainty scaling factor was already applied to the shown error bars (see Section~\ref{sec:est_pars} for details).}
\label{fig:calibration_sources}
\end{figure*}

\begin{figure}
\centering
    \includegraphics[width=0.9\columnwidth]{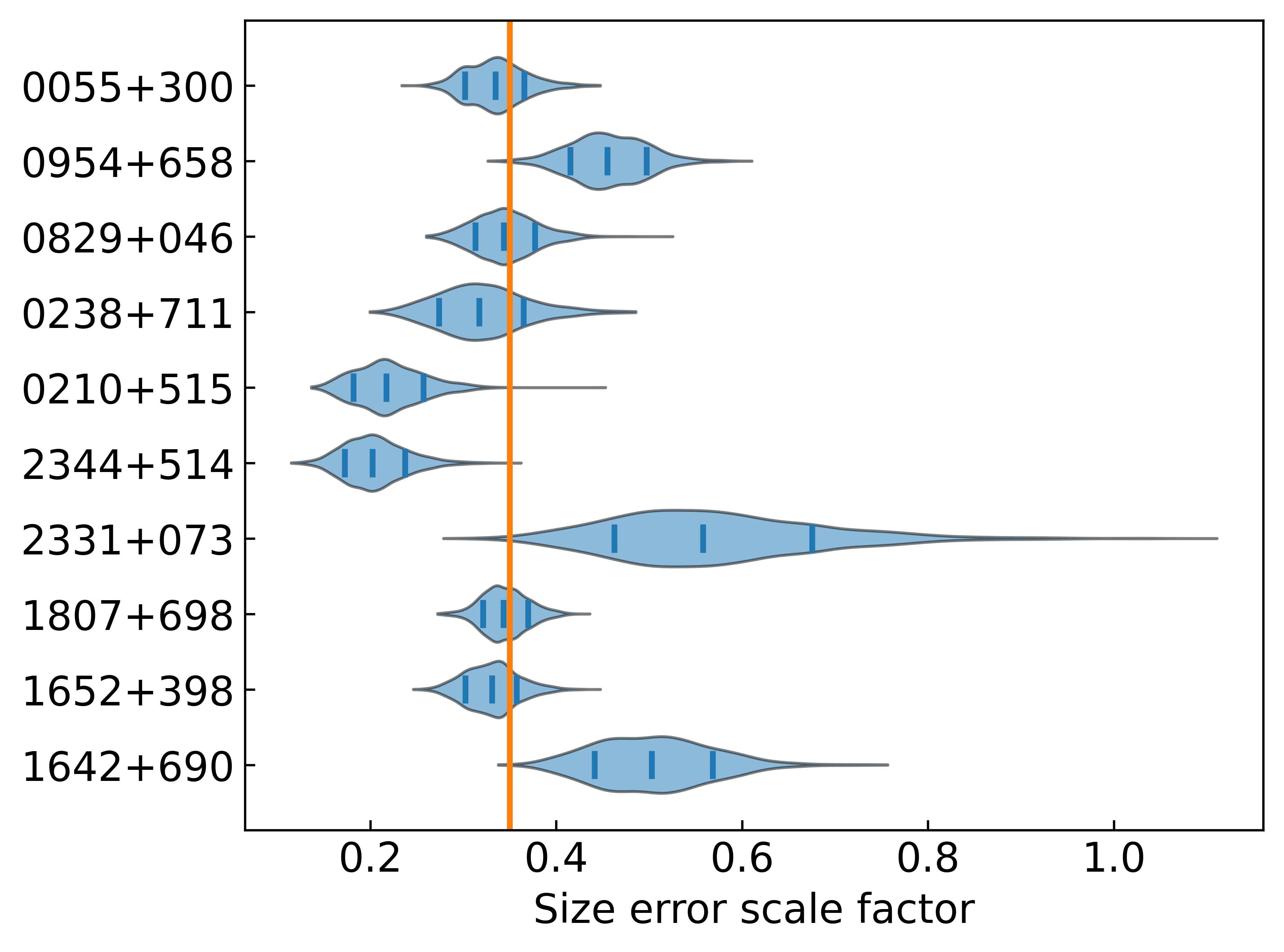}
    \caption{Posterior distributions (blue violin plots) of the size uncertainty scale factor for sources in Section~\ref{fig:calibration_sources}. Vertical orange line shows the median value used in the analysis.}
    \label{fig:scale_factors}
\end{figure}

Here we present the brightness temperature $T_{\rm b}$ - component size $R$ dependence for ten sources (Figure~\ref{fig:calibration_sources}) used to calibrate the uncertainty estimates of \cite{2008AJ....136..159L}. The resultant size error scaling factor is shown in Fig.~\ref{fig:scale_factors}.

%% file: Apndx_other.tex
\begin{figure}
    \includegraphics[width=0.9\columnwidth]{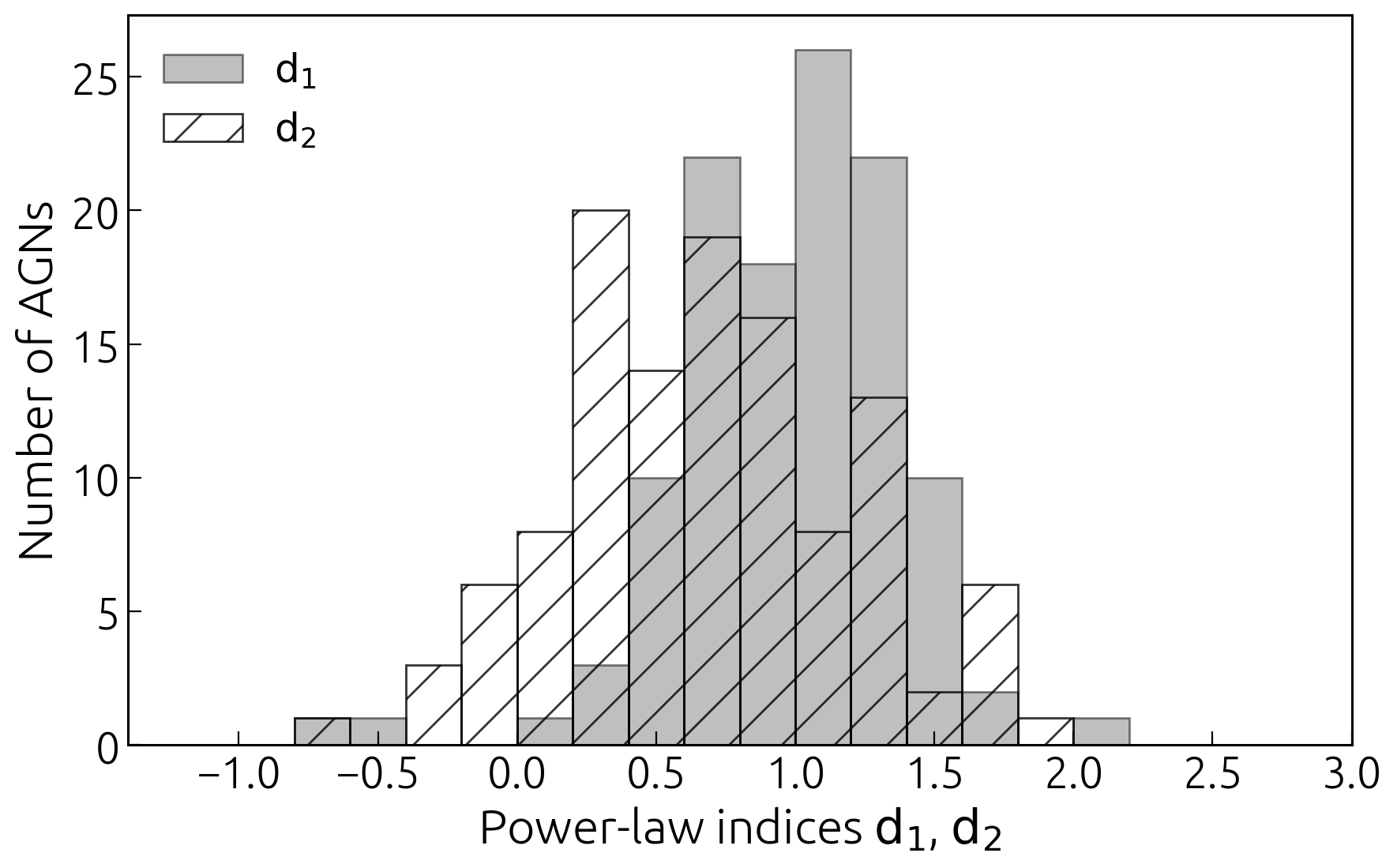}
    \caption{Histograms of the power-law indices $d_1$ and $d_2$ in radial dependence of the component size $R(r)$ for the 117 jets with strong evidence of a break.}
    \label{fig:d_ind_break}
\end{figure}

\section{Testing the integrated path length vs projected radial
distance}
\label{sec:test-realpath}

In the case of the bent jets, the radial displacement of a component is smaller than the total path it travels along the jet. This can introduce bias into our calculations, since component sizes will grow faster with $r$, and the $T_{\rm b}(r)$ dependence will appear steeper than the true (underlying) one.
To estimate magnitude and to test if the zigzag patterns are not produced by this effect, we make the following analysis. We selected three sources: 0738$+$313, 1716$+$686 and 2230$+$114, which have curved jets (fig.~2 of \citet{2017MNRAS.468.4992P}) and show complex distribution of their brightness temperature and component size (see Fig.~\ref{fig:srgrad}). We construct their ridgelines following the same procedure described in \citet{2017MNRAS.468.4992P}, using stacked 15\,GHz Stokes $I$ images.
The distance to the component from the core was then estimated as integrated length along the ridgeline to the projection point. The resultant difference between the radial distance and the integrated path length reaches $\thicksim2.5$\,mas for 0738$+$313, because its jet makes a sharp bend of about 60\degr{} (in projection on the sky plane) at a distance of 3\,mas from the core (see Section~\ref{sec:0738} for details).

We applied the same automated fit to these data. 
No significant difference between the power-law index estimates was found for 1716$+$686 and 2230$+$114. For 0738$+$313, the indices $d$ and $s$ changed by 10 per cent.
About 8 per cent the the sample sources have strongly curved jets.
We conclude that this effect has no significant influence on our statistical results but needs to be accounted for in highly curved jets.

\section{Power-law index distributions for the broken profiles}
\label{sec:apndx_distr_break}

The distribution of the power-law indices $d_1$ and $d_2$ of the sources that are better fit by a double power-law dependence (equation~\ref{eq:Rr}) are shown in Fig.~\ref{fig:d_ind_break} (See details in section~\ref{sec:rd_break}).
The median $d_1$ and $d_2$-values for different spectral classes are listed in Table~\ref{tab:d_ind_break}.
The same for the $T_\mathrm{b}(r)$ dependencies is shown in Fig.~\ref{fig:s_ind_break}.

\begin{figure}
    \includegraphics[width=0.9\columnwidth]{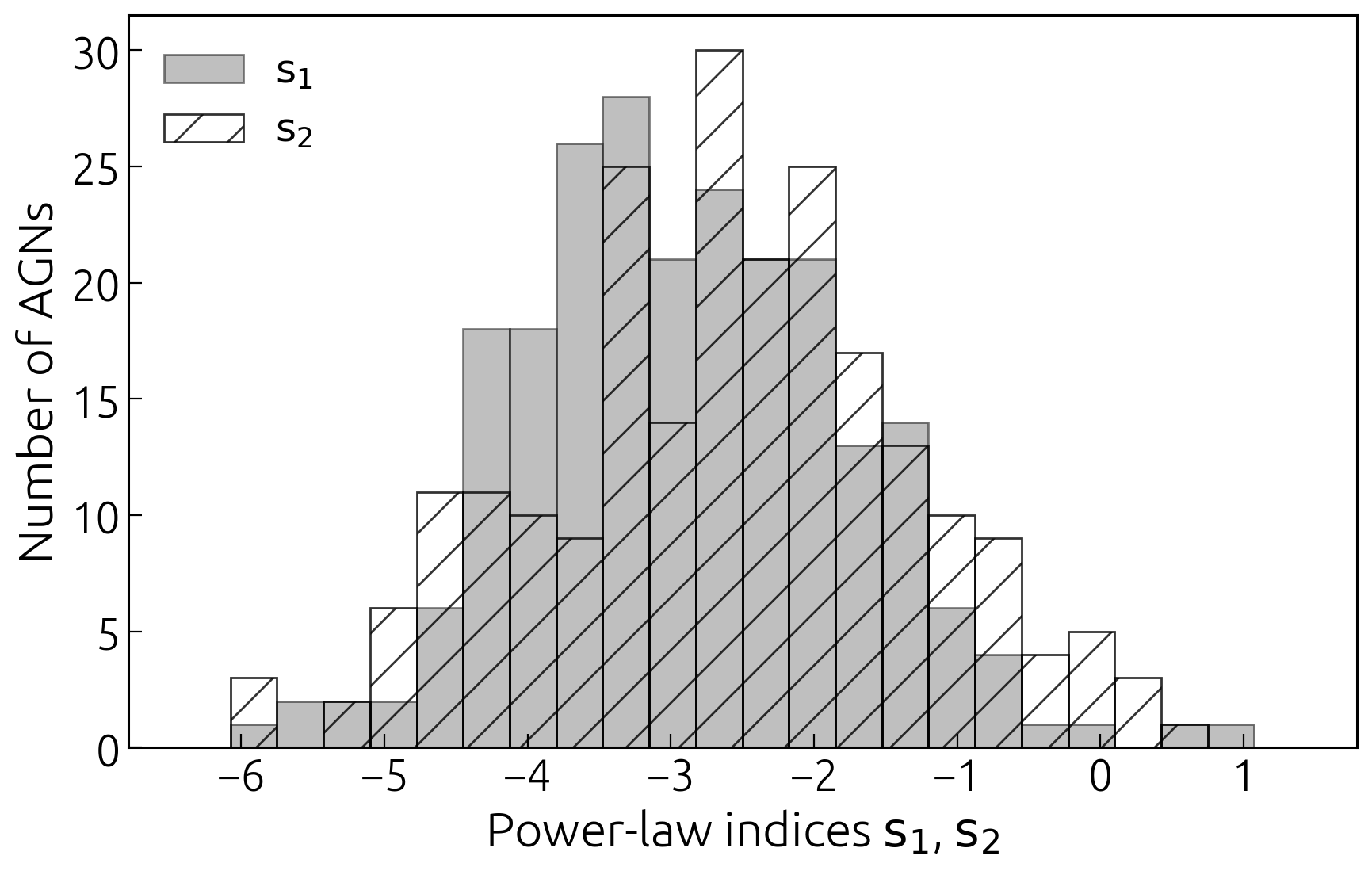}\\
    \caption{Histograms of the power-law indices $s_1$ and $s_2$ in radial dependence of the brightness temperature $T_\mathrm{b}(r)$  for the 171 jets with strong evidence of a break.}
    \label{fig:s_ind_break}
\end{figure}

\begin{table*}
    \centering
    \caption{Median values of the fitted $d$, $s$ and $\hat{s}$-indices for a single and double power-law models. The full data set comprises 447 AGNs, 271 flat spectrum radio quasars, 135 BL Lacertae objects, 25 radio galaxies and 5 radio-loud NLSY1. Number of sources with the significant broken profile is given.}
    \label{tab:d_ind_break}
    \begin{tabular}{lcrcccrcccrcc}
        \hline
        Opt. class & $d$ & N & $d_1$ & $d_2$ & $s$ & N & $s_1$ & $s_s$ & $\hat{s}$ & N & $\hat{s}_1$ & $\hat{s}_2$\\
        \hline
        Quasars        &0.95 & 72& 1.03&0.55 &-2.83&144&-3.02&-2.54 & -2.99&108&-3.00&-2.44\\
        BL~Lac objects &1.12 & 34& 0.98&0.79 &-2.90& 68&-3.03&-2.59 &  -2.59&48&-2.84&-2.22\\
        Radio galaxies &0.77 &  8& 0.88&0.90 &-2.59& 17&-2.15&-2.06 &  -2.83&14&-2.38&-1.85\\
        NLSY1          &1.16 &  2& 1.14&0.60 &-3.04& 2 &-3.15&-1.98 &  -2.78&1&-3.49&-2.25\\
        \hline
        All            &1.02 &117& 1.03&0.65 &-2.82&233&-2.95&-2.53 & -2.87&172&-2.97&-2.35\\
        \hline
    \end{tabular}
\end{table*}

\begin{figure}
    \includegraphics[width=0.9\columnwidth]{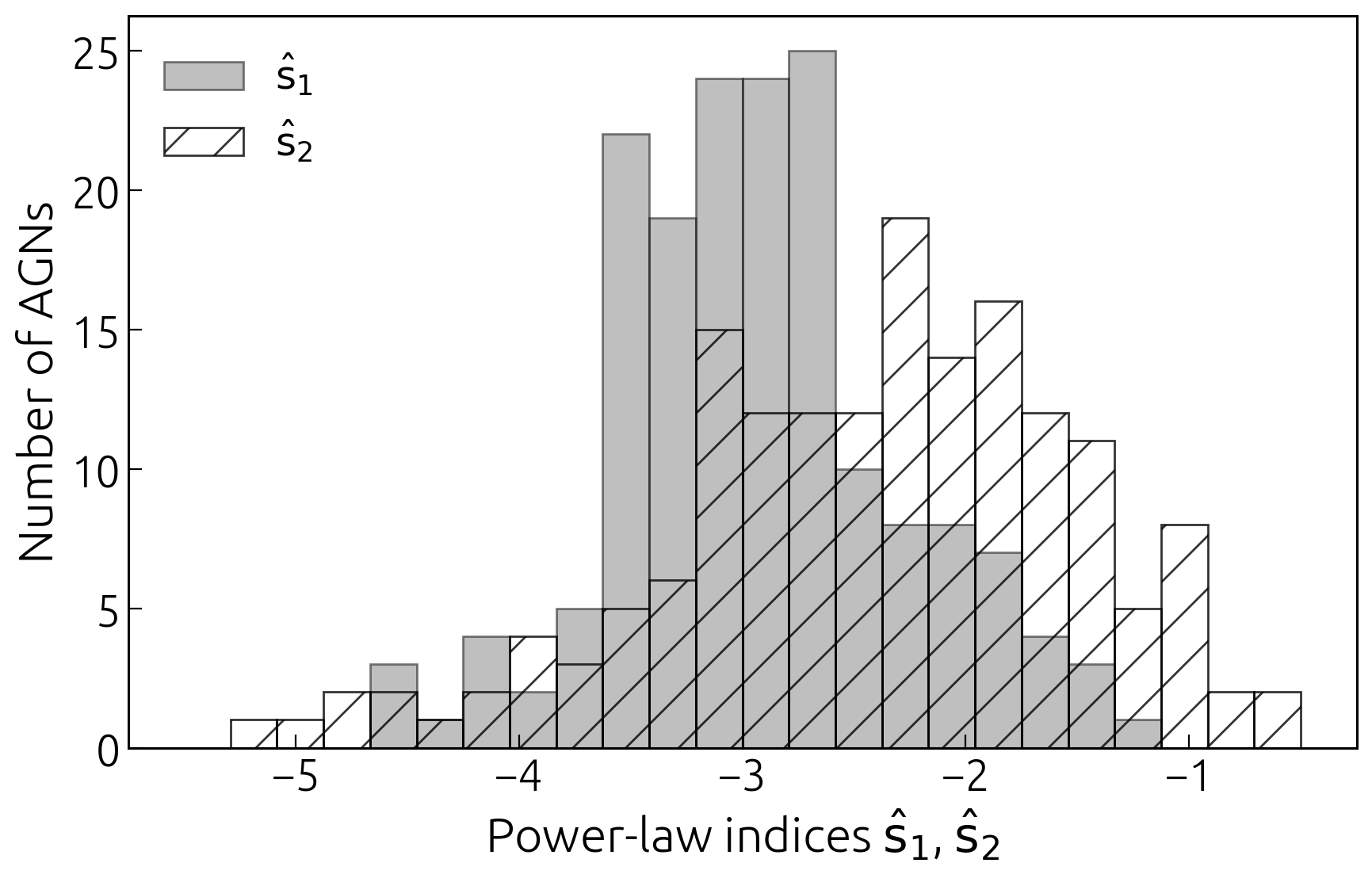}\\
    \caption{Histogram of the power-law indices $\hat{s}_1$ and $\hat{s}_2$ in the dependence of the brightness temperature from the component size $T_\mathrm{b}(R)$ for the 233 jets with strong evidence of a break.} 
    \label{fig:s_ind_break}
\end{figure}

%% file: Apndx_IndivS.tex
\section{Notes on individual sources}
\label{sec:individual}

\subsection*{3C~66A (0219$+$428)}
Both the size and the brightness temperature distribution show complex behaviour at two distinct locations: around 2--3~mas and beyond 8~mas from the core.
Variations at 2--3\,mas were seen before by \citet{2005AJ....130.1418J} for $R(r)$ and by \citet{2005ApJ...631..169B} for $T_{\rm b}(r)$ as well.
\citet{2005AJ....130.1418J} associated this non-monotonic change with the double-shock structure supported by the kinematic analysis which yields multiple stationary components within 3~mas of the core \citep{2017ApJ...846...98J, 2019ApJ...874...43L}.
Meanwhile, \citet{2005ApJ...631..169B} suggested that the jet bend gives rise to these variations.
The second break at 8~mas coincides with an apparent jet bend of $\thicksim$25\degr{} \citep{2017MNRAS.468.4992P}. This may give a change of the Doppler factor, and, therefore, may explain the deviation of the radial gradient in the brightness temperature and the component size from the single power-law.

\subsection*{NRAO~140 (0333$+$321)}
Both $T_{\rm b}(r)$ and $R(r)$ distributions exhibit clear zigzag profiles. 
This is one of a few sources with a detected transverse RM gradient, which together with the RM-corrected orientation of polarization vectors suggests a helical magnetic field in both the jet and its surrounding spine \citep{2008ApJ...682..798A}.
As in the case of the well-studied quasar 3C~273 (see below), the observed complex distributions could originate due to stratified emission and an asymmetric total intensity profile \citep[see also][]{2021AA...654A..27B}.
Alternatively, the first variation in the distributions coincides with the stationary component around 0.5\,mas seen from kinematics \citep{2021ApJ...923...30L}, and thus can be explained by a shock scenario.
The second wave around 6--7\,mas is associated with enhanced emission (an individual blob on the stacked Stokes $I$ image) and flat spectral index \citep{2014AJ....147..143H}, and, therefore, may be produced by a number of factors, e.g. a shock or interaction of the jet with the surrounding medium.

\subsection*{3C~111 (0415$+$379)}
\citet{2020MNRAS.495.3576K} report on a jet geometry transition at $7.0\pm0.5$~mas, indicating that the jet is freely expanding beyond this region. This coincides well with our measures of $r_{\rm b, R}=7.16\pm0.18$~mas and $r_{\rm b, T}=7.81\pm0.0.03$~mas.
\citet{2018AA...610A..32B} detect a sudden decrease of the feature size at a distance of about 3\,mas, which is accompanied by an increase of $T_{\rm b}(r)$ and polarized flux density, which was interpreted as a recollimation shock.
Due to the significantly large amount of data over multiple epochs, it is difficult to spot a localized bump around 3-4\,mas. If we mark the trajectories of individual jet features (Fig.~\ref{fig:pa}), then a zigzag profile in this region becomes apparent.
This variation coincides with a clear change in the jet PA by about 20\degr{}.
The slope of $T_\mathrm{b}(r)$ after the break at 7\,mas is $s=-4.36\pm0.24$. Such a steep gradient may indicate  a shocked decelerating jet after recollimation.

Additionally, \citet{2017ApJ...846...98J} and \citet{2021ApJ...923...30L} detected multiple stationary components within  1\,mas of the core.
Due to a large spread of the data points, we are unable to track any notable variations in distributions at this position, if they are present.

\subsection*{3C~120 (0430$+$052)}
Thanks to the long-term monitoring, this radio galaxy can be studied in detail. First of all, this is one of the jets which shows a geometry transition \citep{2020MNRAS.495.3576K} at a position of $2.7\pm0.4$\,mas.
We note that \citet{2000Sci...289.2317G, 2008ApJ...681L..69G} report on an interaction between the jet components and the external medium or a cloud at the same jet position, consistent with the idea of a precessing jet.
Due to the large scatter of the data points, our automated routine yields $r_{\rm b, R}=5.0\pm0.4$~mas and $r_{\rm b, T}=12.47\pm0.10$~mas.
If we consider individual components and PA variations, a bump in the $T_\mathrm{b}(r)$ around 3\,mas is clearly visible.

There is a notable zigzag pattern around 10~mas in the jet. \citet{2011ApJ...733...11G} reported significantly increased emission at a position  12\,mas from the core. They observed a sign reversal in the Faraday rotation measure at this position when compared to the remaining jet. The RM-corrected direction of linear polarization is the same as for the whole jet, i.e. perpendicular to the jet axis. The stacked image shows that the 3C~120 jet becomes near-edge-brightened at this position \citep{2017MNRAS.468.4992P}. The kinematics show a hint for a stationary component at this position \citep{2021ApJ...923...30L}. Interesting to note, the polarization and the total intensity images of 3C~120 are reminiscent of those of 3C~273 (component Q1 at around 10~mas, \citealt{2005ASPC..340..171A}), as well as the brightness temperature distribution. Thus, a helical instability pattern or jet rotation could produce variation in our distributions.
Beside the breaks discussed above, \citet{2015ApJ...808..162C} and \citet{2017ApJ...846...98J} reported on  multiple stationary components within  1\,mas of the core, which can be traced from enhanced $T_{\rm b}(r)$.

\subsection*{0738$+$313}
\label{sec:0738}
The $T_{\rm b}(r)$ and $R(r)$ distributions show one of the most pronounced zigzag profiles in our sample. In sections~\ref{sec:nonradial} and \ref{sec:test-realpath}, we  mentioned that its jet is sharply bent by $\thicksim40$\degr{} at 3\,mas downstream the 15\,GHz core, accompanied by the flattening of a spectral index \citep{2014AJ....147..143H}.
At the position of the bend, we detect a break in both distributions at $r_{\rm b,T}=3.83\pm0.03$ and $r_{\rm b,R}=3.23\pm0.07$. The Faraday RM value \citep{2012AJ....144..105H} significantly decreases from $(-373\pm84)$\,rad~m$^{-2}$ before the bend to $(-46\pm103)$\,rad~m$^{-2}$ downstream.
This is consistent with a scenario where the jet makes a large angle to the LOS before the bend, and, therefore, more external medium and a higher value of the rotation measure is observed there. After the bend, the jet becomes more aligned with the LOS and smaller RM is seen.
An increase of $T_{\rm b}(r)$ and a decrease of $R(r)$ after the break support this interpretation.

\subsection*{0836$+$710}
$R(r)$ and $T_{\rm b}(r)$ show complex behaviour beyond 2\,mas from the 15\,GHz core, with two extrema around 3 and 10\,mas.
\citet{2012ApJ...749...55P} observed helical structures propagating at relativistic speeds in this jet, suggesting the presence of helical and elliptical surface modes of the Kelvin-Helmholtz instability. 
Displacements of an emitting pattern in the jet relative to the line of sight would yield boosting (or de-boosting) which could explain the complex $R(r)$ and $T_{\rm b}(r)$ distributions.

\subsection*{1055$+$018}
This is the first source where the spine-sheath structure was observed \citep{1999ApJ...518L..87A}. We see significant deviations from a power law in $R(r)$, $T_{\rm b} (r)$ and $T_{\rm b} (R)$ at 5\,mas from the 15\,GHz core.  The 5 and 8.4\,GHz polarized images are determined by a sheath downstream at 4\,mas, which is not visible in the spine-dominated upstream region \citep{2005MNRAS.356..859P}. 
Also, the structure of the jet upstream of this break is represented by two stationary jet features \citep{2017ApJ...846...98J, 2021ApJ...923...30L}.

\subsection*{3C~273 (1226$+$023)}
\label{sec:3c273}
This quasar shows the most intriguing profiles.
Using our automated routine, we found a break at $r_{\rm b,R}=(4.7\pm0.2)$\,mas and $r_{\rm b,T}=(5.22\pm0.03)$\,mas. Although, if we highlight individual components and plot their position angles (see Fig.~\ref{fig:pa}), it appears that different jet features travel along their own zigzag trajectories on the $R(r)$ and $T_\mathrm{b}(r)$ planes. Extrema in these distributions are seen already at distances about 2\,mas from the core.
This coincides well with the region where the jet emission changes from being center-brightened to near-edge-brightened \citep{2012IJMPS...8..265G}.
At the same time, the analysis of the pc-to-kpc jet structure \citep{2022ApJ...940...65O} indicates a shape transition from  parabolic to conical streamlines beyond 15~mas.
Although, \citet{2022ApJ...940...65O} excluded the $\thicksim7-15$~mas region from their analysis because of a change in PA of about 20\degr{}. This is the region where different jet features follow their own trajectories in our distributions, but due to a significantly large number of observations, we can trace PA variations in excess of 35\degr{}.
In view of this complex behaviour, it is difficult to draw firm conclusions about the region of jet recollimation. Our estimate of $d=0.57\pm0.02$ up to $\thicksim5$~mas agrees well with that of \citet{2022ApJ...940...65O}, $d=0.66$.

The 3C~273 jet is also a well-established case of a transverse Faraday RM gradient \citep{2002PASJ...54L..39A, 2021ApJ...910...35L}. 
Recently, \cite{2021AA...654A..27B} detected a limb-brightened structure using 1.6\,GHz {\it RadioAstron} observations.
The transverse asymmetric Stokes $I$ profiles were seen before in the 3C\,273 jet at 43\,GHz \citep{2006A&A...446...71S, 2012IJMPS...8..265G}.
All these observations can be explained by a stratification in total and linearly polarized emission across the jet width due to a helical magnetic field. 
Alternatively, these could be the result of plasma instability patterns developing at the jet boundary \citep{2001Sci...294..128L}.
One of the plausible scenarios which can produce evolving zigzag profiles (Fig.~\ref{fig:pa}) is the rotation of the jet around its axis. Helical patterns then will make different angles to the LOS at different epochs of observations, and their Doppler factors will differ.

\subsection*{3C~279 (1253$-$055)}
This quasar exhibits one of the most pronounced zigzag profiles in $T_{\rm b}(r)$ and $R(r)$, similar to those of 3C~273 (Fig.~\ref{fig:tbgrad}).
On the stacked total intensity image \citep{2017MNRAS.468.4992P}, the jet of 3C~279 becomes far-edge-brightened exactly at the position of $r_{\rm break}\approx3$~mas.
Recent  1.3~mm VLBI observations have shown a systematic change in the source structure, suggesting travelling shocks or instabilities in a bent, possibly rotating the jet \citep{2020A&A...640A..69K}. This is consistent with the continuous variation of the jet PA seen at mas-scales \citep{2021ApJ...923...30L}.
The behaviour of the source at optical and radio wavelengths supports either a predominantly helical magnetic field or motion of the radiating plasma along a spiral path \citep{2020MNRAS.492.3829L}. 
Therefore, the break in our distributions can be explained by the change in the LOS of the emitting regions.
This model would also explain different enhancement of $T_{\rm b}$ at about 3~mas for different jet components, which progressively decreases with time (Fig.~\ref{fig:pa}).

\citep{2003ApJ...589L...9H} argued that component 1 changed in apparent projected direction, becoming more aligned with our line of sight after a bend. This event can be associated with the bump in $T_{\rm b}$ vs. distance seen in Fig.~\ref{fig:pa}, characterized by a steeper slope after the break and supporting the LOS scenario.

\subsection*{OR~186 (1551$+$130)}
The source exhibits complex profiles with apparent multiple zigzag patterns. The jet appears helically twisted in the stacked image \citep{2023MNRAS.520.6053P} accompanied by oscillating PA variations of different jet features.

\subsection*{3C~345 (1641$+$399)}
The brightness temperature gradient along the jet reveals a broken power-law behaviour with a break distance of $\sim 5$ mas and the two slopes $d_1 =-1.85 \pm 0.03$ and $d_2=-2.5\pm 0.3$.
Using the spectral evolution of a jet component, \citet{1999ApJ...521..509L} found evidence of a change from the synchrotron to the adiabatic stage at a distance of 1.2--1.5~mas from the core, which suggests a transition from Compton to a synchrotron stage.
Notable variations in our distributions of different jet features are visible in this jet region.

The recent analysis of the long-term 43\,GHz evolution of the 3C~345 jet components showed that they move on helical paths \citep{2024A&A...684A.211R}. This, for example, can be threads produced by a Kelvin-Helmholtz instability which were recently revealed in the \textit{RadioAstron} observations of the 3C~279 jet \citep{2023NatAs...7.1359F}. 
Non-monotonic gradient of the $T_\mathrm{b}(r)$ then can be produced by an interaction of a Kelvin-Helmholtz thread with a shock \citep{1995Ap&SS.234...49R}.

\subsection*{1716$+$686}
\label{sind:1716}
\begin{figure}
\centering
    \includegraphics[width=0.9\columnwidth]{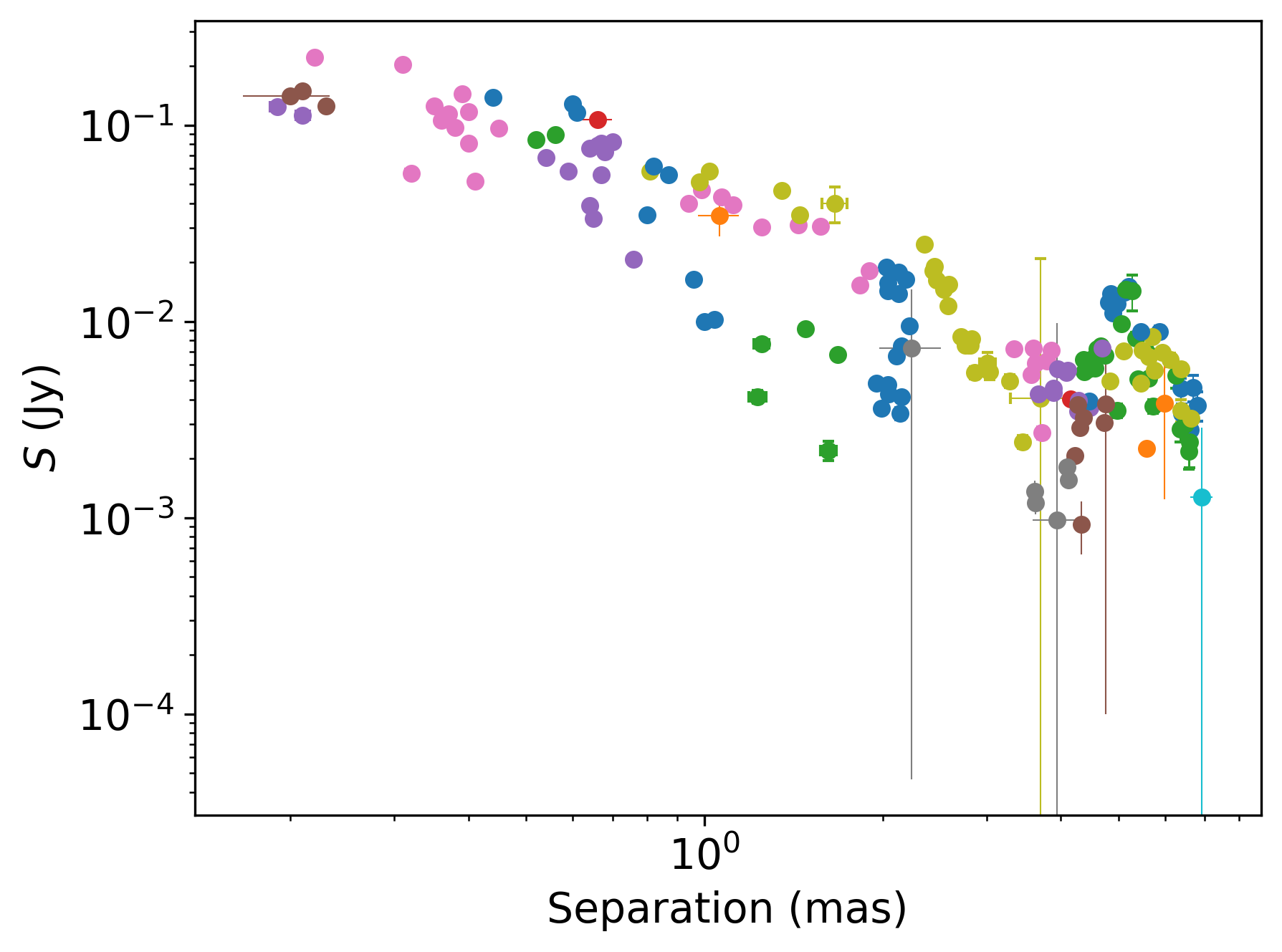}
    \caption{Flux density of the jet components versus the radial distance from the 15\,GHz core in quasar 1716+686. Different colours denote individual components.}
    \label{fig:1716flux}
\end{figure}

This quasar shows the best example of zigzag profiles with multiple humps. The jet appears helical in the stacked Stokes $I$ image \citep{2023MNRAS.520.6053P}. To stress, if $T_{\rm,b}(r)$ variations are not solely due to change in the component size, in Fig.~\ref{fig:1716flux}, we plot the flux density gradient along the jet for different components. It is clear that $T_{\rm b}(r)$, $R(r)$ and $S(r)$ change coherently and are likely being produced by a change of the Doppler factor.
\citet{2021ApJ...923...30L} reported on multiple non-radial trajectory components in this jet. This indicates directional changes of the jet features.

\subsection*{3C~380 (1828$+$487)}
This quasar exhibits one of the complex distributions, reminiscent of a zigzag pattern for individual jet features. Its stacked Stokes $I$ image shows an asymmetric edge-brightened profile beyond 2\,mas, where emission starts to leak from the near toward the far side of the jet. Additionally, \citet{2014MNRAS.438L...1G} reported a transverse RM gradient in the jet. These facts point toward a helical magnetic field, which, in turn, can explain the observed distribution variations.

\subsection*{3C~395 (1901$+$319)}
The variation of both $T_{\rm b}(r)$ and $R(r)$ distributions is connected with the distant stationary component, at a distance of about 15\,mas. 

\subsection*{Cygnus~A (1957$+$405)}
We see significant evidence for a break in all considered distributions. For the component size, we estimate $r_{\rm b,R}=1.68\pm0.09$\,mas, $d_1\approx0.0$ and $d_2=0.81\pm0.11$. Using the brightness temperature distribution, we obtain $r_{\rm b,T}=2.72\pm0.03$\,mas, $s_1=-2.8\pm0.5$ and $s_2=-2.2\pm0.3$.
These values are quite consistent with the results of \citet{2016AA...585A..33B} and \citet{2019ApJ...878...61N}, who observed the narrowing of the jet at 2\,mas and the formation of a stationary feature at this position. They explain these changes by active collimation.
Meanwhile, these authors estimated $d_1=0.55$ within the inner 2\,mas and $d_2=0.56$ beyond.
Most likely, this discrepancy with our estimates of the jet width profile arises because we lack jet features at distances of 3-5\,mas in our data set.

\subsection*{BL~Lac (2200$+$420)}
\label{sind:2200}

\citet{2020MNRAS.495.3576K} reported a geometry transition in this blazar jet at $\thicksim2.5$\,mas using 1.4 and 15\,GHz VLBA observations. They estimated the indices $d_1\thicksim0.53$ and $d_2\thicksim1.1$ before and after the break, respectively.
Our jet expansion profile given in Fig.~\ref{fig:srgrad} is well consistent with the result of \citet{2020MNRAS.495.3576K}. 
The automated fit yields $r_{\rm b,R}=2.43\pm0.02$\,mas and $r_{\rm b,T}=1.33\pm0.03$\,mas. Visually, the jet reconfines in between these estimates, at $\lesssim 2$\,mas.
We defined $d_1=0.75\pm0.02$ and $d_2=1.05\pm0.06$. 
Meanwhile, the power-law index values of the brightness temperature evolution before and after the break are well consistent with each other within the errors, $s_{1,2}\thicksim-4$, consistent with adiabatic losses being the dominant energy loss mechanism.
Recent high-resolution 43 and 86\,GHz VLBI data analysis provides a hint for a more complex jet profile upstream 1.5\,mas of the jet \citep{2021AA...649A.153C}. This could be created by another process of jet reconfinement, ending up with the formation of a recollimation shock at around 1.5\,mas. 
The \textit{RadioAstron} polarimetric space VLBI observations \citep{2016ApJ...817...96G} suggest that the jet of BL~Lac contains another set of recollimation shocks within the 40--250~$\mu$as upstream of the
radio core.
Such a pattern of recollimation shocks is consistent with the scenario where the jet propagates through an ambient medium with a decreasing pressure gradient \citep{2016ApJ...817...96G}.

\subsection*{CTA~102 (2230$+$114)}
The variation in the brightness temperature along the jet is complex and reflects the change in the component size: the slopes of distributions change in different jet regions within 1--10\,mas.
These profiles are well consistent with the results of a detailed study by \citet{2013AA...551A..32F}, who interpreted the radial evolution of the jet parameters by recollimation shocks which produce a transverse pressure gradient and lead to a jet bend. This is supported by a deceleration and change in the viewing angle of the flow. 
\citet{2013AA...551A..32F} suggested expansion and deceleration upstream of the jet, which ends up in the stationary component at around 2\,mas. Then there is a new process of jet recollimation within the $4\leq r\leq 8$\,mas region.
From our $R(r)$ profiles, it is unclear if the jet width decreases at $\thicksim2$\,mas, but it is clear that the jet reconfines at $\thicksim7$\,mas.
Besides, \citet{2012AJ....144..105H} reported on transverse RM gradient in the quasar jet, which points toward a helical magnetic field. This can partially explain the observed significant curvature of the jet \citep{2017MNRAS.468.4992P}.

\subsection*{3C~454.3 (2251$+$158)}
The $T_{\rm b}(r)$ and $R(r)$ distributions beyond $\thicksim 2$\,mas are complex, accompanied by a notable bump at $\thicksim 6$\,mas. This position coincides with the region of a significant jet bend. 
\citet{2013MNRAS.436.3341Z} showed that the asymmetric transverse structure of the quasar jet provides convincing evidence that it is threaded by a large-scale, ordered, helical magnetic field. This scenario would explain the  broken profiles.

\subsection*{4C~+45.51 (2351$+$456)}
The significant break in $R(r)$ at $r_{\rm b,R}=2.8\pm0.5$, which is accompanied by the break in $T_{\rm b}(R)$ at the same jet location, can be associated with the stationary component number 2 located at a mean distance of $r \approx 3.8$~mas. 